\pdfoutput=1
\documentclass[useAMS,usenatbib,iop,numberedappendix]{mn2e} 

\usepackage{mathptmx}
\usepackage{amsmath,amsfonts,amssymb}
\usepackage{mathrsfs}
\usepackage{makeidx}
\usepackage{graphicx}
\usepackage{epstopdf}
\usepackage[colorlinks=True]{hyperref}
\usepackage{multirow}
\usepackage{rotating}
\usepackage{color}
\usepackage{tablefootnote}
\usepackage{lscape}
\usepackage{subcaption}
\usepackage{tablefootnote}
\usepackage[T1]{fontenc}
\usepackage{siunitx}

\definecolor{inon}{rgb}{1.00,0.27,0.00}

\author[Chiu et al.]{
\parbox{\textwidth}{
\LARGE
I-Non Chiu$^{1}$,
Keiichi Umetsu$^{1}$,
Ryoma Murata$^{2,3}$,
Elinor Medezinski$^{4}$,
\\
Masamune Oguri$^{2,3,5}$
}
\vspace{0.4cm}
\\
\parbox{\textwidth}{
$^{1}$ Academia Sinica Institute of Astronomy and Astrophysics (ASIAA), 11F of AS/NTU Astronomy-Mathematics Building, No.1, Sec. 4, Roosevelt Rd, Taipei 10617, Taiwan\\
$^{2}$ Kavli Institute for the Physics and Mathematics of the Universe (WPI), The University of Tokyo Institutes for Advanced Study (UTIAS), The University of Tokyo, 5-1-5 Kashiwanoha, Kashiwa-shi, Chiba, 277-8583, Japan\\
$^{3}$ Department of Physics, University of Tokyo, 7-3-1 Hongo, Bunkyo-ku, Tokyo 113-0033 Japan\\
$^{4}$ Department of Astrophysical Sciences, Princeton University, Princeton, NJ 08544, USA\\
$^{5}$ Research Center for the Early Universe, University of Tokyo, Tokyo 113-0033,
Japan
}
}

\newcommand{\LCDM}{\ensuremath{\Lambda\textrm{CDM}}}
\newcommand{\OmegaM}{\ensuremath{\Omega_{\mathrm{M}}}}

\newcommand{\Hnow}{\ensuremath{H_{0}}}
\newcommand{\seight}{\ensuremath{\sigma_{8}}}
\newcommand{\Msun}{\ensuremath{\mathrm{M}_{\odot}}}

\newcommand{\Rfiveoo}{\ensuremath{R_{500}}}
\newcommand{\Mfiveoo}{\ensuremath{M_{500}}}

\newcommand{\redshift}{\ensuremath{z}}
\newcommand{\dif}{\ensuremath{\mathrm{d}}}

\newcommand{\HST}{\emph{HST}}

\newcommand{\zd}{\ensuremath{z_{\mathrm{d}}}}

\newcommand{\zcl}{\ensuremath{z_{\mathrm{cl}}}}

\newcommand{\shearone}{\ensuremath{\gamma_{1}}}
\newcommand{\sheartwo}{\ensuremath{\gamma_{2}}}
\newcommand{\shear}{\ensuremath{\gamma}}
\newcommand{\sigmam}{\ensuremath{\Sigma_{\mathrm{m}}}}
\newcommand{\sigmacrit}{\ensuremath{\Sigma_{\mathrm{c}}}}
\newcommand{\rhocrit}{\ensuremath{\rho_{\mathrm{c}}}}
\newcommand{\dl}{\ensuremath{D_{\mathrm{l}}}}
\newcommand{\ds}{\ensuremath{D_{\mathrm{s}}}}
\newcommand{\dls}{\ensuremath{D_{\mathrm{ls}}}}

\newcommand{\MPIV}{\ensuremath{M_{\mathrm{piv}}}}
\newcommand{\ZPIV}{\ensuremath{z_{\mathrm{piv}}}}

\newcommand{\rich}{\ensuremath{N}}
\newcommand{\mcut}{\ensuremath{m_{\mathrm{cut}}}}

\newcommand{\lowz}{low-\ensuremath{z}}
\newcommand{\hghz}{high-\ensuremath{z}}

\newcommand{\Arich}{\ensuremath{A_{N}}}
\newcommand{\Brich}{\ensuremath{B_{N}}}
\newcommand{\Crich}{\ensuremath{C_{N}}}
\newcommand{\Drich}{\ensuremath{\sigma_{N}}}

\newcommand{\rtm}{\ensuremath{N}--\ensuremath{M}}

\newcommand{\nd}{\ensuremath{n_{\mathrm{d}}}}
\newcommand{\nzero}{\ensuremath{n_{0}}}

\newcommand{\slope}{\ensuremath{\alpha}}

\newcommand{\fcl}{\ensuremath{f_{\mathrm{cl}}}}
\newcommand{\fmk}{\ensuremath{f_{\mathrm{mask}}}}
\newcommand{\fem}{\ensuremath{f_{\mathrm{res}}}}

\newcommand{\percent}{\ensuremath{\%}}

\newcommand{\appropto}{\mathrel{\vcenter{
  \offinterlineskip\halign{\hfil$##$\cr
    \propto\cr\noalign{\kern2pt}\sim\cr\noalign{\kern-2pt}}}}}

\title[Weak Lensing Magnification in the Subaru HSC Survey]{
The Richness-to-Mass Relation of CAMIRA Galaxy Clusters from Weak-lensing Magnification in the Subaru Hyper Suprime-Cam Survey
}

\hypersetup{linkcolor=blue,filecolor=magenta,citecolor=cyan}

\begin{document}
\pdfpageheight 11.7in
\pdfpagewidth 8.3in

%%%%%%%%%%%%%%%%%%%%%%%%%%%%%%%%%%%%%%%%%%
%
% TITLE
%
%%%%%%%%%%%%%%%%%%%%%%%%%%%%%%%%%%%%%%%%%%

\maketitle 

%%%%%%%%%%%%%%%%%%%%%%%%%%%%%%%%%%%%%%%%%%
%
% ABSTRACT
%
%%%%%%%%%%%%%%%%%%%%%%%%%%%%%%%%%%%%%%%%%%

\begin{abstract}
We present a statistical weak-lensing magnification analysis on an optically selected sample of 3029 \texttt{CAMIRA} galaxy clusters with richness $\rich>15$ at redshift $0.2\leq\redshift<1.1$ in the Subaru Hyper Suprime-Cam (HSC) survey. 
We use two distinct populations of color-selected, flux-limited background galaxies, namely the \lowz\ and \hghz\ samples at mean redshifts of $\approx1.1$ and $\approx1.4$, respectively, from which to measure the weak-lensing magnification signal by accounting for cluster contamination as well as masking effects. 
Our magnification bias measurements are found to be uncontaminated according to validation tests against the ``null-test'' samples for which the net  magnification bias is expected to vanish.
The magnification bias for the full \texttt{CAMIRA} sample is detected at a significance level of $9.51\sigma$, which is dominated by the \hghz\ background. 
We forward-model the observed magnification data to constrain the normalization of the richness-to-mass (\rtm) relation for the \texttt{CAMIRA} sample with informative priors on other parameters.
The resulting scaling relation is $N\propto {\Mfiveoo}^{0.92\pm0.13} (1 + \redshift)^{-0.48\pm0.69}$, with a characteristic richness of $N=\left(17.72\pm2.60\right)$ and intrinsic log-normal scatter of $0.15\pm0.07$ at $\Mfiveoo = 10^{14}h^{-1}\Msun$.
With the derived \rtm\ relation, we provide magnification-calibrated mass estimates of individual \texttt{CAMIRA} clusters, with the typical uncertainty of $\approx39\percent$ and $\approx32\percent$ at richness$\approx20$ and  $\approx40$, respectively.
We further compare our magnification-inferred \rtm\ relation with  those from the shear-based results in the literature, finding good agreement. 
\end{abstract}

%%%%%%%%%%%%%%%%%%%%%%%%%%%%%%%%%%%%%%%%%%
%
% KEYWORDS
%
%%%%%%%%%%%%%%%%%%%%%%%%%%%%%%%%%%%%%%%%%%


%%%%%%%%%%%%%%%%%%%%%%%%%%%%%%%%%%%%%%%%%%
%
% INTRODUCTION
%
%%%%%%%%%%%%%%%%%%%%%%%%%%%%%%%%%%%%%%%%%%

\section{Introduction}
\label{sec:introduction}

Galaxy clusters, as local peaks of cosmic density perturbations, are powerful cosmological tools because they provide a representative view
of non-linear growth of structure over cosmic time. 
Cosmological probes based on galaxy clusters enable independent tests to examine any viable cosmological models and constrain fundamental properties of the universe, such as the amplitude of cosmic inhomogeneity, the growth of cosmic structure, and the equation of state of dark energy \citep[e.g .,][]{wang1998,holder01b}.
Measurements of the cluster abundance as a function of redshift in large cluster surveys have been used in cosmological studies with promising
success \citep{benson13,planck2015clstrcnts,bocquet15,deHaan16,bocquet19}, demonstrating that the constraining power of galaxy clusters is as competitive as other complementary probes, such as observations of the cosmic microwave background (CMB) and the large-scale clustering of galaxies.
However, one of the most substantial limitations of utilizing galaxy clusters as a feasible cosmological tool is to obtain an accurate and precise observable-to-mass scaling relation.
This is because the cluster mass is not directly observable and must be inferred from a mass proxy---an observed quantity that is well correlated with the underlying cluster mass.

Various observable properties of galaxy clusters can be used as a mass proxy, such as the X-ray luminosity and temperature due to the Bremsstrahlung emission from the hot intra-cluster medium (ICM). 
In the context of cluster galaxy populations, the richness \rich---the overdensity of red-sequence galaxies---is often used as a mass proxy for galaxy clusters selected in optical imaging surveys.
Although optical richness serves as a low-scatter mass proxy \citep{rykoff12}, an unbiased calibration of the cluster mass is needed to accurately anchor the richness-to-mass (\rtm) scaling relation.  
With the goal of achieving an accurate mass calibration, a number of observational approaches are available, including those based on the velocity dispersion of member galaxies \citep{saro13}, the dynamical Jeans analysis \citep{capasso19a,capasso19b}, X-ray observations assuming hydrostatic equilibrium \citep{vikhlinin06,vikhlinin09a,martino14}, weak gravitational  lensing \citep[hereafter weak lensing;][]{umetsu14,vonderlinden14a,hoekstra15,melchior17,dietrich19,mcclintock18,murata18}, and so on. 
Among all these probes, weak lensing is regarded as the most direct approach to calibrating the cluster mass with near zero bias, because it is free from any assumption about the dynamical state of clusters and is only sensitive to the underlying mass distribution.
Accordingly, weak lensing has been extensively used as a direct mass
probe for cosmological studies, albeit with large scatter \citep{mantz15,grandis18,bocquet19}.

Gravitational lensing due to mass inhomogeneities along the line of sight deflects light from distant background sources, resulting in various observable effects, such as strong lensing, weak-lensing shear, and weak-lensing magnification.
The weak shear effect introduces a small but coherent change in the observed ellipticities of background images.
The effect of lensing magnification enlarges (reduces) the observed solid angle on the sky, which results in an increase (decrease) of the total flux of background sources \citep[for more details, see][]{bartelmann01}.  

Over the last two decades, tremendous efforts have been made to standardize the weak shear effect as a reliable mass calibrator through systematic oberving campaigns \citep[e.g.,][]{vonderlinden14a, umetsu14, hoekstra15, okabe16, sereno17, stern18, dietrich19,miyatake19}, with an aid of intensive image simulations \citep{massey07,bridle10,mandelbaum15,hoekstra17}. 
On the other hand, there have been relatively less attempts on using lensing magnification for cluster mass measurements, which is mainly due to the fact that the signal-to-noise ratio of magnification measurements is significantly less than those from shear-based measurements \citep{schneider00}. 
Although the signal-to-noise ratio of lensing magnification is low on an individual cluster basis, stacking a sizable sample of galaxy clusters allows us to overcome this problem, providing a precise mass calibration. 
It is worth emphasizing that this stacking strategy is becoming progressively valuable and competitive because of ongoing and forthcoming large cluster surveys \citep[e.g., the Dark Energy Survey;][]{DES05}, as demonstrated in recent studies of  CMB cluster lensing \citep[e.g.,][]{baxter18} and dynamical analysis \citep[e.g.,][]{capasso19b}.
Alternatively, lensing magnification can be used in combination with weak lensing shear to perform a joint reconstruction of the cluster mass distribution 
\citep{schneider00,umetsu08,umetsu13,umetsu18,chiu18b}, effectively breaking degeneracies inherent in a standard shear-only analysis \citep{bartelmann01}.

Recently, high signal-to-noise measurements of lensing magnification have been obtained by stacking over large cluster samples, from which to calibrate the cluster mass with better precision and accuracy \citep[e.g.,][]{hildebrandt09,ford12,ford13,chiu16b}.
It is also worth mentioning that the magnification-induced change in the size and flux of background sources has been used to constrain the mass of foreground lens \citep{schmidt12, duncan16}, offering a new window to the cluster mass calibration.
In contrast to the standard shear-based analysis, measuring the effect of lensing magnification does not require source galaxies to be spatially resolved.  However, it does require accurate photometry and a stringent flux limit against incompleteness effects \citep{umetsu13}. 
Hence, lensing magnification has been extended to measure the mass of galaxy clusters at high redshift \citep[$\redshift\approx1.4$;][]{tudorica17}, for which the shape measurement of faint source galaxies required by shear-based methods is extremely challenging with ground-based observations. 
In this context, lensing magnification is unique and attractive for  mass calibration of galaxy clusters at high redshift,  especially for ongoing and upcoming wide and deep lensing surveys. 

In this study, we aim to extract the lensing magnification signal around a large statistical sample of optically selected galaxy clusters from the ongoing Hyper Suprime-Cam (HSC) survey \citep{aihara18a}.   
We then calibrate the cluster mass scale of the selected sample over a wide range of richness and redshift using lensing magnification alone.  
Here we focus on the effects of ``flux magnification bias'' (see more details in Section~\ref{sec:basics}), where we extend the approach of \cite{chiu16b}. 
Specifically, we select two distinct populations of background galaxies behind the clusters in color-color space and measure their  projected number density contrast relative to random fields. 
The most notable improvement over the previous work is that we use a forward-modelling approach (Section~\ref{sec:sr_modelling}) to jointly constrain the underlying observable-to-mass relation, namely the \rtm\ relation.

This paper is organized as follows.
A brief overview of cluster lensing magnification is given in Section~\ref{sec:basics}. 
Section~\ref{sec:data} describes the HSC survey and data.
The cluster sample we use is described in Section~\ref{sec:cluster_sample}. 
We detail the analysis of lensing magnification and the mass calibration procedures in Section~\ref{sec:analysis}.
We present our results in Section~\ref{sec:results}.
Discussion of systematic uncertainties is given in Section~\ref{sec:sys}. 
Finally, a summary and conclusions are provided in Section~\ref{sec:conclusions}.

Throughout this paper, we assume a flat \LCDM\ cosmology with
$\OmegaM=0.3$,
$\Hnow = h\times100$\,km\,s\,$^{-1}$\,Mpc$^{-1}$ with
$h=0.7$, and $\seight = 0.8$, the rms amplitude of linear mass fluctuations in a sphere of comoving radius $8h^{-1}$\,Mpc. 
We adopt the standard notation $\Mfiveoo$ to denote the mass enclosed within a sphere of radius $\Rfiveoo$ within which the mean overdensity equals to $500$ times the critical density $\rhocrit(\redshift)$ of the
universe at the cluster redshift.
Unless otherwise stated, all quoted errors represent the $68\percent$ confidence level (i.e., $1\sigma$). We use the AB magnitude system in photometry.
The notation $\mathcal{N}(x,y^2)$ stands for a normal distribution with the mean $x$ and the standard deviation $y$. 

%%%%%%%%%%%%%%%%%%%%%%%%%%%%%%%%%%%%%%%%%%
%
% Theory
%
%%%%%%%%%%%%%%%%%%%%%%%%%%%%%%%%%%%%%%%%%%

\section{The Basics of Gravitational Lensing}
\label{sec:basics}

A brief review of gravitational lensing, with emphasis on the weak lensing effect of magnification bias around galaxy clusters, is provided in this section.
For more details, we refer the reader to \cite{bartelmann01}, \cite{umetsu10} and \cite{hoekstra13}.

In the limit of thin lens approximation, a galaxy cluster at redshift \zd\ is considered as a single lens embedded in a homogeneous universe where background sources behind the cluster are all gravitationally lensed. 
To the first order, the true and observed angular positions of the source, denoted by $\vec{\beta}$ and $\vec{\theta}$ at the source and observed planes, respectively, can be related to each other by the lensing Jacobian matrix $J$, defined by
\begin{equation}
\label{eq:jacobian}
J(\vec{\theta}) \equiv \frac{\partial\vec{\beta}}{\partial\vec{\theta}} = 
\begin{pmatrix}
1 - \kappa - \shearone & -\sheartwo \\
-\sheartwo & 1 - \kappa + \shearone
\end{pmatrix},
\end{equation}
where $\kappa$, $\shearone$, and $\sheartwo$ are linear combinations of second derivatives of the effective lensing potential.  
The lensing convergence, $\kappa(\vec{\theta})$, is a dimensionless version of the projected surface mass density $\sigmam(\vec{\theta})$, defined by
\begin{equation}
\label{eq:kappa_def}
\kappa(\vec{\theta}) = \frac{\sigmam(\vec{\theta})}{\sigmacrit},
\end{equation}
where $\sigmacrit$ is the critical surface density defined as $\sigmacrit\equiv\frac{c^2}{4\pi G} \frac{\ds}{\dl\dls}$, which depends on the angular diameter distances of the observer-to-cluster (\dl), observer-to-source (\ds), and the cluster-to-source (\dls) pairs, respectively; $G$ is the Newton's constant, and $c$ is the speed of light. 
The surface mass density $\sigmam(\vec{\theta})$ is the projected density at the angular position of $\vec{\theta}$ integrated along the line of sight.

In the subcritical-lensing regime, lensing magnification changes the flux of background sources by a factor of $\mu$, where $\mu$ is the inverse determinant of the Jacobian matrix:
\begin{equation}
\label{eq:mu_def}
\mu = \frac{1}{\left(1 - \kappa\right)^2 - |\shear|^2} \, .
\end{equation}
In the limit of $\kappa\approx|\shear|\ll 1$ and $\kappa>0$ as
expected in an over-dense environment around clusters, the observed surface density of the sources above a flux threshold increases because of flux amplification.  
On the other hand, magnification also reduces the observed area on the source plane given a solid angle, such that the surface number density of sources given an observed solid angle is effectively decreased.
As a result, lensing magnification alters the surface number density of a ``flux-limited'' sample of background sources. 
Moreover, the net change depends on the intrinsic slope of the source luminosity function. 
This change in number counts caused by gravitational lensing is known as magnification bias \citep{broadhurst95,taylor98,umetsu13}.

The effect of magnification bias can be measured using cumulative number counts of a flux-limited sample of background galaxies. 
Assuming that the luminosity function of background galaxies can be locally approximated by a power-law function of flux around the flux limit, the magnification bias can be described by
\begin{equation}
\label{eq:magnification_bias}
\frac{\nd(<m)}{\nzero(<m)} = \mu^{\slope - 1} \, ,
\end{equation}
where $\nd(<m)$ is the lensed surface number density of background galaxies that are brighter than the magnitude limit $m$, $\nzero(<m)$ is its unlensed version, and \slope\ is the logarithmic count slope,
\[ 
\slope\equiv 2.5\frac{\dif\log n_{0}(<m)}{\dif{m}} \, .
\]
In the weak-lensing limit ($\kappa\ll1$ and $|\shear|\ll1$), equation~(\ref{eq:magnification_bias}) reduces to
\begin{equation}
\label{eq:magnification_bias_linear}
\frac{\nd(<m)}{\nzero(<m)} - 1 \approx 2(\slope - 1)\kappa \, .
\end{equation}
In the case of $\slope>1$, lensing magnification results in a net enhancement of the number density, and vice versa. 
When $\slope=1$, lensing magnification gives no net change in the source counts, even though background galaxies are magnified.
This is because the depletion of souce counts due to geometric expansion of the observed sky is in balance with the increase of souce counts due
to flux amplification. 
It should be noted that we use equation~(\ref{eq:magnification_bias}) without assuming the weak-lensing limit when interpreting the observed magnification profiles with the standard halo modeling (equation~(\ref{eq:magni_model})).

We note that a more advanced analysis can be used to apply different weights $w(m)$ to the source galaxies, which are experiencing different levels of magnification bias at different magnitudes \citep[see more details in][]{menard02}.
The approach used in this work is effectively identical to the case with no weights (or $w=1$).
In this case, the signal is dominated by the faint end where the number of galaxies is at least ten times more than the one at the bright end, as it is in this work.

To summarize, we can constrain the lensing convergence $\kappa$, and thus the cluster mass, by measuring the number density contrast of background galaxies with respect to random fields, once the unlensed count slope $\slope(m)$ as a function of the limiting magnitude $m$ is known. 
It is worth stressing again that this method does not require source galaxies to be fully resolved, thus providing an independent way to calibrate the cluster mass, complementary to the standard shear-based method.

\begin{figure}
\centering
\resizebox{0.5\textwidth}{!}{
\includegraphics[scale=1]{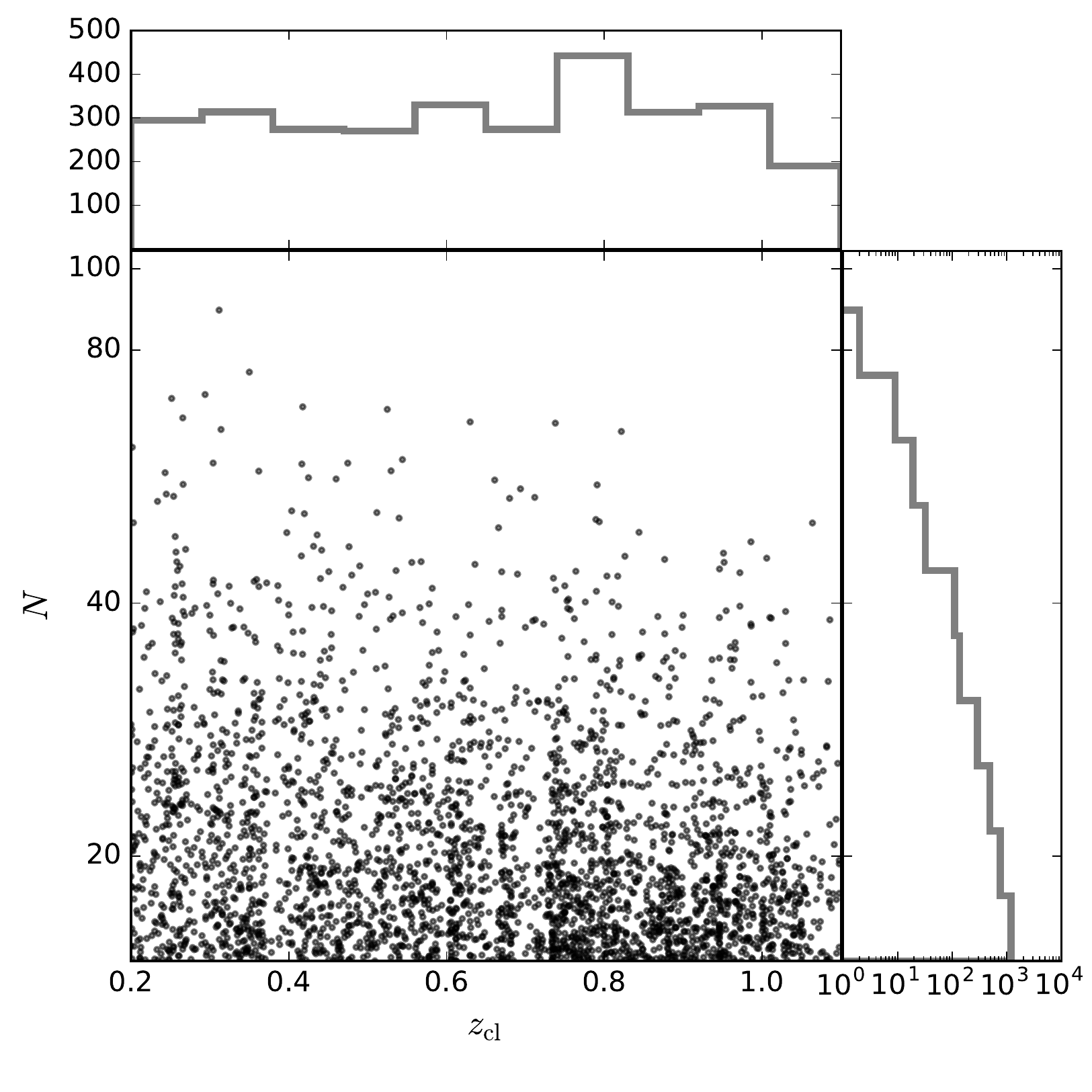}
}
\caption{
The cluster sample with $\rich\geq15$ at redshift $0.2\leq\redshift<1.1$ used in this work.
In total, there are $3029$ clusters.
The photometric redshift and the richness of the clusters are shown in the x- and y-axis, respectively. 
The histograms of the richness and redshift are presented in the right and upper panels, respectively.
}
\label{fig:sample}
\end{figure}
%

%%%%%%%%%%%%%%%%%%%%%%%%%%%%%%%%%%%%%%%%%%
%
% Data
%
%%%%%%%%%%%%%%%%%%%%%%%%%%%%%%%%%%%%%%%%%%

\section{Data}
\label{sec:data}

We use the optical photometry obtained from the Hyper Suprime-Cam (HSC) Survey \citep{aihara18a}, which is carried out in the framework of a Subaru
Strategic Program (SSP) using the newly installed wide-field camera Hyper Suprime-Cam \citep{miyazaki15} on the 8.2\,m Subaru Telescope.
A brief summary of the HSC survey is given below, and we refer the reader to \cite{aihara18a} for more details.

The HSC SSP is a five-year mission initialized in 2014 to survey a large sky area of 1400\,deg$^2$ in five broadband filters ($grizy$), with the goal of performing state-of-the-art weak lensing studies 
preparing for the era of the Large Synoptic Survey Telescope \citep[LSST;][]{ivezic08}.
The HSC survey is composed of three layers: Wide, Deep and Ultradeep.
We only use data from the Wide layer in this work, because it is designed to cover the area of 1400~deg$^2$ that is significantly larger than the Deep ($\approx25$~deg$^2$) and Ultradeep ($\approx3.5$~deg$^2$) layers.
The exposure time of the full depth in the Wide layer  is 10 (20) min for $g$- and $r$-band ($i$-, $z$-, and $y$-band).
The $5$ sigma limiting magnitudes around a $2\arcsec$ aperture are $26.5$~mag, $26.1$~mag, $25.9$~mag, $25.1$~mag, and $24.4$~mag for $g$-, $r$-, $i$-, $z$-, and $y$-band, respectively.
As a result, this represents the deepest optical multi-band survey over an area of more than one thousand square degrees to date. 
This unique combination of area and depth enables us to search for and characterize galaxy clusters out to a redshift of $z\approx1$ or beyond with an unprecedentedly statistical power.
In particular, the imaging of $i$-band is specifically taken under the good seeing condition ($<0.8\arcsec$), resulting in a mean seeing of $0.58\arcsec$ \citep{mandelbaum18}, which is excellent for weak-lensing studies, by design.

The imaging reduction and catalog construction of the HSC survey are processed by the \texttt{hscPipe v5.4} \citep{bosch18}, a precursor pipeline of the data reduction and management for the LSST survey \citep{axelrod10,juric15}.
Details of the \texttt{hscPipe} are fully given in \cite{bosch18}, and we only briefly summarize the key steps, as follows.

The source detection and cataloging are carried out in two phases.
In the first phase, \texttt{hscPipe} processes the single-epoch images in each band, involving several basic reductions, such as applying the flat-fielding correction and removing signatures of instrumental defects, followed by the initial calibration of astrometrics, photometry and the Point Spread Function (PSF) using the bright sources only.
In the second phase, the source detection is performed on the coadd images, and the photometry is measured using the ``forced'' mode with the $i$-band as the reference band.
The \texttt{cmodel} photometry \citep{lupton10}, which is the model-fitting photometry estimated by a composite template of an exponential profile and a de Vaucouleurs profile convolving with the locally registered PSF, is used in this work.
The performance of the \texttt{cmodel} photometry has been intensively verified in \cite{huang18}, showing that it can robustly deliver the unbiased estimates of color and magnitude for galaxies.
Since our analysis does not include the cluster cores, we do not use the PSF-matched aperture photometry, which is designed to improve the estimation of colors in extremely crowded fields.
More discussion about the PSF-matched aperture photometry can be found in \cite{aihara18b}.

The bright star masks with the version of \texttt{Arcturus} are applied, as described in \cite{coupon18}.
We only apply the star masks in the $i$- and $z$-bands.
The flag of $\mathtt{i\_extendedness\_value}==1$ is applied to separate galaxies from stars.
To construct the Full-Depth-Full-Color (FDFC) catalog, we apply the flags of $\mathtt{g[r]countinputs}\geq4$ and $\mathtt{i[zy]countinputs}\geq6$.
We further apply a magnitude cut in the $z$-band by the flag of $\mathtt{z\_cmodel\_mag}-\mathtt{a\_z}<26$ to remove the extremely faint objects, which do not have reliable photometry measurements and are not of interest in this work; we have confirmed that this cut does not affect our magnification measurements.
Similar to \cite{oguri18}, we also apply the quality cuts to discard the objects whose photometry measurements are severely affected by the defected pixels or cosmic rays.
A summary of the \texttt{sql} query is given in Table~\ref{tab:sql}.

The HSC survey is planned to be completed by the end of 2019.
There are three Public Data Release (PDR) for the HSC survey:
The first PDR took place in February 2017 \citep{aihara18b}, containing the initial FDFC footprint of an area of $\approx140$\,deg$^2$ taken up to 2016.
The second and third PDR are scheduled to be in 2019 and 2021, respectively.
In this work, we use the FDFC data taken up to 2017 (s17A) with an area of $\approx380$\,deg$^2$ (excluding the star-masked regions).

%%%%%%%%%%%%%%%%%%%%%%%%%%%%%%%%%%%%%%%%%%
%
% Sample
%
%%%%%%%%%%%%%%%%%%%%%%%%%%%%%%%%%%%%%%%%%%

\section{Cluster Sample}
\label{sec:cluster_sample}

In this work, the galaxy clusters are identified using the \texttt{CAMIRA} algorithm \citep[Cluster finding Algorithm based on Multi-band Identification of Red-sequence gAlaxies;][]{oguri14}.
We summarize the \texttt{CAMIRA} algorithm as follows, and refer the reader to \cite{oguri14} and \cite{oguri18} for more details.

\texttt{CAMIRA} is a matched-filter cluster finder to search for the overdensity of red-sequence galaxies based on the stellar population synthesis model, which is calibrated against a sample of  spectroscopically observed galaxies.
Each identified cluster is assigned with a photometric redshift estimate with accuracy better than $1\%$ and a mass proxy of richness (\rich) that is well correlating with the underlying cluster mass \citep{oguri18}.
We use the cluster catalog produced by the \texttt{CAMIRA}, which is run on the s17A FDFC footprint with an area of $\approx380$\,deg$^2$ excluding the star-masked regions.
This results in a sample of $3029$ clusters with $\rich\geq15$ at redshift of $0.2\leq\redshift<1.1$.
Figure~\ref{fig:sample} shows the cluster sample we use in this work.

We use the location of the Brightest Cluster Galaxy (BCG) as the cluster center.
Using a subset of clusters with available X-ray imaging, the center offset between the BCG and the X-ray peak is generally small ($\lesssim0.1$ Mpc/$h$) with a long tail toward the high offset, suggesting that miscentering is not significant \citep{oguri18}.
Moreover, lensing magnification is probing the surface mass density $\Sigma$, instead of the differential surface mass density $\Delta \Sigma$, thus is less sensitive to the miscentering effect \citep{ford13, ford16}.
Therefore, we ignore the miscentering effect in this work.

We use the photometric redshift estimation (\zcl) for each cluster, even for those clusters with available spectroscopic redshift $z_{\rm{spec}}$, in the interest of uniformity. 
With the HSC data, the performance of photometric redshift is excellent: the bias, scatter, and the outlier rate of $(\zcl - z_{\rm{spec}})/(1 + z_{\rm{spec}})$  are $-0.0013$, $0.0081$, and $1.7\percent$, respectively \citep{oguri18}.
Given this quality, hereafter we ignore the photometric redshift uncertainty, which has negligible impact on the final result.

\begin{figure*}
\centering
\resizebox{\textwidth}{!}{
\includegraphics[scale=1]{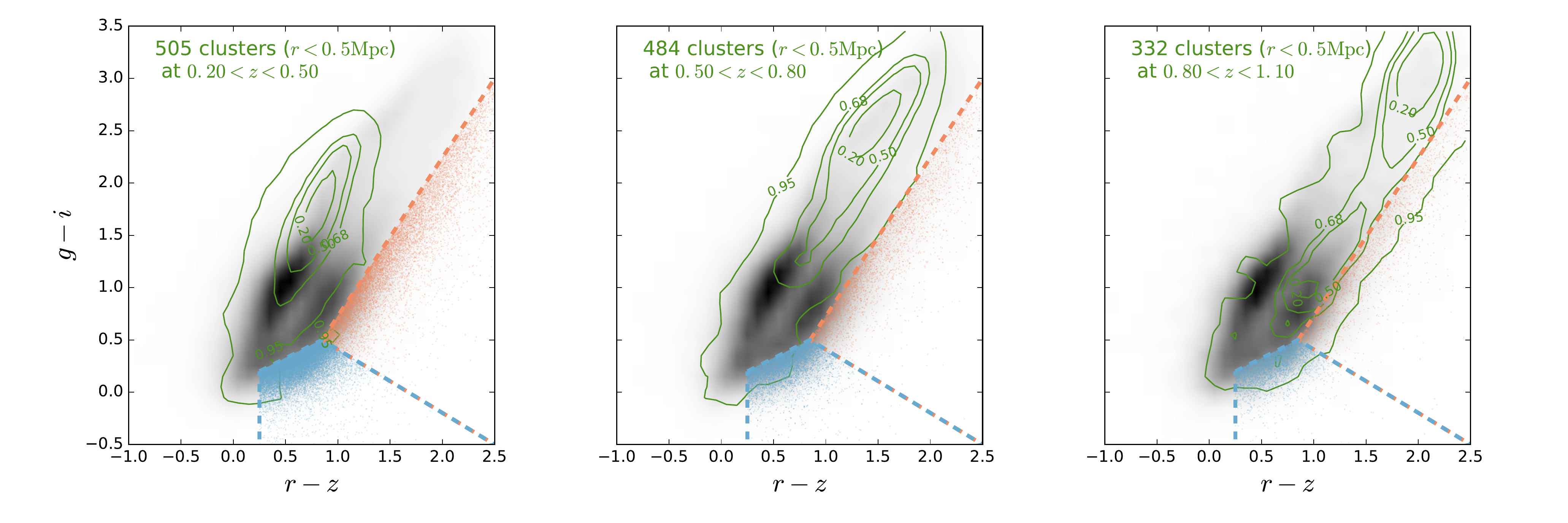}
}
\caption{
The color-color diagrams in $g-i$ vs $r-z$ showing 
(1) the selected \lowz\ and \hghz\ background populations in red and blue, respectively, 
(2) the distribution of the cluster galaxies as the green contours, and
(3) the smoothed distribution of galaxies in the random fields as the underlying grey image.
The cluster galaxies that lie projected within the clustercentric radius of $0.5$~Mpc around the BCGs are used to derive the distribution of cluster galaxies (green contours), after the statistical background subtraction using the random fields.
The random fields are randomly drawn from the HSC FDDFC footprint.
The labels in the green contours represent the enclosed percentage of the selected cluster galaxies.
The color-color selections for the \lowz\ and \hghz\ backgrounds are shown by the red and blue dashed lines, respectively.
The left to right panels show the results of lower, medium, and higher redshift bins, respectively.
These plots are produced by stacking the galaxies with $i$-band magnitude smaller than $24.5$~mag
in the cluster fields and the same amounts of the randomly drawn apertures, for which the redshift range and the number of the pointing are labelled in the upper-left corner in each panel.
We find clear concentration of cluster galaxies---which is the cluster red sequence---moving toward to the upper-right corner in the color-color diagrams as increasing cluster redshift.
On the other hand, a clear enhancement of galaxy concentration starts to appear at the colors of $g-i\approx1$ and $r-z\approx1$ for the high-redshift clusters at $\redshift\gtrsim0.8$, as an indication of increasing blue members in clusters.
}
\label{fig:sources}
\end{figure*}
\begin{figure*}
\centering
\resizebox{\textwidth}{!}{
\includegraphics[scale=1]{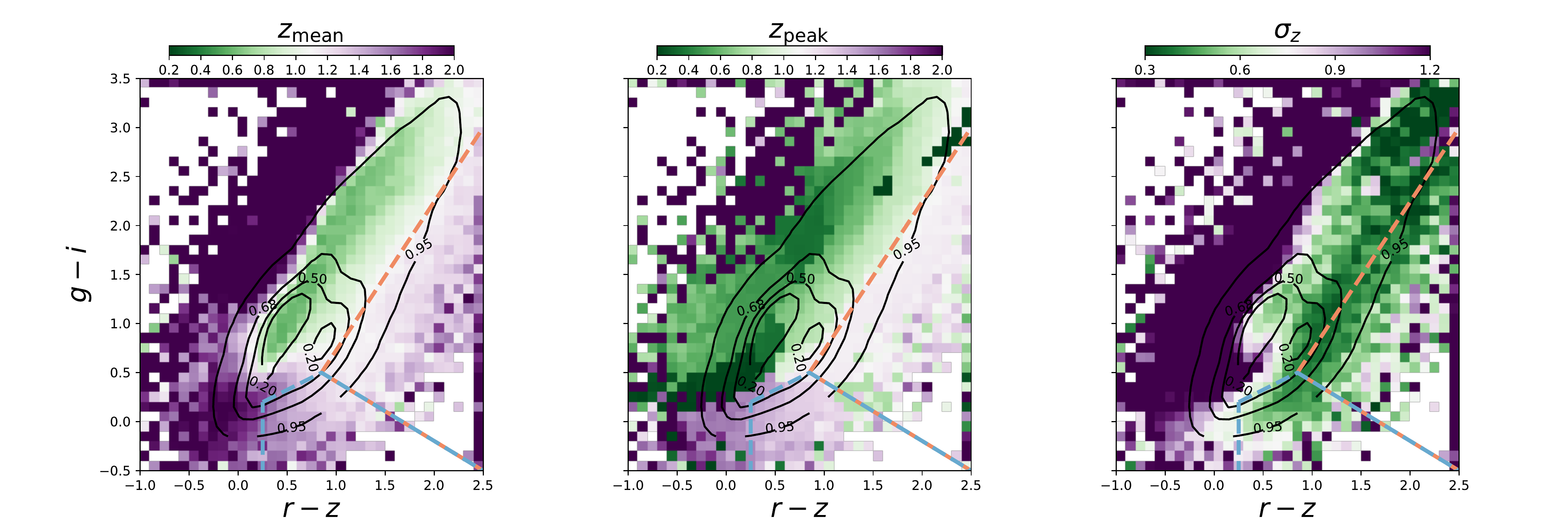}
}
\caption{
The color-color, $g-i$ vs $r-z$, diagrams showing the mean $\redshift_{\mathrm{mean}}$ (left), peak $\redshift_{\mathrm{peak}}$ (middle), and the scatter $\sigma_{\redshift}$ (right) of the stacked $P(\redshift)$ of each color-color cell. 
These color-color cells are defined as the boxes with a width of $0.1$~mag.
The black contours in each panel represent the normalized distribution in the random fields, for which the labels show the enclosed percentage of the selected galaxies.
These plots are produced by stacking the $P(\redshift)$ of the galaxies with $i$-band magnitude smaller than $24.5$~mag in the randomly drawn apertures.
The color-color selections for the \lowz\ and \hghz\ backgrounds are, respectively, shown by the red and blue dashed lines.
In this way, we can avoid the redshift-confusion regions (see the text).
}
\label{fig:zmap}
\end{figure*}
\begin{figure}
\centering
\resizebox{0.48\textwidth}{!}{
\includegraphics[scale=1]{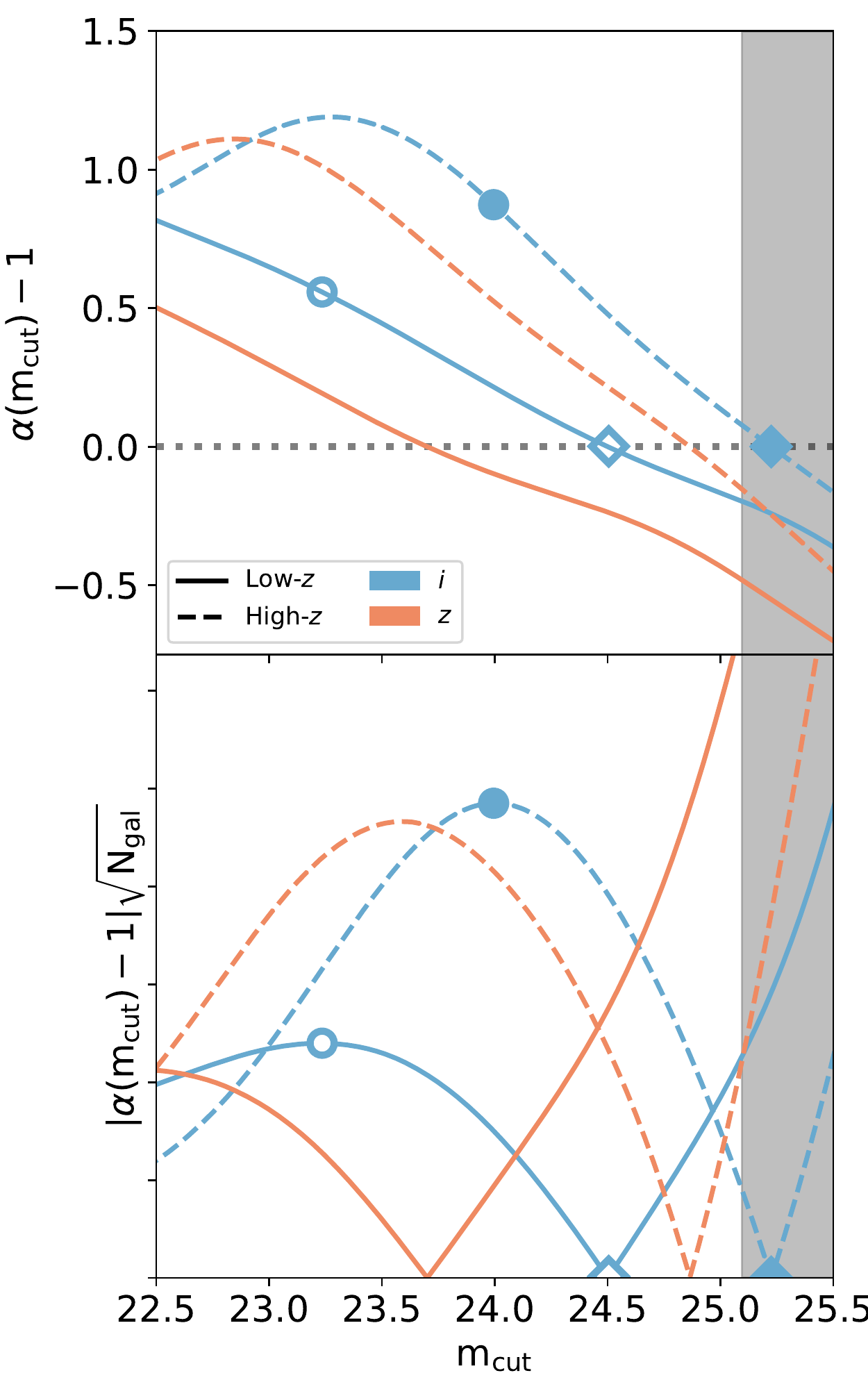}
}
\caption{
The expected signals (upper panel) and the signal-to-noise ratios (lower panel) of the magnification bias as a function of magnitude.
The estimations of the \lowz\ and \hghz\ backgrounds are indicated by the solid and dashed lines, respectively.
We show the results of the $i$-band in blue, while we also additionally present the results of the $z$-band in red, as a comparison.
The dotted grey lines show the case of $\slope-1=0$, where the net magnification bias vanishes. 
The open (solid) circle and diamonds are the \mcut\ used for selecting the flux-limited ``lensing-cut'' and ``null-test'' samples, respectively, for the \lowz\ (\hghz) background.
The grey area presents the regime where the depth of $i$-band is below the $10\sigma$ limiting magnitude.
}
\label{fig:slope}
\end{figure}
%

%%%%%%%%%%%%%%%%%%%%%%%%%%%%%%%%%%%%%%%%%%
%
% METHODS
%
%%%%%%%%%%%%%%%%%%%%%%%%%%%%%%%%%%%%%%%%%%

\section{Analysis}
\label{sec:analysis}

Our goal is to characterize the \rtm\ scaling relation given the observables of the clusters.
For each cluster, we have two observables: the richness \rich\ and the observed magnification profile.
The richness \rich\ is estimated by the $\texttt{CAMIRA}$ algorithm (see Section~\ref{sec:cluster_sample}).
In what follows, we will detail the extraction of the magnification profile for each cluster and the modelling of the \rtm\ scaling relation.

\begin{figure*}
\centering
\resizebox{0.31\textwidth}{!}{
\includegraphics[scale=1]{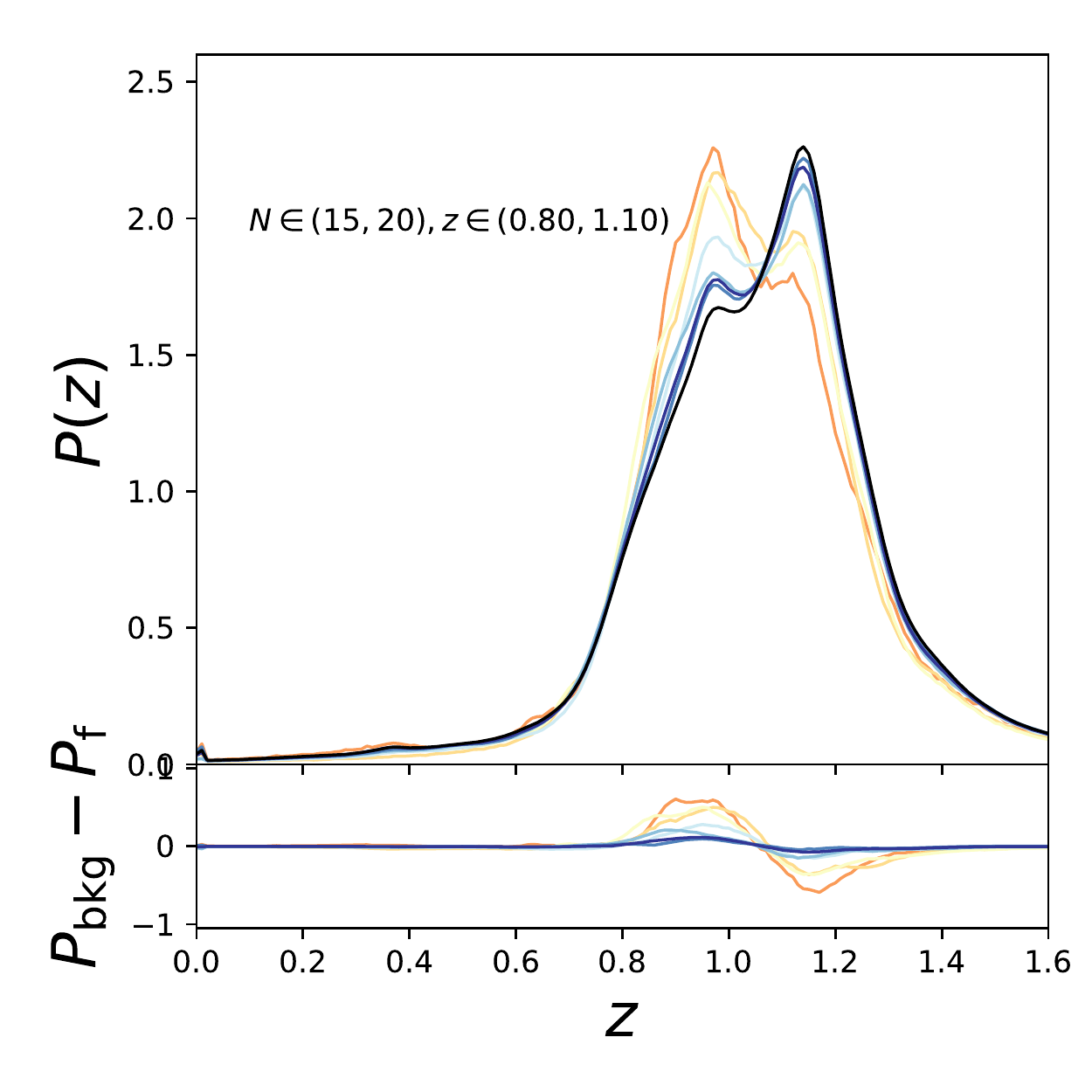}
}
\resizebox{0.31\textwidth}{!}{
\includegraphics[scale=1]{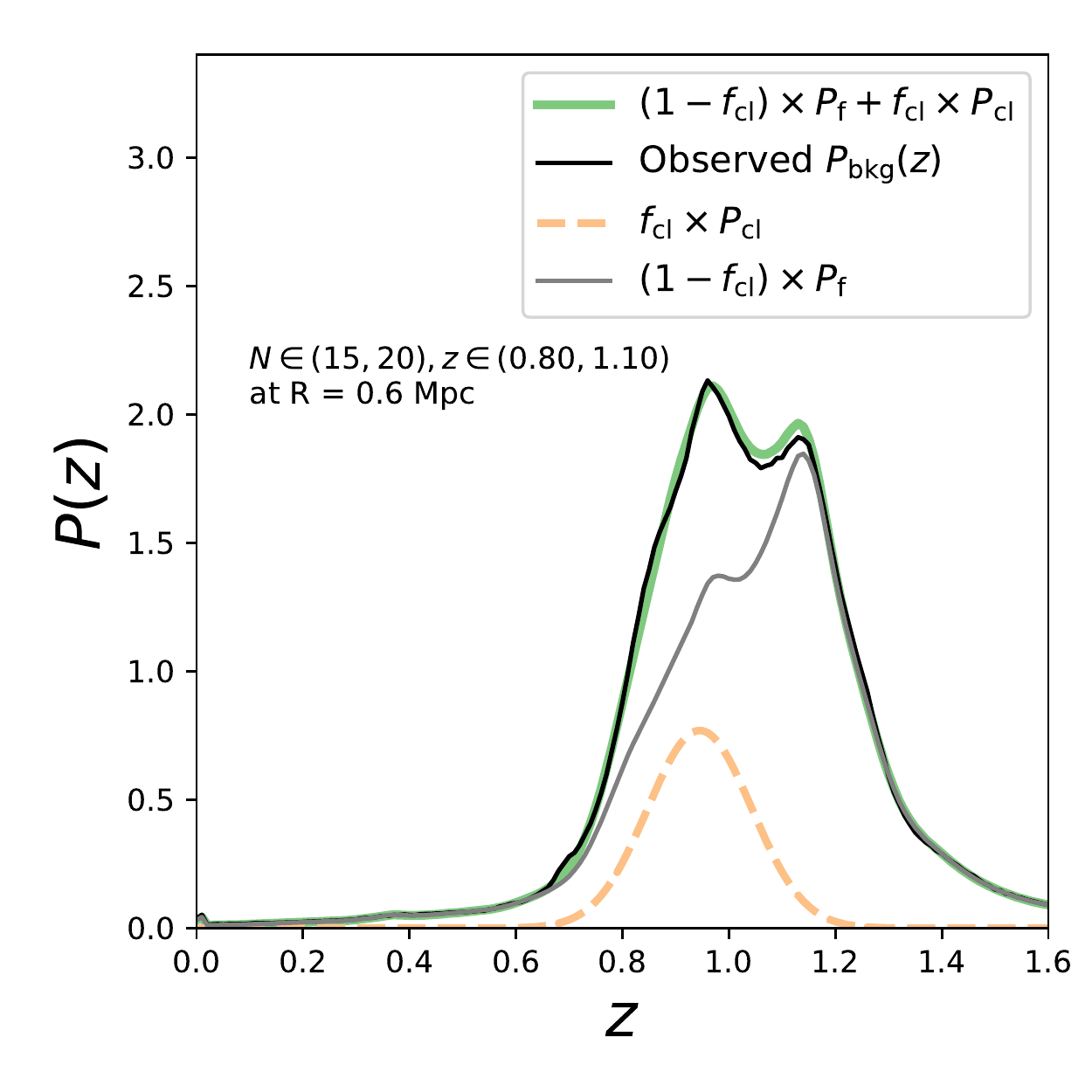}
}
\resizebox{0.31\textwidth}{!}{
\includegraphics[scale=1]{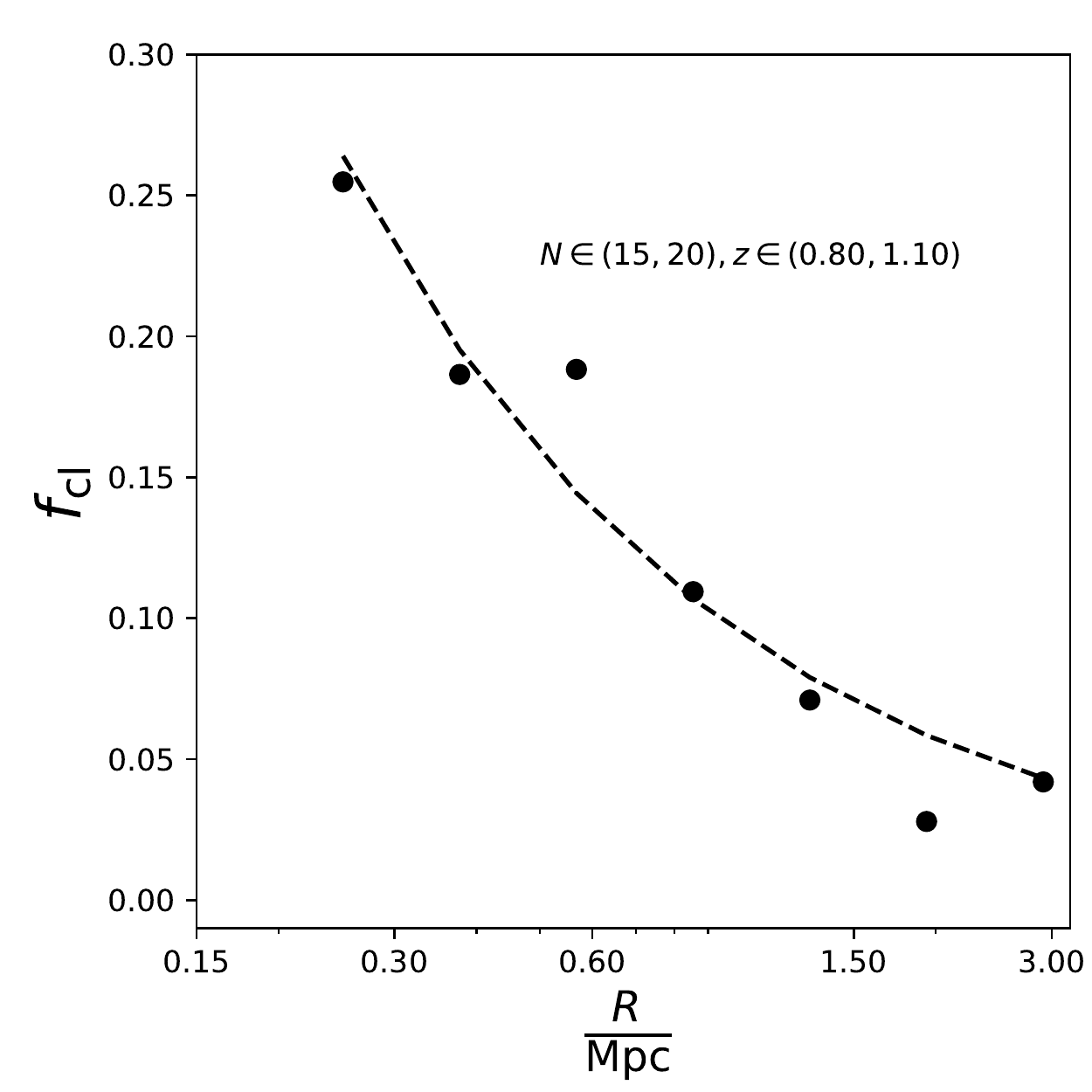}
}
\caption{
The plots demonstrating the $P(\redshift)$-decomposition method \citep{gruen14} in estimating the cluster contamination.
We show the results of the \lowz\ background for the clusters with richness of $15\leq\rich<20$ at redshift $0.8\leq\redshift<1.1$, as an example (see Section~\ref{sec:fcl}).
The upper subplot of the left panel shows the stacked probability redshift distributions $P(\redshift)$ of the selected source sample, with redder color as decreasing clustercentric radius; the one from the random fields in shown in black.
To highlight the cluster contamination in terms of $P(\redshift)$, we also show the difference between the observed clusters $P_{\mathrm{bkg}}(\redshift)$  and the random fields $P_{\mathrm{f}}(\redshift)$ in the lower subplot of the left panel.
The enhancement of probability in the cluster redshift range $0.8\leq\redshift<1.1$ is clearly seen in the cores of clusters.
The middle panel demonstrates the $P(\redshift)$-decomposition method in the radial bin of $R\approx0.6$~Mpc: the observed redshift distribution $P_{\mathrm{bkg}}$ is in black; the redshift distributions of the random fields $P_{\mathrm{f}}$ and the clusters $P_{\mathrm{cl}}$ normalized by the 
$\left(1 - \fcl\right)$ and \fcl\ 
are in grey and yellow lines, respectively; the best-fit redshift distribution is shown by the green line.
The right panel shows the results of the derived cluster contamination profile $\fcl(R)$ with the best-fit power law indicated by the dashed line.
}
\label{fig:pz_radii}
\end{figure*}

\subsection{Background Selection}
\label{sec:background_selection}

We apply a color-color selection to obtain the background galaxies as the source sample used in this work.
This selection is made in the color-color space of $g-i$ vs $r-z$, for which this combination of colors has been demonstrated as an efficient way to deliver a sample of background galaxies with high purity \citep{medezinski18a}.
This color-color combination is designed to remove the foreground and cluster galaxies based on the synthetic evolutionary tracks of various types of galaxies in the color-color space \citep{medezinski10}.
Interested readers are referred to \cite{medezinski18a} for more details.

In this work, the color-color cuts are defined with the guidance from 
\begin{enumerate}
\item the distribution of the cluster galaxies, and
\item the redshift distribution among the color-color cells,
\end{enumerate}
in the space of $g-i$ vs $r-z$.
The former can be derived as follows.
First, we stack the distribution of the galaxies that lie projected within the clustercentric radius of $0.5$~Mpc around the clusters in the $g-i$ vs $r-z$ space.
Second, we also derive the reference distribution by carrying out the same stacking process on the random fields, and subtract the reference distribution from 
the stacked cluster field.
In this way, we can obtain the distribution of the cluster galaxies in the space of $g-i$ vs $r-z$ by statistically removing the fore/background.
We only use the CAMIRA clusters with richness of $\rich>20$ and the galaxies with magnitude smaller than $24.5$ for the procedure above.
We repeat this exercise for three redshift bins ($0.2<\redshift\leq0.5$, $0.5<\redshift\leq0.8$, and $0.8<\redshift\leq1.1$).
The results of three redshift bins are shown in Figure~\ref{fig:sources}:
The green contours are the distributions of the cluster galaxies after the statistical fore/background subtraction, and the underlying greyscale images represent the smoothed galaxy densities drawn from the random fields.

As the cluster redshift increases from the left to right panel in Figure~\ref{fig:sources}, it is clear that the cluster red sequence is moving toward the redder regime in both colors and that an increasing population of the cluster blue members appears at $r-z\approx0.9$ and $g-i\approx1$.
Our goal is to define the color-color cuts to avoid these two regions. 
Meanwhile, we want to select the background galaxy population with high purity and high number density.
To do so, we further follow the guidance (ii) by making use of the photometric redshift.
Specifically, we stack the probability distribution functions $P(\redshift)$ of the photometric redshift of all galaxies that lie within each color-color cell, 
and calculate the mean, peak and the scatter of the stacked $P(\redshift)$.
These color-color cells are defined as boxes with with a width of $0.1$~mag in the space of $g-i$ vs $r-z$.
This procedure is performed on the random fields to avoid the clusters, for which the galaxy population has a biased distribution in the color-color space.

We use the 
$P(\redshift)$
estimated by \texttt{DemP} \citep{hsieh14}, a machine-learning based code that has been extended to robustly obtain various properties of galaxies, including the redshift and stellar mass \citep{lin17}.
The 
$P(\redshift)$ estimated by \texttt{DemP} are very well calibrated \citep{tanaka18}, suggesting no signs of bias according to the tests of the probability integral transform (PIT) and the continuous ranked probability score \citep{polsterer16}.
In addition, the point estimates of the photometric redshift are accurate to better than $1\percent$ in term of $\left\langle\Delta\redshift/(1+\redshift)\right\rangle$ with scatter of $\approx0.04$ and an outlier rate of $\approx8\percent$ for galaxies with $i<24$~mag \citep{tanaka18}. 
More details of the calibration of photometric redshift can be found in \cite{tanaka18}.

In Figure~\ref{fig:zmap}, we show the mean (left), peak (middle) and scatter (right) of the stacked $P(z)$ in the space of $g-i$ vs $r-z$.
The black contours in Figure~\ref{fig:zmap} represent the normalized galaxy density with the labels showing the percentages of the enclosed galaxies.
By comparing the left and middle panels in Figure~\ref{fig:zmap}, it is clear that there is severe degeneracy of the photometric redshift estimation between the redshift of $\approx0.5$ and $\approx2$ at the colors of $g-i\approx0.7$ and $r-z\approx0$. 
This is also reflected 
in the right panel that the scatter of the stacked $P(\redshift)$ is significantly larger in the whole upper-left region. 
This suggests that either 
the true redshift distributions are wide in these color-color cells, or the photometric redshift is ill-constrained in these regions.
On the other hand, the ideal background populations of the galaxies can be identified at the lower-right corner of the color-color space with mean redshift of $1.1\lesssim\redshift\lesssim1.6$ and small dispersion, avoiding the redshift-confusion regions.

Based on the guidance (i) and (ii)  as well as the information of the galaxy number density in the color-color space, 
we select two background populations, as referred to the \lowz\ and \hghz\ backgrounds, respectively. 
The \lowz\ and \hghz\ backgrounds are shown by the red and blue points in Figure~\ref{fig:sources}, respectively.
In Figure~\ref{fig:sources} and Figure~\ref{fig:zmap}, we also mark the color-color selections by the dashed lines.
These color-color cuts are defined such that we can minimize the regions overlapping with the clusters, while maximizing the source density with low dispersion in the stacked $P(\redshift)$.

Although these color-color selections are optimized to select the sources with high purity, the selected populations are still likely to be contaminated by the cluster members, due to the fact that the color-color distribution of the cluster field is highly biased with respect to the random field.
We will quantify this cluster contamination in Section~\ref{sec:fcl}, and correct for them in Section~\ref{sec:sr_modelling}.

After the color-color selection, we need to further apply a magnitude cut because we measure the magnification bias using a flux-limited sample.
As seen in equation~(\ref{eq:magnification_bias}), the signal of magnification bias is a function of the logarithmic count slope \slope, which generally depends on the magnitude and the choice of the passband.
The count slope \slope\ is larger than $1$ at bright end and is monotonically decreasing as increasing magnitude, thus we expect a density enhancement (depletion) for a cut at bright (faint) end.
To anticipate a high signal of density enhancement, the slope \slope\ is required to be much higher than $1$, which is typically at very bright end where the Poisson noise is large.
Conversely, a fainter magnitude cut results in a source sample with a larger size, thus the Poisson noise decreases.
However, the signal is expected to become lower because of a smaller \slope\ at the faint end.
Therefore, one needs to optimize the magnitude cut (\mcut) at the appropriate passband to achieve the highest signal-to-noise ratio.
Motivated by this, we show the slopes and the relative signal-to-noise ratios as functions of magnitude in the passbands in the upper and lower panels of Figure~\ref{fig:slope}, respectively.
We note that we only show $i$- and $z$-band in Figure~\ref{fig:slope} for simplicity.

To measure the slope $\slope\equiv 2.5\times \dif\log N(<m)/\dif m$, we
first calculate the cumulative source counts $N(<m)$ as a function of magnitude
between 17 and 30.5\,mag with a bin width of 0.01~mag. 
This results in a total of 1350 bins.
Then, for each magnitude bin $m_i$, we fit a polynomial function of the
form $0.5 a \times m^2 + b \times m + c$ to the data points of
$\log[N(<m_k)]$ ($k=1,2,...$),
with $|m_k - m_i| < 0.5~\mathrm{mag}$ for the $i$th bin.  
We thus have $\dif\log N(<m)/\dif m = a m + b$, where $a$ and $b$ are
the best-fit parameters in each magnitude bin. 
We repeated this procedure for all 1350 magnitude bins, resulting in 1350 estimates of $\dif\log N(<m)/\dif m$.
Last, we linearly interpolate these 1350 slope data points to obtain $\slope(m)$ at an arbitrary magnitude.
Note that the resulting slopes are not sensitive to the chosen bin size.
The resulting slope functions $\alpha(m_\mathrm{cut})$ are shown in the
upper panel of Figure~\ref{fig:slope}.

To anticipate the relative signal-to-noise ratios of magnification bias (as shown in the lower panel of Figure~\ref{fig:slope}), we simply consider the Poisson noise as the only  source of uncertainty.
Here, the signal of magnification bias is defined as the contrast of the galaxy density, namely, $\delta_{n}\equiv\frac{\nd}{\nzero} - 1$.
Based on equation~(\ref{eq:magnification_bias_linear}), the signal-to-noise ratio of magnification bias is proportional to 
$|\mu^{\alpha-1}-1|/\left(\sqrt{N_{\mathrm{Ngal}}}/N_{\mathrm{Ngal}}\right) =
  |\mu^{\alpha-1}-1|\sqrt{N_{\mathrm{Ngal}}} \appropto
  |\alpha-1|\sqrt{N_{\mathrm{Ngal}}}$
for a given cluster, where $N_{\mathrm{Ngal}}$ is the total counts of the background sources of interested.
Note that we construct this signal-to-noise estimator based on the assumption of the weak-lensing regime (i.e., $\kappa\approx|\shear|\ll1$ in equation~(\ref{eq:magnification_bias_linear})), while we use equation~(\ref{eq:magnification_bias}) in our modelling in Section~\ref{sec:sr_modelling}.
Then, we estimate $N_{\mathrm{Ngal}}$ of the \lowz\ and \hghz\ backgrounds using all the sources located in the FDFC footprint to produce the lower panel of Figure~\ref{fig:slope}.
The results of the \lowz\ and \hghz\ populations are the solid and dashed lines, respectively.

Although slightly higher signal-to-noise ratios are expected for bluer bands (e.g., the $g$- and $r$- bands), we stress that we do not use them to select the flux-limited sample for the analysis of magnification bias.
This is because the variation of seeing in both $g$- and $r$- bands is large ($\approx0.5\arcsec-1.3\arcsec$), which introduces significant non-uniformity across the field.
On the other hand, the HSC observing strategy specifically requires the $i$-band imaging to be taken under a good seeing condition, resulting in a much more uniform seeing distribution across the field \citep{mandelbaum18}.
In this regard, we only use $i$-band for the analysis of magnification bias in this work.
We apply a magnitude cut of $m<\mcut$ in the $i$-band after the color-color selection to construct the flux-limited sample, such that the maximized signal-to-noise ratio of the magnification bias measurement is expected. 
Specifically, we use $\mcut=23.2$~mag and $\mcut=24$~mag for the \lowz\ and \hghz\ populations, which are marked by the solid and open circles in Figure~\ref{fig:slope}, respectively.
We refer to these samples as the ``lensing-cut'' samples.

On the other hand, there exists a special magnitude where $\slope=1$, such that we do not expect the magnification bias signal with the magnitude cut of $m<\mcut$ 
(see Section~\ref{sec:basics}).
With this sample, we can verify our magnification bias measurements of the ``lensing-cut'' sample, and quantify the residual biases, if any.
For the \lowz\ (\hghz) background, this limiting magnitude is $\mcut=24.5$~mag ($\mcut=25.2$~mag), which is marked by the open (solid) diamond in Figure~\ref{fig:slope}.
Therefore, we independently construct the flux-limited samples with these magnitude cuts for both \lowz\ and \hghz\ backgrounds, for which we refer to as the ``null-test'' samples.
We will use the ``null-test'' samples to validate our magnification bias measurements (see Section~\ref{sec:em}).

Following the same procedure in \cite{chiu16b}, we quantify the completeness in the $i$-band at the imposed magnitude limits.
Specifically, we assume that the source detection in the $i$-band between the magnitude of 20 and 22 is $100\percent$ complete and can be described by a power law as a function of magnitude.
Then, we derive the ratio of observed source counts to the reference model of $100\percent$ completeness that is extrapolated by the best-fit power law at the magnitude  fainter than 22~mag.
Last, we fit completeness function defined as
\[
F_{\mathrm{c}}(m) = \frac{1}{2} - \frac{1}{2}\mathrm{erf} \left( \frac{m - m_{50}}{\sigma_\mathrm{m}} \right) \, ,
\]
to the ratios, where $m_{50}$ and $\sigma_{\mathrm{m}}$ are free parameters to be fitted.
As a result, the completeness is close to $100\percent$ for both ``null-test'' and ``lensing-cut'' samples in the \lowz\ background and for the ``lensing-cut'' sample in the \hghz\ background.
At the $10\sigma$ depth ($25.1$~mag), the completeness reaches $90\percent$.
Our limiting magnitude of $25.2$~mag for the ``null-test'' sample in the \hghz\ background is fainter than  the $10\sigma$ depth, resulting in the completeness of $86\percent$.
There is no sign showing that this incompleteness of $86\percent$ is correlated with clusters, as we do not observe significant radial patterns in the galaxy distribution for the ``null-test'' sample in the \hghz\ background (see Section~\ref{sec:em} and Figure~\ref{fig:fem_append}).
Therefore, we conclude that our results are not affected by the incompleteness due to the background noises.

We stress that we do not impose any cut at the bright end, as opposed to
what is typically done in a weak shear analysis in order to minimize
the contamination from unlensed cluster members and foreground galaxies  \citep[e.g.,][]{medezinski18a}. 
This is because the bright-end cut will modify the signal of magnification
bias, which neeeds to be accounted for in the calculation of the
logarithmic count slope $\alpha$ (see
Appendix~\ref{sec:brightendcut_slope} for more details).
The resulting number densities are $0.51$ and $0.92$ galaxies per square arcmin for the \lowz\ and \hghz\ populations, respectively.
The mean redshift of the \lowz\ and \hghz\ populations are $\approx1.1$ and $\approx1.4$, respectively.

In this work, we focus on the density enhancement of magnification bias at relatively bright end with $\slope-1\gg0$.
It is worth mentioning that we cannot probe the density depletion of magnification bias in the regime of $\slope\ll1$, where the detection is significantly suffering from the shot noise at very faint end ($i\gtrsim25.5$~mag).
The grey area in Figure~\ref{fig:slope} shows the magnitude fainter than $25.1$~mag, which is the $10\sigma$ depth of the $i$-band, indicating that the source detection starts to be dominated by the shot noise.

\subsection{Cluster Member Contamination}
\label{sec:fcl}

One of the most critical bias in cluster lensing using a photometry-selected source sample is the cluster contamination. 
That is, cluster members could leak into the source sample, causing bias in lensing signals.
For example, the average tangential shear of the selected background sample is diluted by the cluster contamination, thus resulting in a biased-low mass estimate. 
In the case of magnification bias, the leaked cluster members result in a density enhancement that is not due to gravitational lensing but the clustering of member galaxies, returning a biased high (low) magnification-inferred mass for the case of $\slope>1$ ($\slope<1$).
Moreover, the cluster contamination is expected to vary with the clustercentric radius, cluster redshift and the cluster mass.
In this regard, this is necessary to quantify and account for the cluster contamination, if any.

In this work, the cluster contamination is quantified by the method developed in \cite{gruen14}, which has been widely used in other cluster lensing studies \citep{chiu16b,melchior17,dietrich19, medezinski18a, mcclintock18}.
This demonstrates a robust approach to successfully estimate and correct for the cluster contamination \citep{varga18}.
In what follows, we briefly summarize this method and refer the reader to \cite{gruen14} for more details.

The basic idea of this method is to decompose the photometric redshift distribution, $P_{\mathrm{bkg}}(R,\redshift)$, of the selected background sample observed at the clustercentric radius of $R$ into two components, the photometric redshift distributions of the source sample $P_{\mathrm{f}}(\redshift)$ and the cluster members $P_{\mathrm{cl}}(\redshift)$, using a linear relation of 
\begin{equation}
\label{eq:fcl}
P_{\mathrm{bkg}}(R,\redshift) = (1 - \fcl(R)) P_{\mathrm{f}}(\redshift) + \fcl(R) P_{\mathrm{cl}}(\redshift) \, ,
\end{equation}
where $\fcl(R)$ is the cluster contamination at the radius $R$, and $0\leq\fcl(R)\leq1$, by definition.
Here, $P_{\mathrm{f}}(\redshift)$ is estimated by the source galaxies selected in the apertures randomly drawn from the field, and $P_{\mathrm{cl}}(\redshift)$ is characterized by a normal distribution with the mean $\mathbb{Z}_{\mathrm{cl}}$ and standard deviation $\sigma_{\mathrm{cl}}$ that need to be constrained.
In this way, the cluster contamination $\fcl(R)$ can be derived for the observed $P_{\mathrm{bkg}}(R,\redshift)$ at the projected radius $R$.

In practice, we have to stack the clusters to obtain the meaningful constraints of \fcl, because the the observed photometric redshift distribution is too noisy to carry out this method on a basis of individual clusters.
Specifically, we stack clusters into three redshift bins ($0.2\leq\redshift<0.5$, $0.5\leq\redshift<0.8$, $0.8\leq\redshift<1.1$) and three richness bins ($15\leq\rich<20$, $20\leq\rich<30$, $30\leq\rich$), with nine bins in total.
Again, we use the probability distribution functions of the photometric redshift estimated by \texttt{DemP} to estimate the cluster contamination.

For a radial bin in each richness and redshift bin, there are three parameters (\fcl, $\mathbb{Z}_{\mathrm{cl}}$, $\sigma_{\mathrm{cl}}$) to be fitted.
Similarly to \cite{melchior17}, we jointly fit the mean $\mathbb{Z}_{\mathrm{cl}}$ and standard deviation $\sigma_{\mathrm{cl}}$ of the normal distribution (representing $P_{\mathrm{cl}}(\redshift)$ in equation~(\ref{eq:fcl})) for the same redshift bin, while the cluster contamination \fcl\ is varying in each richness and radial bin.
The result of the binning of $15\leq\rich<20$ at $0.8\leq\redshift<1.1$ is shown in Figure~\ref{fig:pz_radii}, as the example of highly contaminated regimes.
In the left panel, we show the $P_{\mathrm{bkg}}(\redshift)$ of the \lowz\ background at different clustercentric radii (the redder, the inner).
There is an increasing enhancement in the stacked $P_{\mathrm{bkg}}(\redshift)$ at the cluster redshift range of $0.8\lesssim\redshift\lesssim1.1$ as decreasing clustercentric radius, clearly indicating the cluster contamination. 
In the middle panel, the decomposition of $P_{\mathrm{bkg}}(\redshift)$ at the radius of $r\approx0.6$~Mpc is shown.
Moreover, this cluster contamination can be well modelled by a normal distribution (the yellow dashed line).

Motivated by \cite{applegate14}, we further fit a power-law profile in radius normalized at $1$~Mpc,
\[
\fcl(R) = f_{\mathrm{cl},1\mathrm{Mpc}}\times \left(\frac{R}{1~\mathrm{Mpc}}\right)^{\Gamma_{\mathrm{cl}}} \, ,
\]
to the derived \fcl\ profile to statistically account for the cluster contamination as a function of radius.
The fitting of the normalization and power-law index $(f_{\mathrm{cl},1\mathrm{Mpc}}, \Gamma_{\mathrm{cl}})$ is performed for every richness and redshift bins, which is sufficient to describe the radial trend of \fcl, as shown by the dashed line in the right panel of Figure~\ref{fig:pz_radii}.
We have tried the fitting with varying $\mathbb{Z}_{\mathrm{cl}}$ and $\sigma_{\mathrm{cl}}$ in each bin (instead of jointly fitting), which returns a consistent result.

There are some caveats of the cluster contamination estimated by the \cite{gruen14} method, in which the fundamental assumption is that the observed 
$P(\redshift)$
of the selected background population in the cluster fields is statistically identical to that observed in the random fields.
This assumption might be broken due to the degraded performance of photometry in crowded fields, such as clusters.
The assessment using the \texttt{hscPipe} around the cluster fields (see Section~\ref{sec:masking})
quantifies that less than $5\percent$ of the sources are heavily blended with the member galaxies or even missed in the cluster fields.
This suggests that photometry for majority of the source galaxies is not significantly affected by the cluster members.
The photometry performance of the \texttt{hscPipe} around the cluster fields will be further investigated in the forthcoming paper (Murata in prep.).

Another concern regarding cluster fields is the presence of Intra-Cluster Light (ICL), which could bias the photometry, photo-\redshift, and/or mass modelling.
\cite{gruen19} estimate the impact of the ICL on the mass modelling, suggesting a bias of $\lesssim 2\percent$ in the differential surface mass profile at $r\gtrsim 300~k\mathrm{pc}$ for clusters with $\Mfiveoo\lesssim 3\times 10^{14}\Msun$.
This amount is negligible compared to the current statistical uncertainty in this work, therefore we ignore this effect.

To validate the derived cluster contamination, one typically needs a very detailed and customized simulation \citep{varga18} or the spectroscopic observations of the cluster field to further identify the cluster members, which are not currently available to our sample.
However, we can in principle empirically validate the magnification measurements of the ``lensing-cut'' sample based on the ``null-test'' sample (see Section~\ref{sec:background_selection}), if there exists any residual bias caused by the inaccurate correction for the cluster contamination.
We will return to this in Section~\ref{sec:em}.

\begin{figure}
\centering
\resizebox{0.45\textwidth}{!}{
\includegraphics[scale=1]{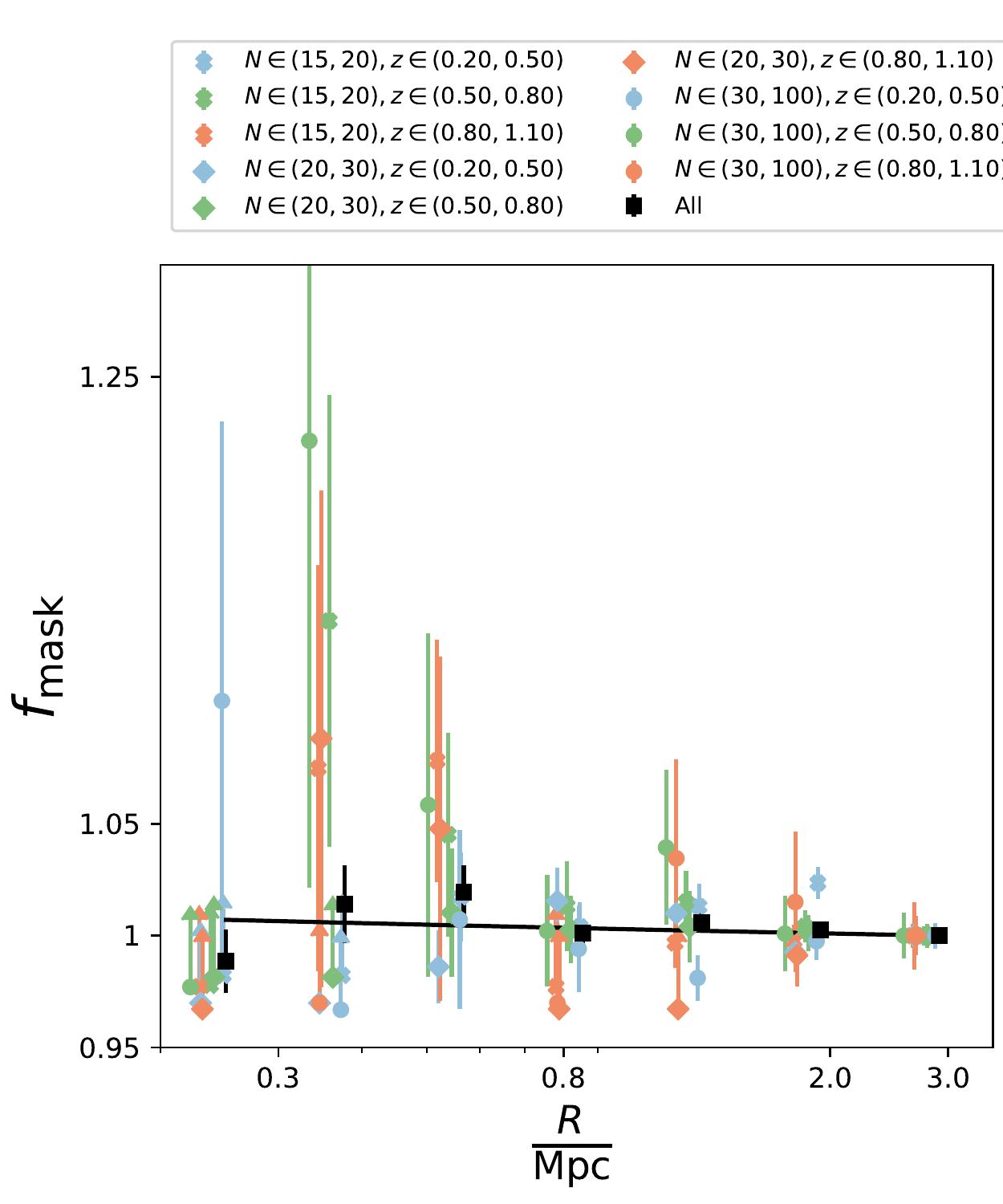}
}
\caption{
The corrections for the masking effect in the \hghz\ background populations by using the synthesis galaxies generated by the \texttt{Synpipe} code.
The resulting \fmk\ in the richness bins of $15\leq\rich<20$, $20\leq\rich<30$, and $\rich\leq30$ are shown by thick crosses, diamonds, and circles, respectively.
These markers are color-coded by blue, green, and red for the redshift bins of $0.2\leq\redshift<0.5$, $0.5\leq\redshift<0.8$, and $0.8\leq\redshift<1.1$, respectively.
The black squares are the masking correction estimated by stacking all clusters.
The black curve is the best-fit power law, which is ultimately used to statistically correct for the masking effect.
We note that we normalize the masking correction to the outer most bin $R_{\mathrm{out}}$ (see the text in Section~\ref{sec:masking}), such that $\fmk(R_{\mathrm{out}})=1$ to remove the effect of random masking.
In some cases that we do not have masked synthesis galaxies in the radial bins, we mark the data points as the lower bounds shown by the arrows.
}
\label{fig:fmk}
\end{figure}
\begin{figure*}
\centering
\resizebox{0.95\textwidth}{!}{
\includegraphics[scale=1]{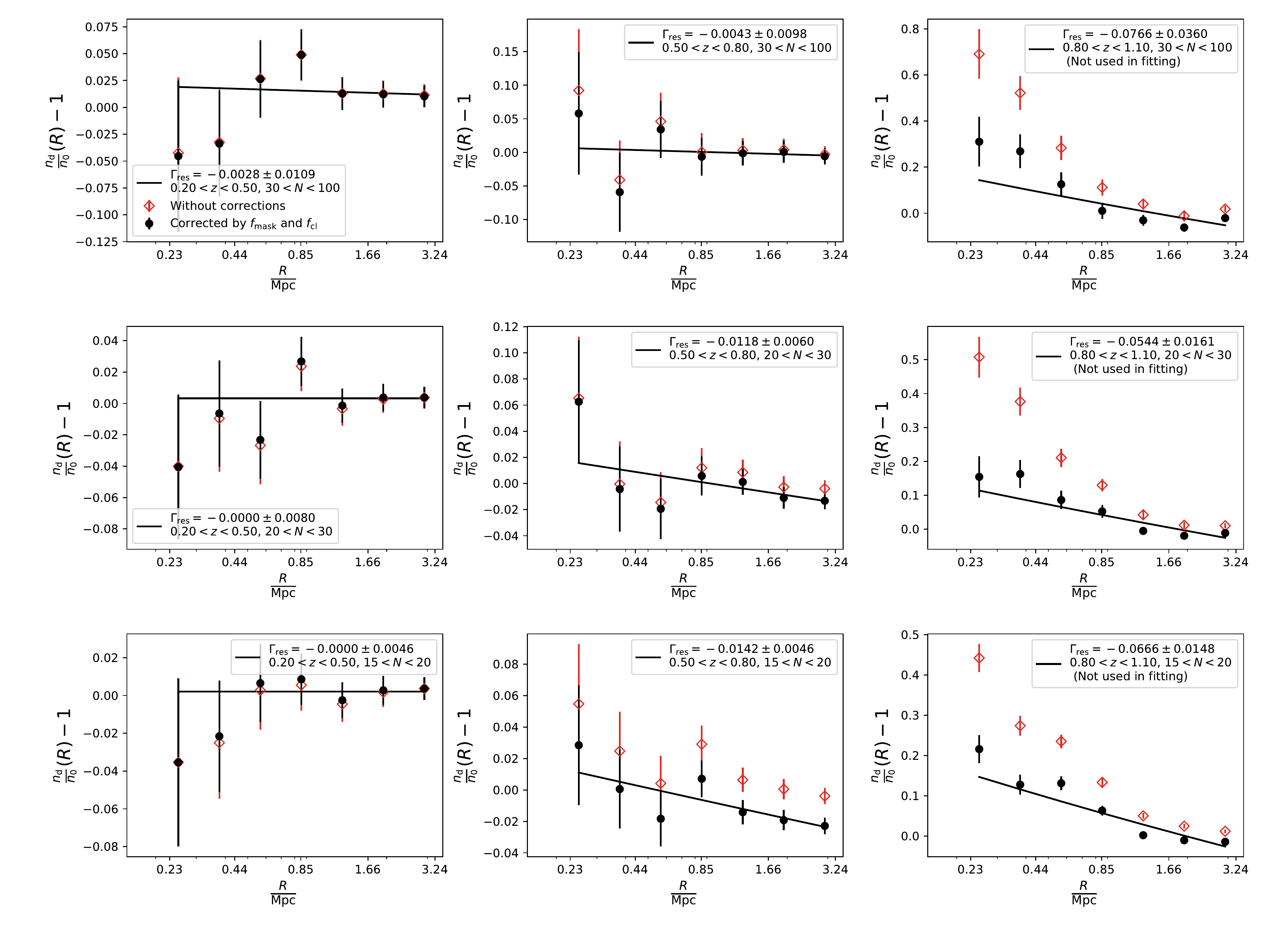}
}
\caption{
The stacked density contrast of the ``null-test'' sample (see Section~\ref{sec:background_selection}) for the \lowz\ background as functions of clustercentric radius in three richness (lower to upper) and three redshift (left to right) bins.
The red diamonds are the direct measurements of $\omega\left(R\right)$ without any correction, while the black circles are measurements after the corrections for the cluster contamination \fcl\ and masking \fmk.
The deviation of $\frac{\nd}{\nzero}-1$ from zero indicates the residual bias.
The best-fit power-law models (equation~(\ref{eq:fem})) are indicated by the black curves with the best-fit radial indices ${\Gamma}_{\mathrm{res}}$ shown in the upper-right corners.
As seen in this plot, we observe significant residual bias in the highest redshift bin ($0.8<\redshift<1.1$), for which we ignore this redshift range in our analysis.
These estimations of residual bias are used to quantify the systematic uncertainty of our final results.
}
\label{fig:fem}
\end{figure*}

\subsection{Masking Correction}
\label{sec:masking}

Another complexity that could bias the observed magnification bias around clusters is the masking effect:
Bright member galaxies could mask the sources behind clusters, effectively mimicking the density depletion as the decreasing clustercentric radius.
This bias can be approximated by calculating the fractional angular area occupied by the bright member galaxies in each radial bin, as described in \cite{umetsu11}.
Another approach is to run realistic image simulations of cluster fields to access the successful rate of the source detection as a function of clustercentric radius, as demonstrated in \cite{chiu16b}.
Nevertheless, simulating the realistic galaxy population in clusters requires the prior knowledge of cluster mass and the \rtm\ relation, for which this approach is circular and thus not ideal for this work.

In this work, we improve the method in \cite{chiu16b} by directly simulating galaxies of interest in the real imaging.
This method is similar to \cite{suchyta16}  and \cite{chiu16d}, where they embedded synthesis galaxies to the real imaging to quantify various observational systematics.
Specifically, we embed synthesis galaxies in the observed images, processed by the identical pipeline, to access the source detection around the cluster field.
This is done by using the \texttt{Synpipe} \citep{huang18}, a \texttt{Python} package which embeds synthesis galaxies and runs the end-to-end \texttt{hscpipe} pipeline to validate the performance of the photometric measurements.

A brief summary of \texttt{Synpipe} is given as follows.
The \texttt{Synpipe} was first introduced in \cite{huang18} to investigate the accuracy of photometry in the HSC survey by embedding the synthesis galaxies in one tract of the HSC footprint (corresponding to area of $\approx1.5$~deg$^{2}$).
To support a detailed investigation of systematic uncertainties in weak-lensing analysis, a follow-up study (Murata in prep.) expands the scale of the \texttt{Synpipe} run to the whole GAMA09h field with area of $\approx43$~deg$^{2}$.
The goal of Murata in prep. is to specifically study how blending could effect the shape measurement in the HSC survey, as presented in \cite{mandelbaum18}.
In this work, we take the resulting ``\texttt{Synpipe} catalog'' of these synthesis galaxies in Murata in prep. and independently investigate the masking effect raised from the galaxy clusters, as a standalone result.
In what follows, we briefly describe how the synthesis galaxies are embedded by the \texttt{Synpipe} in Murata in prep..

The real galaxies in the COSMOS field \citep{capak07,ilbert09}, which are modelled by a single-Sersic galaxy template \citep{lackner12} based on the observed \HST/ACS images, are used as the input of the synthetic galaxies.
The input catalog of the synthesis galaxies is flux-limited with $I_{\mathrm{F814}}\leq25.2$~mag, which is also implemented in the simulation toolkit  \texttt{Galsim} \citep{rowe13}.
The injection of the synthesis galaxies is carried out on a basis of single-epoch images, convolved with the locally measured PSF, followed by the realistic Poisson noise added.
The synthesis galaxies are embedded in grids with a separation of $15\arcsec$ to avoid the self-blending.
Nine identical inputs are used for every $3\times3$ grid.
Due to the extremely large demand for computation, \texttt{Synpipe} is only run on the $i$-band imaging in the GAMA09h field, corresponding $\approx43$~deg$^{2}$.
We stress that these synthesis galaxies are embedded in grids in the GAMA09h field as a whole, regardless of the distribution of the clusters.
That is, the spatial distribution of these synthesis galaxies is statistically uniform with respect to the centers of clusters at a fixed redshift.
For example, the maximal physical radius of $R=3.5$~Mpc studied in this work is $\approx440\arcsec$ for a cluster at $\redshift=1$.
This corresponds to an aperture with angular area of $\approx160$~$\mathrm{arcmin}^2$ that would be spanned by $160\times60^2/15^2\approx2500$ synthesis galaxies per cluster, on average.
For the small scale at $0.2\lesssim R/\mathrm{Mpc}\lesssim0.3$, this corresponds to $\approx11$ synthesis galaxies for a cluster at $\redshift=1$.
In the highest richness and redshift bin ($30\leq\rich$ at $0.8\leq\redshift<1.1$), there are $\approx10$ clusters in the GAMA09h footprint, corresponding to a total number of $110$ synthesis galaxies  (or $\approx12$ different inputs) that would be used in investigating the masking effect in the smallest radial bin, as the worst scenario.
The total number and the input type of the synthesis galaxies embedded in the GAMA09h footprint is sufficient to statistically investigate the masking effect if stacking all clusters as a whole, although the area of the \texttt{Synpipe} run needs to be increased in order to fully resolve the most severe regime mentioned above.

After re-detecting the embedded synthesis galaxies, the resulting catalog is matched to the HSC-Wide-depth COSMOS catalog, which is also observed by the HSC survey, via their unique IDs to obtain the five-band HSC photometry.
That is, the final \texttt{Synpipe} catalog contains not only the realistic colors from the HSC survey itself, but also the information of source detection that is subject to the galaxy type and the local properties of the footprint.
Because the real galaxies observed in the COSMOS field are used for the input synthesis catalog of \texttt{Synpipe}, we stress that the embedded galaxies naturally capture the underlying correlation between the galaxy properties (e.g., the morphology) and the magnitude. 

Next, we apply the same background selections (as in Section~\ref{sec:background_selection}) to the resulting \texttt{Synpipe} catalog.
Then, we stack the galaxy number profiles around the CAMIRA clusters to derive the detection rate of the embedded source galaxies.
In this way, the masking correction at the clustercentric radius $R$ is defined as
\begin{equation}
\label{eq:fmk}
\fmk(R) = \frac{1}{1 - \frac{ N_{\mathrm{syn,mask}}(R) }{ N_{\mathrm{syn,tot}}(R)} } = \frac{ N_{\mathrm{syn,tot}}(R) }{ N_{\mathrm{syn,tot}}(R) - N_{\mathrm{syn,mask}}(R) }  \, ,
\end{equation}
where $N_{\mathrm{syn,mask}}(R)$ and $N_{\mathrm{syn,tot}}(R)$ are the numbers of synthesis galaxies (after the background selection) at the projected radius $R$ that are masked and embedded, respectively.
To derive a more precise masking correction, we need to stack clusters.
In the stacking procedure, we evaluate $N_{\mathrm{syn,tot}}(R)$ and $N_{\mathrm{syn,mask}}(R)$  by summing the synthesis galaxies at the clustercentric radius $R$ around all clusters in the bin of interest, i.e.,
\[
N_{\mathrm{syn,mask}}(R) = \sum_{i\mathrm{-th~cluster}\in\mathrm{bin}} {N_{\mathrm{syn,mask}}}_{i}(R) \, ,
\]
and
\[
N_{\mathrm{syn,tot}}(R) = \sum_{i\mathrm{-th~cluster}\in\mathrm{bin}} {N_{\mathrm{syn,tot}}}_{i}(R) \, .
\]
We note that the $\fmk\left(R\right)$ correction is then normalized to one at the outermost radial bin $R_{\mathrm{out}}$, i.e., $\fmk\left(R\right)\rightarrow\fmk\left(R\right)/\fmk\left(R_{\mathrm{out}}\right)$, because we do expect that the source detection suffers from random masking even without clusters.
That is, we only consider the radial trend of the masking effect raised from clusters, effectively removing the effect of random masking by re-normalization.
For the \lowz\ and \hghz\ background populations, the random masking is $\approx2\percent$.

There are 491 CAMIRA clusters with $\rich\geq15$ at $0.2\leq\redshift<1.1$ in the GAMA09h field that is contained in the resulting \texttt{Synpipe} catalog.
We divide the sample
as stated in Section~\ref{sec:fcl}, to investigate the possible dependence of \fmk\ on the cluster richness or redshift.
The result of the \lowz\ and \hghz\ backgrounds are similar, therefore we only show the latter in Figure~\ref{fig:fmk}, as a demonstration.
In the case of $N_{\mathrm{syn,mask}}(R)=0$ at some small radii, we mark the data point as the lower bound in Figure~\ref{fig:fmk}.
By stacking all clusters, the masking correction \fmk\ (the black points) shows a near zero radial trend, with $\approx1.5\percent$ in the cluster cores ($r\lesssim 0.3$~Mpc).
In addition, there is no strong indication for the dependence on richness or redshift, although the uncertainties remain large.
Thus, we derive the final masking correction based on the \fmk\ of stacking all clusters (the black points).
We note that we re-calculate $N_{\mathrm{syn,mask}}(R)$ and $N_{\mathrm{syn,tot}}(R)$ in equation~(\ref{eq:fmk}) for each binning in Figure~\ref{fig:fmk}, instead of directly stacking the values of $\fmk(R)$ to obtain the black points.
The final masking correction is obtained by fitting equation~(\ref{eq:fmk_def}) to the \fmk\ of stacking all clusters regardless of the richness and redshift. 
\begin{equation}
\label{eq:fmk_def}
\fmk(R) = f_{\mathrm{mk,1Mpc}} \times \left(\frac{R}{0.1~\mathrm{Mpc}}\right)^{\Gamma_{\mathrm{mk}}} \, .
\end{equation}
The best-fit model for \fmk\ is shown by the black curve in  Figure~\ref{fig:fmk}.
We note that this correction is marginally smaller than the one estimated in \cite{tudorica17}, for which the masking correction for the $u$-dropout sources is $\approx1.05$ at $r\approx 0.3$~Mpc monotonically decreasing to $\lesssim 1.035$ at large radii ($r\gtrsim 3$~Mpc).
Moreover, their masking correction rapidly increases to a value of $\gtrsim9\percent$ at $R\lesssim0.2$~Mpc.
This might be explained by a different methodology used in \cite{tudorica17} to estimate the \fmk, in which they approximated the masking effect by the fraction of angular area spanned by the cluster members.

We note that there are two assumptions implicitly made in this approach:
(1) The masking effect of the clusters estimated in the GAMA09h field is the same as the other fields, statistically.
This is a reasonable assumption, given that the seeing distribution of the GAMA09h field is a good representative for the other fields \citep{mandelbaum18}.
(2) Once the sources are successfully detected by \texttt{hscPipe}, we assume that their photometry performance is not significantly degraded because of the masking effect.
That is, to first order we only consider the masking effect on the detectability of the sources, and assume that the pipeline can successfully deblend the fluxes coming from the neighboring objects as long as the sources are detected.
Therefore, we can directly obtain the photometry (colors and magnitude, specifically) relying on the external catalog of the HSC-Wide-depth COSMOS field, where no masking effect due to massive clusters  exists.
However, the masking indeed could affect the photometry in practice.
Moreover, this might be a chromatic effect, which possibly depends on the galaxy type of cluster members and, therefore, the cluster mass and redshift as well.
To access this, a detailed run of \texttt{Synpipe} on the whole HSC survey with the synthesis galaxies embedded in the five bands is clearly warranted.
In this work, we choose to empirically validate our  magnification bias signals based on the ``null-test'' sample (see Section~\ref{sec:em} for more details), if any residual bias of the masking effect persists.

\begin{figure*}
\centering
\resizebox{0.95\textwidth}{!}{
\includegraphics[scale=1]{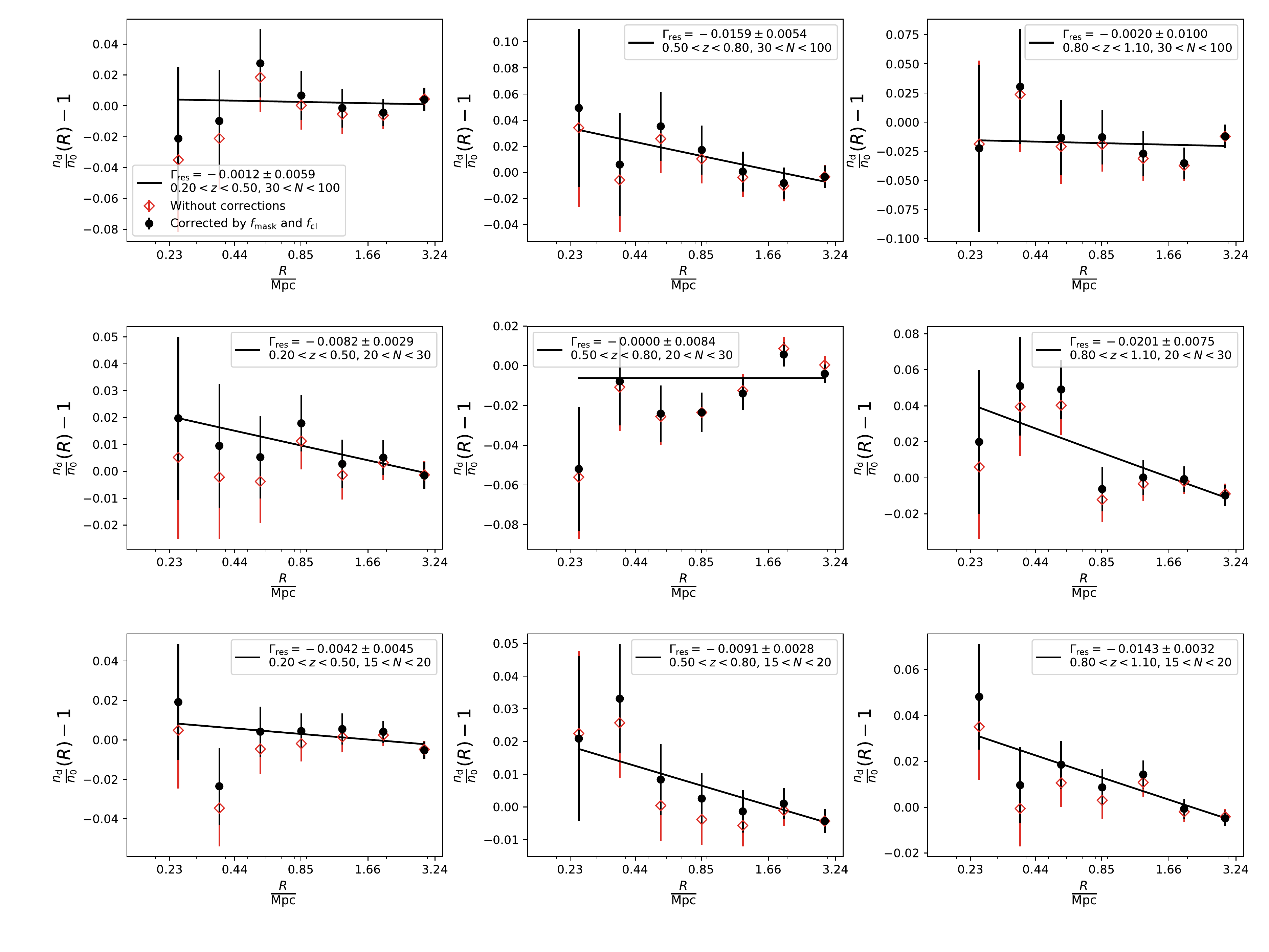}
}
\caption{
The validation of the magnification measurements for the \hghz\ background.
This plot is produced in the same way as in Figure~\ref{fig:fem}.
We do not observe significant residual bias in the case of using \hghz\ background, therefore all redshift and richness bins of magnification bias are used in fitting the scaling relation.
}
\label{fig:fem_append}
\end{figure*}

\subsection{Validating the Measurements of Magnification Bias}
\label{sec:em}

Our goal is to obtain the unbiased measurements of magnification bias and use it to constrain the underlying \rtm\ scaling relation. 
To achieve this, we have statistically corrected for the known biases, such as the cluster contamination existing in the photometrically selected source samples (see Section~\ref{sec:fcl}) and the masking effect due to the cluster member galaxies (see Section~\ref{sec:masking}).
However, these corrections could be insufficient due to the over-simplified assumptions, as follows.

For example, we assume that the  performance of the photometric redshift estimates in the crowded fields of clusters is statistically the same as the random fields, such that we can exploit the method of $P(\redshift)$-decomposition to extract the cluster contamination.
If the photometric redshift distribution is 
different in the cluster fields, then the extracted cluster contamination would be inevitably biased.
Another caveat is that we assume that the photometry in the cluster fields are not severely degraded.
Therefore, we can investigate the masking effect based on the photometry obtained from the external COSMOS catalog.
Relating to the photometry in the crowded fields, it is also known that the current \texttt{hscpipe} tends to over-deblend the bright objects \citep{huang18}, which could result in an artificial enhancement in the number density around bright cluster members.

Among of the concerns above, the most important one is probably the (deblended) photometry in the cluster fields, for which the bias in photometry could be propagated to the photo-\redshift\ estimation and the analysis afterward.
The photometry of the selected sources could be biased due to imperfect deblending:
If the flux of cluster members leaks into the neighboring source at the background, then this would result in a biased-high flux of the source and, therefore, a biased-high number density of the source sample above the flux threshold.
This would mimic the density enhancement of magnification bias.
Vice versa, if the flux of the background source is lost due to a failure in deblending, then the number density of the flux-limited sample would be biased-low, introducing the density depletion that is not due to gravitational lensing.
In addition, this bias is expect to depend on the clustercentric radius.
On the other hand, the blending effect could exist in multi-wavelength, resulting in biased color that could further sabotage the color-color source selection.
To this end, the blending effect could also depend on the broadband filters and the galaxy types, as well as the relative sizes and brightness of the blended sources and cluster members.
This implies that the masking effect might correlate with the cluster mass and redshift.
In a nutshell, the crowded field is expected to have impact on the source selection, photometry and therefore the performance of the photometry redshift estimation in a very subtle but complex way, which is very hard to quantify without a realistic and end-to-end simulations. 
A detailed investigation is clearly warranted (Murata in prep.).

In this work, we choose to empirically validate for the magnification measurements.
Specifically, we carry out the same end-to-end analysis on the ``null-test'' sample, 
for which the net magnification effect vanishes because of $\slope -1=0$ (see Section~\ref{sec:basics} for more details).
We use the ``null-test'' sample to empirically access the residual bias that could exist in the measurements of the ``lensing-cut'' sample.
The only assumption made in this approach is that we assume the residual bias is the same between these two samples.
This is a reasonable assumption, to the first order, given that the ``null-test'' sample is selected as the population with the same color as the ``lensing-cut'' sample, except generally $\approx1$~mag deeper. 
To be exact, the ``null-test'' sample is $\approx1.3$~mag and $\approx1.2$~mag deeper than the ``lensing-cut'' sample for the \lowz\ and \hghz\ background populations, respectively.
As a fact, we also do not observe any significantly difference in the derived cluster contamination \fcl\ and masking correction \fmk\ between the ``null-test''  and ``lensing-cut'' samples. 
This suggests that the residual bias, if exist, is expected to be consistent between these two.

To extract the residual bias, we perform the identical analysis on the ``null-test'' samples to derive the profiles of the density contrast $\frac{n_{\mathrm{d}}}{n_{0}}(R)$.
We use the same binning scheme according to the cluster richness and redshift (as in Section~\ref{sec:fcl} and Section~\ref{sec:masking}), and then apply the corrections of \fcl\ and \fmk\ that are re-derived based on their \mcut.
The results of the ``null-test'' sample of the \lowz\ background is demonstrated in Figure~\ref{fig:fem}, where the red diamonds and black circles represent the profiles of $\frac{n_{\mathrm{d}}}{n_{0}}(R)-1$ before and after the corrections (both \fcl\ and \fmk), respectively. 
To be more quantitative, we fit the density contrast $\frac{n_{\mathrm{d}}}{n_{0}}(R)$ of each richness and redshift bin by a power-law relation,
\begin{equation}
\label{eq:fem}
\fem(R) =  f_{\mathrm{res,1Mpc}} \left(\frac{R}{1~\mathrm{Mpc}}\right)^{\Gamma_{\mathrm{res}}} \, ,
\end{equation}
where  $f_{\mathrm{res,1Mpc}}$ is the normalization, and $\Gamma_{\mathrm{res}}$ is the radial index (as shown in Figure~\ref{fig:fem}).

In Figure~\ref{fig:fem}, we clearly observe the biased-high density enhancement (the black circles) toward the cluster centers for the high-redshift clusters at $0.8\leq\redshift<1.1$; the derived $\Gamma_{\mathrm{res}}$ are all deviating from zero with high significance.
Moreover, the high-richness clusters with $30\leq\rich<100$ show larger residual bias than the low-richness clusters ($15\leq\rich<20$ and $20\leq\rich<30$), especially in the cluster cores.
The fact that the severe residual bias exists in our magnification measurements of the ``null-test'' sample---even after the corrections for the cluster contamination and masking---for the clusters at high redshift ($\redshift\geq0.8$) suggests a failure related to the photometry in the extremely crowded fields.
On the other hand, there is no clear indication that the magnification measurements are significantly corrupted for other richness-redshift bins.
This suggests that the \lowz\ background is a highly pure sample for studying the clusters at $\redshift<0.8$. 
Therefore, we choose not to use the magnification measurements from the \lowz\ background for the clusters at $0.8\leq\redshift<1.1$.

The results of the \hghz\ background are shown in Figure~\ref{fig:fem_append}.
For the case of the \hghz\ background, we find that the density contrast $\frac{n_{\mathrm{d}}}{n_{0}}(R)-1$ is all statistically consistent with zero (within $2\sigma$) for each richness and redshift bin.
This suggests that we can obtain the magnification measurements free from the residual bias for the \hghz\ background.
Therefore, the magnification measurements derived from the \hghz\ background are used for all clusters.

We note that the derived residual bias is typically consistent with zero at a level $\lesssim2\sigma$ for both background populations, except for the highest redshift bin ($0.8\leq\redshift\leq1.1$) in the \lowz\ background.
Moreover, the \fem\ is a noisy estimator of the residual bias for the ``lensing-cut'' sample, because it is estimated by using the ``null-test'' sample of galaxies that are much fainter, smaller and harder to measure.
In this work, the \fem\ is used to empirically calibrate our magnification measurements assuming that the contamination to the ``lensing-cut'' sample (if exists) is consistent with that of the ``null-test'' sample.
This empirical calibration using \fem\ results in a change in the normalization of the \rtm\ relation at a level of $\approx1.9\sigma$, which is taken as the systematic uncertainty in this work (see Section~\ref{sec:sys}).

\begin{figure*}
\centering
\resizebox{\textwidth}{!}{
\includegraphics[scale=1]{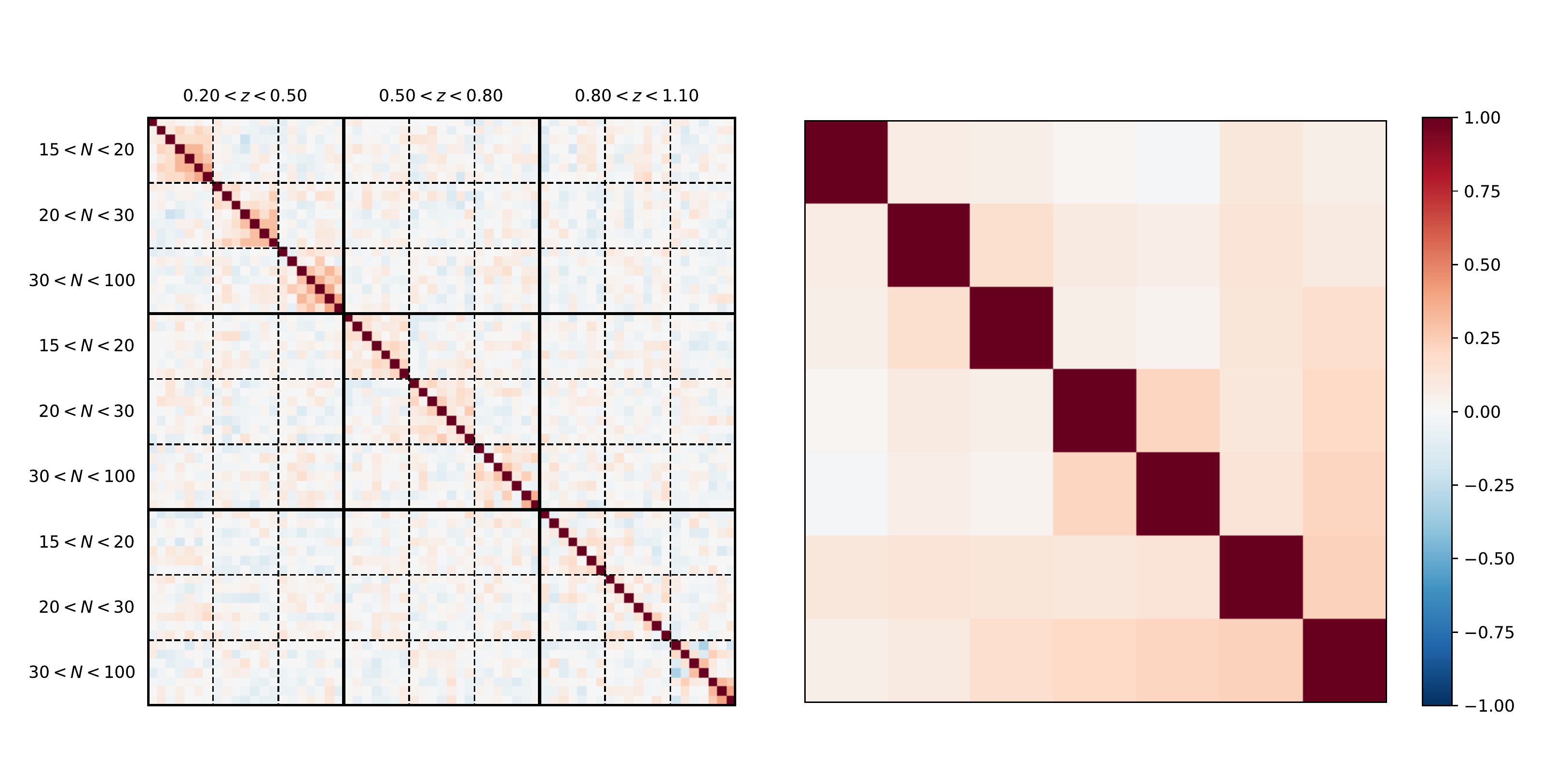}
}
\caption{
The correlation matrix of the magnification bias profiles for the \hghz\ background.
The left panel shows the correlation matrix between different richness and redshift bins, where each box enclosed by the dashed (solid) lines represents the richness (redshift) bins indicated on the left y-axis (top x-axis).
There is no significant correlation of the magnification bias between different richness and redshift bins.
The right panel shows the correlation matrix derived by using all clusters without any richness and redshift binning.
}
\label{fig:corre}
\end{figure*}
\begin{table*}
\centering
\caption{
The summary of the parameters that are constrained in the scaling relation fitting.
The first column records the names of the parameters, which are defined in equations~(\ref{eq:richness_to_mass}),~(\ref{eq:scatter}), and~(\ref{eq:magni_model}).
The second column shows the priors. 
The third to the fifth columns are the resulting constraints of the parameters using the \lowz\, \hghz\ and joint backgrounds, respectively.
The sixth column presents the result including the two background populations of all clusters while applying the correction $\fem(R)$ for the residual bias according to equation~(\ref{eq:correct_em}), in order to access the systematic uncertainty (see the text in Section~\ref{sec:sys}).
The seventh column is the result discarding the innermost radial bin, showing good agreement with our fiducial analysis.
}
\label{tab:params}
\begin{tabular}{rrrrrrr}
\hline
Parameter & Priors & \multicolumn{5}{c}{Constraints} \\[3pt]
\hline\hline
                &           & Low-$z$ & High-$z$ & Joint & Joint  & Joint  \\[3pt]
                &           & & & & Include all clusters   & Discard the  \\
                &           & & & & with the \fem\ correction   & innermost bin \\[3pt] \hline
$A_{\lambda}$      & $(0,100)$                  & $16.93 \pm 7.01$ & $19.35 \pm 3.04$ & $17.72 \pm 2.60$ & $22.77 \pm 3.63$ & $18.43 \pm 3.31$ \\
$B_{\lambda}$      & $\mathcal{N}(0.7, 0.2^2)$ & $0.85 \pm 0.17$ & $0.89 \pm 0.14$ & $0.92 \pm 0.13$ & $0.86 \pm 0.14$ & $0.90 \pm 0.14$ \\
$C_{\lambda}$      & $\mathcal{N}(0.0, 1^2)$   & $-0.55 \pm 0.98$ & $-0.24 \pm 0.71$ & $-0.48 \pm 0.69$ & $-0.05 \pm 0.70$ & $0.18 \pm 0.76$ \\
$\sigma_{\lambda}$ & $\mathcal{N}(0.15, 0.09^2)$ \& $\left(0.01, \inf \right)$   & $0.15 \pm 0.08$ & $0.14 \pm 0.07$ & $0.15 \pm 0.07$ & $0.14 \pm 0.07$ & $0.16 \pm 0.08$ \\
$\delta_{\mathrm{low-}z}$      & $\mathcal{N}(-0.0053,0.0023^2)$   & $-0.0059 \pm 0.0020$ & $\cdots$ & $-0.0060 \pm 0.0019$ & $-0.0054 \pm 0.0020$ & $-0.0057 \pm 0.0019$ \\
$\delta_{\mathrm{high-}z}$     & $\mathcal{N}(-0.0068,0.0028^2)$   & $\cdots$ & $-0.0051 \pm 0.0018$ & $-0.0053 \pm 0.0018$ & $-0.0035 \pm 0.0018$ & $-0.0051 \pm 0.0018$ \\
[3pt]\hline\hline
\end{tabular}
\end{table*}

\subsection{Magnification Bias Profiles}
\label{sec:magnification_estimator}

In this section, we extract the observable of magnification bias using the contrast of the number density profiles.
In what follows, details are given.

For each cluster, we cross-correlate the position of the BCG, as the cluster center, with the source catalog by the estimator, 
\begin{equation}
\label{eq:cross_estimator}
\omega(R) = \frac{\mathrm{LS(R)}}{\mathrm{LR(R)}} - 1 \, ,
\end{equation}
where the symbols of $\mathrm{LS(R)}$ and $\mathrm{LR(R)}$ are the normalized numbers of the cluster-source and cluster-random pairs with the projected separation of $R$, respectively.
We construct a large random catalog ($\approx100$ times larger than the source catalog) that has the same geometric layout as the HSC FDFC footprint accounting for the brightstar- and defect-mask \citep{coupon18}.
Because we derive the estimator $\omega$ for each cluster separately, we do not need to build the random catalog for the clusters.
We use seven logarithmic radial bins ranging from $0.2$~Mpc to $3.5$~Mpc in the physical unit for each cluster, ensuring that we are probing the same portion of the radial profiles of the clusters at different redshifts.
We have confirmed that discarding the inner most bin does not significantly affect the final results (see the last column of Table~\ref{tab:params}).
Finally, the magnification bias profile $\Delta_{\mu}$ is derived after statistically accounting for the cluster contamination (Section~\ref{sec:fcl}) and the masking effect (Section~\ref{sec:masking}).
Specifically, $\Delta_{\mu}(R)$ at the radius $R$ can be derived as follows.
\begin{equation}
\label{eq:magni}
\Delta_{\mu}(R) 
= \left(\omega(R) + 1\right)\times \left(1 - \fcl(R)\right)\fmk(R) - 1 \, .
\end{equation}
The correction for the cluster contamination (the masking effect) is \fcl\ (\fmk) evaluated at the radius of $R$ in the richness and redshift bin where the cluster locates.
For each cluster, we repeat the whole procedure described above for the \lowz\ and \hghz\ backgrounds with both ``lensing-cut'' and ``null-test'' samples each (see Section~\ref{sec:background_selection} for the definitions  of the ``lensing-cut'' and ``null-test'' selection).

\begin{figure*}
\centering
\resizebox{1.1\textwidth}{!}{
\includegraphics[scale=1]{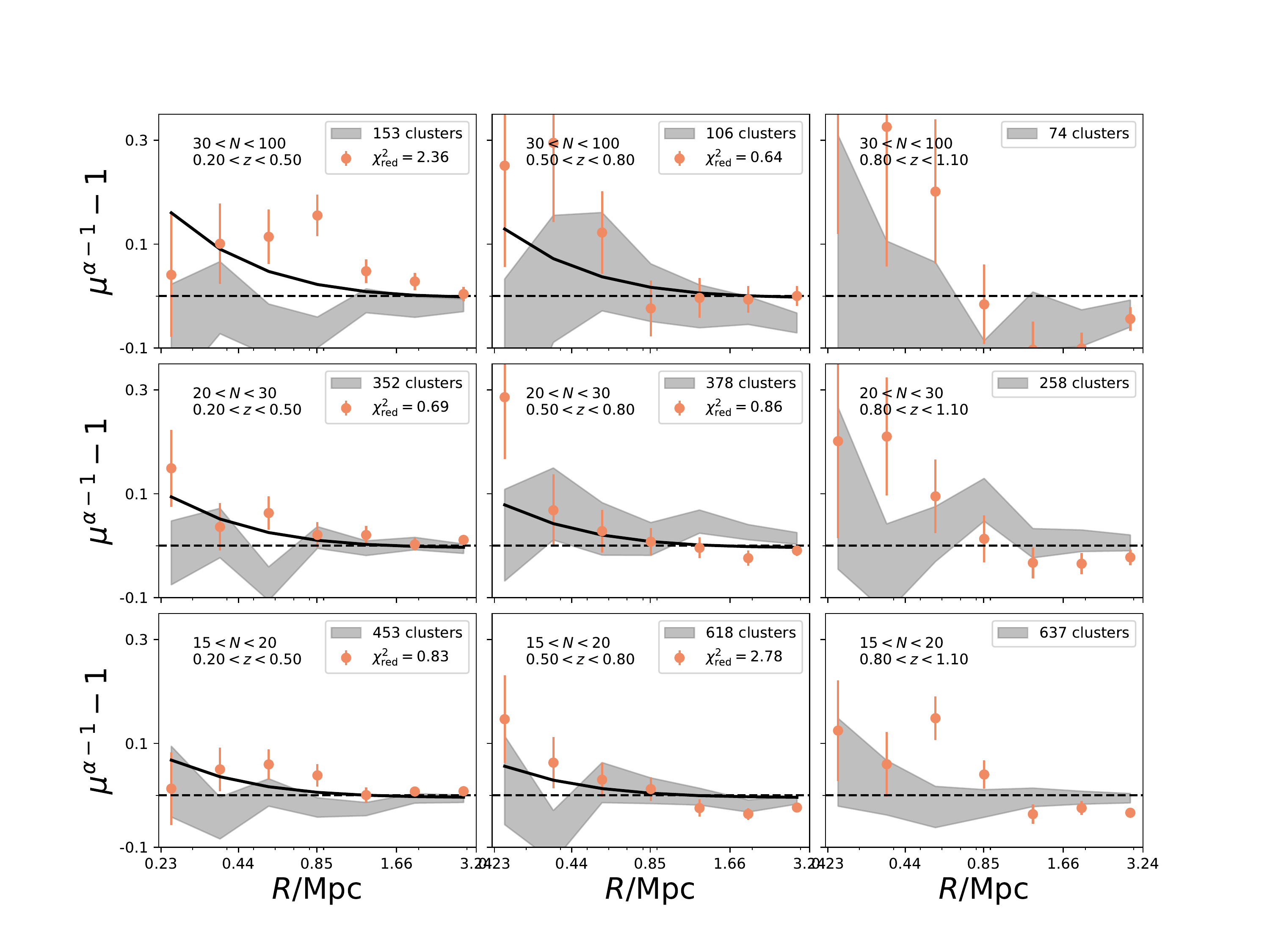}
}
\caption{
The stacked magnification bias profiles of the \lowz\ background in different richness and redshift bins.
The x- and y-axis show the clustercentric radius in the physical unit and the magnification bias measurements, respectively.
The data points are the stacked magnification profiles, while the black curves are the best-fit model predicted by using the joint constraints of the combined background.
The grey area is the measurements by repeating the identical analysis on the random fields.
The dashed lines indicate no signal of magnification bias.
Each panel represents the result of the richness and redshift binning, for which the number of clusters used in the bin and the reduced chi-square with respect to the best-fit model are stated in the upper-right box.
We note that we do not use the clusters at high redshift ($0.8<\redshift<1.1$) in fitting the magnification profiles derived from the \lowz\ background.
}
\label{fig:lowz_prof}
\end{figure*}

\begin{figure*}
\centering
\resizebox{1.1\textwidth}{!}{
\includegraphics[scale=1]{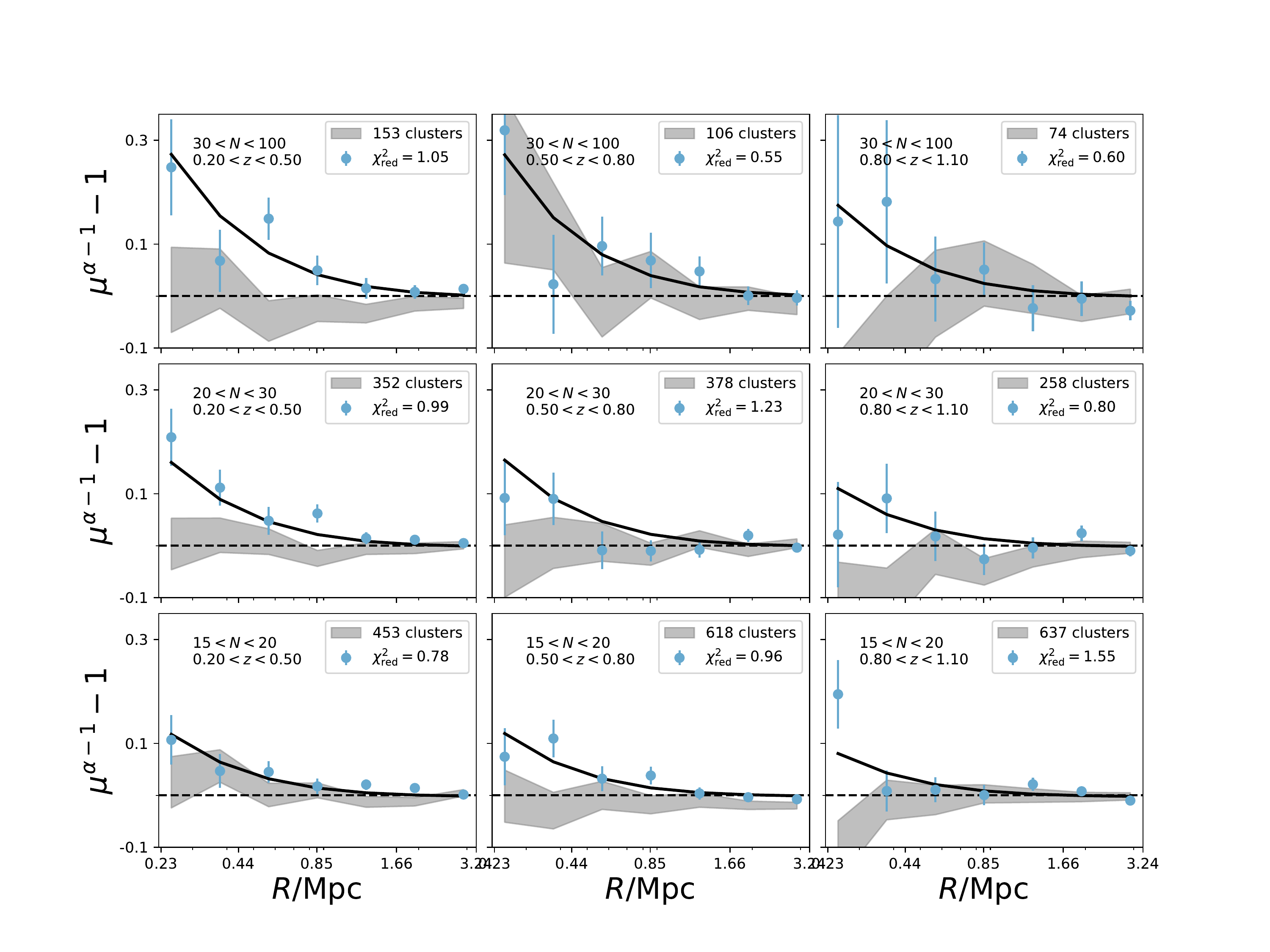}
}
\caption{
The stacked profile made in the same way as Figure~\ref{fig:lowz_prof} but for the case of the \hghz\ background.
}
\label{fig:hghz_prof}
\end{figure*}

We also carry out the same analysis on the random fields to access the possible bias in the estimator of $\omega$.
Specifically, we randomly draw the same amount of apertures in the HSC FDFC footprint with the same physical sizes of each cluster in each richness-redshift bin, and repeat equation~(\ref{eq:magni}) without the corrections for the cluster contamination and the masking effect (i.e., $\fcl(R) = 0$ and $\fmk(R) = 1$).
If the estimator on the HSC fields is unbiased, then we expect that $\omega(R)$ is statistically consistent with zero.
The results are shown as the grey area in Figure~\ref{fig:lowz_prof} and Figure~\ref{fig:hghz_prof} for the \lowz\ and \hghz\ backgrounds, respectively.
Notwithstanding the measurements of the random fields are noisy, we observe a mildly systematic offset deviating from zero that persists out to large radii.
If we stack all clusters together, the global offsets $\delta$ are $-0.0053\pm0.0023$ and $-0.0068\pm0.0028$ for the \lowz\ and \hghz\ backgrounds, respectively.
This offset reflects the imperfect star mask that is failed to capture the outer parts of bright stars at a sub-percent level. 
This results an over-sampled random catalog around the edges of bright stars and, therefore, a resulting biased-low $\omega(R)$.
This needs to be accounted for, because this amount of offset is comparable to the signal of the 2-halo term, which is included in our modelling (see Section~\ref{sec:sr_modelling}).
It is important to note that this global offset is independent of clusters and has no radial dependence.
Thus, we choose to marginalize this bias $\delta$ in our forward-modelling approach (see Section~\ref{sec:sr_modelling}).

The observed and corrected magnification profiles are still deviated from the underlying ones due to the presence of the measurement uncertainty, uncorrelated large-scale structure, and the intrinsic scatter at fixed mass.
In this work, these factors above are characterized and accounted for by the covariance matrices that are directly derived from the data, following the same approach in \cite{melchior17}.
Specifically, we conduct a spatial jackknife technique that divides the footprint into 300 equal-area patches on the sky using a $k$-means algorithm\footnote{https://github.com/esheldon/kmeans\_radec}, followed by the same amount of repetition of deriving the measurements, i.e., equation~(\ref{eq:magni}), while omitting one patch each time.
For the $j$-th repetition with $j$-th patch omitted, we concatenate the measurements as a data vector denoted by ${\Delta_{\mu}}_{(j)}$.
Then, the data covariance matrix is obtained as
\begin{equation}
\label{eq:covar}
\mathbb{C} =  \frac{N_{\mathrm{K}} - 1}{N_{\mathrm{K}} - N_{\mathrm{D}} - 2}
\frac{N_{\mathrm{K}}-1}{N_{\mathrm{K}}}\sum_{j=1}^{N_{\mathrm{K}}} 
\left( {\Delta_{\mu}}_{(j)} - {\Delta_{\mu}}_{(\cdot)} \right)^{\mathrm{T}} \cdot
\left( {\Delta_{\mu}}_{(j)} - {\Delta_{\mu}}_{(\cdot)} \right) \, ,
\end{equation}
where ${\Delta_{\mu}}_{(\cdot)} = \frac{1}{N_{\mathrm{K}}} \sum_{j=1}^{N_{\mathrm{K}}} {\Delta_{\mu}}_{(j)}$, $N_{\mathrm{D}}$ is the total number of the measurements in the data vector, and $N_{\mathrm{K}}$ is the number of the equal-area patches, i.e., $N_{\mathrm{K}} = 300$.
The inclusion of the first term, $\frac{N_{\mathrm{K}} - 1}{N_{\mathrm{K}} - N_{\mathrm{D}} - 2}$, is needed because the noisy covariance matrix tends to underestimate the uncertainty \citep{hartlap07}.
The size of one jackknife patch corresponds to the physical radius of $\gtrsim8$~Mpc for the cluster at $\redshift>0.2$, which is suitable for the radial scale of interest in this work.
We have verified that the resulting covariance matrix is converged and is not sensitive to the current choice of $N_{\mathrm{K}}$. 
The results of normalized covariance matrix (i.e., the correlation matrix) of the \hghz\ background is presented in Figure~\ref{fig:corre}, as an example.

We do not observe significant cross-correlation of the magnification measurements among different richness and redshift bins, as demonstrated in the left panel of Figure~\ref{fig:corre}, although the correlation matrices are noisy. 
Thus, we treat the magnification measurements of different richness-redshift bins as the independent measurements.
We also derive the full covariance matrix by calculating equation~(\ref{eq:covar}) using all clusters without the richness and redshift binning. 
The full correlation matrix is shown in the right panel of Figure~\ref{fig:corre}.
We confirm that there is no significant difference in term of the radial correlation pattern between the subsample and the full sample.
A similar picture is also suggested for the \lowz\ background.

\begin{figure}
\centering
\resizebox{0.5\textwidth}{!}{
\includegraphics[scale=1]{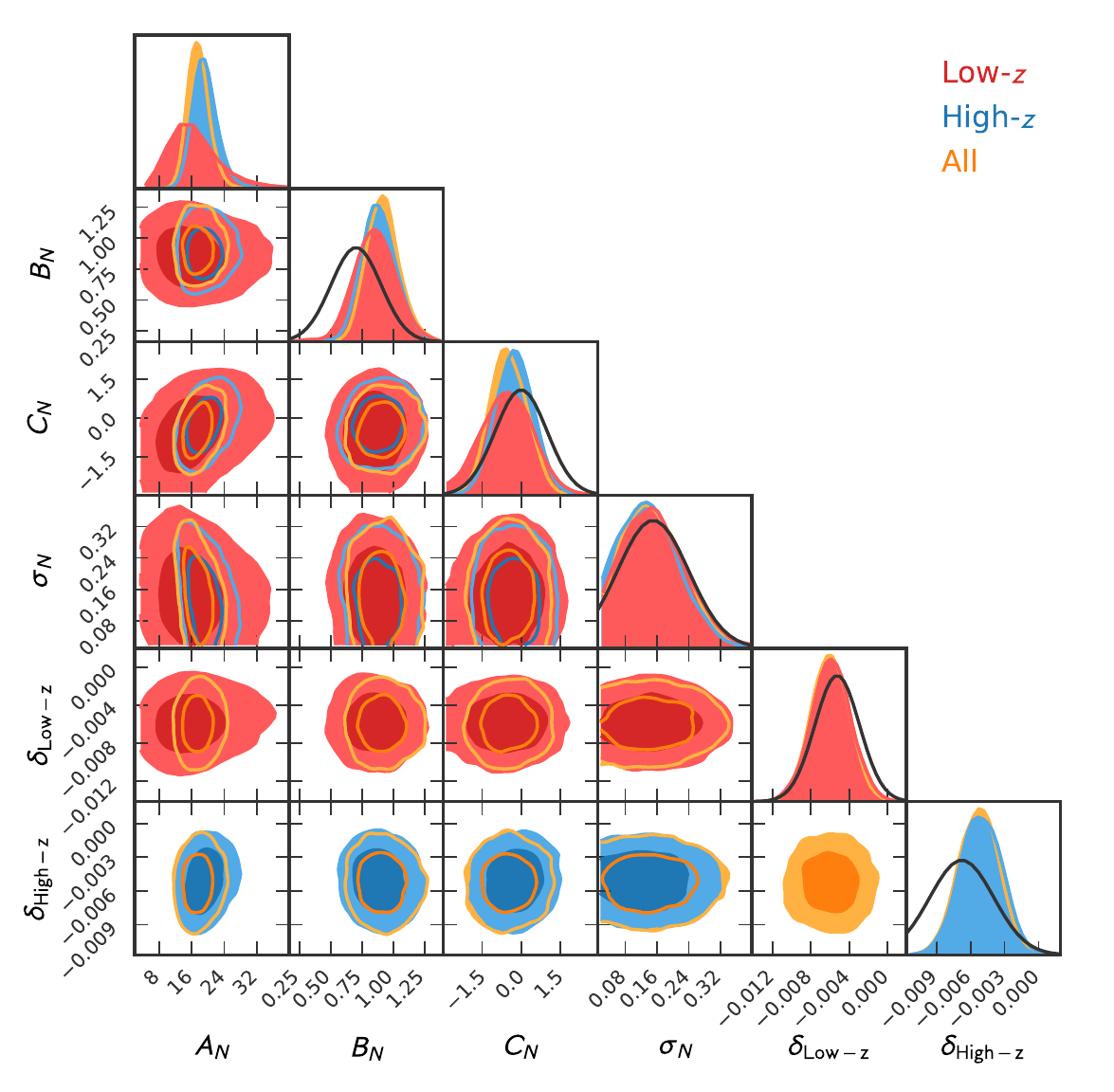}
}
\caption{
The constraints of the \rtm\ scaling relation parameters $\left(\Arich, \Brich, \Crich, \Drich\right)$ and the marginalized global offsets $\left(\delta_{\mathrm{Low}-z}, \delta_{\mathrm{High}-z}\right)$.
The red (blue) contours represent the $1\sigma$ and $2\sigma$ confidence levels for the \lowz\ (\hghz) background.
The resulting constraints of the \lowz\ and \hghz\ backgrounds are statistically consistent with each other, so we combine them to obtain the joint constraints, as shown by the yellow contours.
The black curves are the adopted Gaussian priors on the parameters (see Table~\ref{tab:params}).
We note that we additionally apply a hard cut on the lower bound of the intrinsic scatter parameter to ensure that $\Drich>0.01$.
}
\label{fig:gtc}
\end{figure}

\subsection{Modelling of the Scaling Relation}
\label{sec:sr_modelling}

Our goal is to characterize the \rtm\ relation of the CAMIRA cluster sample using the HSC optical richness and weak-lensing magnification of individual clusters.
To this end, we use a forward-modelling approach to constrain the richness-to-mass relation.
Specifically, we assume an underlying \rtm\ relation characterized by its log-normal intrinsic scatter at fixed mass and evaluate the probability of observing the weak-lensing magnification signal for a given richness.
This method has been widely used in previous work to calibrate the observable-to-mass relation of a cluster sample with a well-defined selection function  \citep{chiu16c,chiu18a,bulbul18}. 
In addition, the effects of Malmquist and Eddington bias are fully accounted for in this approach.  
We have tested and validated the fitting procedure by using mock observations that are  more than ten times larger than our cluster
sample. 
It is found that we can recover the true input parameters of the scaling relation within the $1\sigma$ uncertainties.  
We describe our analysis framework below and refer the interested reader to \cite{liu15a} and \cite{bocquet15} for more details.

For the $i$-th cluster at redshift ${\redshift}_{i}$ with the observed
magnification profile ${\Delta_{\mu}}_{i}$ and the richness
${\rich}_{i}$, we evaluate the likelihood
\begin{eqnarray}
\label{eq:like}
\mathcal{L}( {\Delta_{\mu}}_{i} | {\rich}_{i}, {\redshift}_{i}, \mathbf{p} ) &= &\frac{
P({\Delta_{\mu}}_{i}, {\rich}_{i} | {\redshift}_{i}, \mathbf{p})
}{
P({\rich}_{i} | {\redshift}_{i}, \mathbf{p})
} \nonumber \\
&= &\frac{
\int P({\Delta_{\mu}}_{i}, {\rich}_{i} | \Mfiveoo, {\redshift}_{i},  \mathbf{p}) n(\Mfiveoo, {\redshift}_{i}) \dif \Mfiveoo
}{
\int P({\rich}_{i} |  \Mfiveoo, {\redshift}_{i},  \mathbf{p}) n(\Mfiveoo, {\redshift}_{i}) \dif \Mfiveoo
} \nonumber  \, , \\ 
\end{eqnarray}
where $n(\Mfiveoo, {\redshift}_{i})$ is the mass function at redshift
${\redshift}_{i}$, and $\mathbf{p}$ represents the parameter vector
containing $(\Arich, \Brich, \Crich, \Drich)$ of the \rtm\ relation,
\begin{equation}
\label{eq:richness_to_mass}
\left\langle\ln\rich|\Mfiveoo\right\rangle = \ln\Arich + \Brich\ln\left( \frac{\Mfiveoo}{\MPIV} \right) + \Crich\ln\left(\frac{1 + \redshift}{1 + \ZPIV}\right) \, ,
\end{equation}
with the log-normal intrinsic scatter at fixed mass of 
\begin{equation}
\label{eq:scatter}
\Drich \equiv \sigma_{\ln\rich | \Mfiveoo} \, .
\end{equation}
The pivot mass and redshift are fixed as $\MPIV = 10^{14}h^{-1}\Msun$ and $\ZPIV = 0.6$, respectively.
The inclusion of the mass function in equation~(\ref{eq:like}) is necessary to account for the Eddington bias.
Here we use the mass function of \citet{bocquet16}.

The denominator of equation~(\ref{eq:like}) represents the probability
of obtaining the richness ${\rich}_{i}$ for the $i$-th cluster at
redshift ${\redshift}_{i}$ given the underlying mass \Mfiveoo\ with the
log-normal intrinsic scatter \Drich\ and the measurement uncertainty. 
%for which 
The variance of the observed $\ln\left(\rich\right)$ at fixed
$\Mfiveoo$ can be expressed as 
\begin{equation}
\label{eq:var_rich}
\mathrm{Var}\left(\ln\rich | \Mfiveoo\right) = {\Drich}^{2} + \exp\left(-\left\langle \ln\rich |\Mfiveoo\right\rangle\right) \, ,
\end{equation}
where the second term is due to the Poisson noise.
Hence, the $P({\rich}_{i} |  \Mfiveoo, {\redshift}_{i}, \mathbf{p})$ term
can be evaluated for a given parameter vector
$\mathbf{p}$.

On the other hand, we can decompose the numerator of
equation~(\ref{eq:like}) as 
\begin{equation}
\label{eq:numerator}
P({\Delta_{\mu}}_{i}, {\rich}_{i} | \Mfiveoo, {\redshift}_{i},  \mathbf{p}) = 
P({\Delta_{\mu}}_{i} | \Mfiveoo, {\redshift}_{i},  \mathbf{p}) 
P({\rich}_{i} | \Mfiveoo, {\redshift}_{i},  \mathbf{p}) \, ,
\end{equation}
assuming that there is no correlated intrinsic scatter between $\Delta_{\mu}$ and \rich.
This is a reasonable assumption because the magnification profile is
derived using the background source sample behind clusters, which is
independent of cluster member galaxies, to first order. 
We evaluate the log-probability of the first term in equation~(\ref{eq:numerator}) as
\begin{equation}
\label{eq:magni_prob}
\ln P({\Delta_{\mu}}_{i} | \Mfiveoo, {\redshift}_{i},  \mathbf{p}) = 
-\frac{1}{2} \left({\Delta_{\mu}}_{i} - {\Delta_{\mathrm{model}}}_{i}\right)^{\mathrm{T}} \mathfrak{C}^{-1} \left({\Delta_{\mu}}_{i} - {\Delta_{\mathrm{model}}}_{i}\right) \, ,
\end{equation}
where
$\mathfrak{C}=N_{i\in\rich-\redshift~\mathrm{bin}}\times\mathbb{C}$ is
the covariance matrix in the richness-redshift bin which the $i$-th
cluster belongs to (see Section~\ref{sec:magnification_estimator}),
re-scaled by the number $N_{i\in\rich-\redshift~\mathrm{bin}}$ of
clusters in that bin; 
${\Delta_{\mathrm{model}}}_{i}$ is the model prediction of the
magnification profile for the $i$-th cluster with \Mfiveoo. 
Specifically, the model prediction ${\Delta_{\mathrm{model}}}_{i}$ at the projected radius $R$ is expressed as
\begin{equation}
\label{eq:magni_model}
{\Delta_{\mathrm{model}}}_{i}(R) = \mu_{i}(R)^{\slope - 1} -1 + \delta \, ,
\end{equation}
where the slope \slope\ is fixed to the value at the magnitude cut
\mcut.
Additionally, we include one more free parameter $\delta$ in our
modelling to account for the global offset in our estimator caused by 
residual systematics in the bright-star mask correction, as we
quantified using random field measurements (see
Section~\ref{sec:magnification_estimator}).
To evaluate $\mu_{i}$, we adopt the standard halo model \citep{oguri11}, which is a linear sum of a smoothly truncated version of the Navarro--Frenk--White \citep[hereafter NFW;][]{navarro1997} profile and the 2-halo term:
\begin{eqnarray}
 \begin{aligned}
\label{eq:halomodel}
\rho(r) &= f_{\mathrm{t}}(r)\rho_{\mathrm{NFW}}(r) +
  \rho_\mathrm{2h}(r) \, , \\
  \rho_\mathrm{NFW}(r) &= \frac{\rho_\mathrm{s}}{(r/r_\mathrm{s})\left(1+r/r_\mathrm{s}\right)^2} \, ,\\
  f_{\mathrm{t}}(r) &= \left(\frac{1}{1 + \frac{r^2}{{r_{\mathrm{t}}}^2}}\right)^2 \, , 
  \end{aligned}
\end{eqnarray}
where $\rho_\mathrm{NFW}(r)$ is the NFW density profile specified by the characteristic scale density $\rho_\mathrm{s}$ and the characteristic scale radius $r_\mathrm{s}$, the transition term $f_\mathrm{t}(r)$ characterizes the steepening around a truncation radius, $r_\mathrm{t}$,   and the 2-halo term $\rho_\mathrm{2h}(r)$ is expressed as
\begin{equation}
\label{eq:2halo}
\rho_{\mathrm{2h}}(r) = \rho_{\mathrm{m}}(\redshift) b(\Mfiveoo,\redshift)\xi_\mathrm{m}(r) \, ,
\end{equation}
with $b(\Mfiveoo,\redshift)$ the linear bias factor, $\rho_\mathrm{m}(\redshift)$ the mean matter density of the universe evaluated at the cluster redshift, and  $\xi_\mathrm{m}(r)$ the linear matter correlation function at the cluster redshift.\footnote{We express the correlation function in physical length units.}
We set $r_{\mathrm{t}} = 4.5\Rfiveoo$, which is consistent with the typical value used in the literature \citep{oguri11,umetsu16}. 
We note that equation~(\ref{eq:halomodel}) reduces to the Baltz--Marchall--Oguri model \citep{baltz09} when the 2-halo term is ignored.

We fix the concentration parameter of the NFW model and the linear bias
factor $b(\Mfiveoo,\redshift)$ to those predicted for the $i$-th
cluster with \Mfiveoo\ at the redshift ${\redshift}_{i}$
using the scaling relations of \citet{diemer15} and \citet{tinker10},
respectively.
Next, we calculate the surface mass density $\Sigma_{\mathrm{m}}(R)$ of the cluster at the projected radius $R$ by integrating the mass density $\rho(r)$ along the line of sight (i.e., $\Sigma_{\mathrm{m}}(R)=\int \rho(\sqrt{R^2 + x^2})\dif x$).
Last, the mean lensing efficiency, $\beta \equiv \int \frac{\dls}{\ds} P_{\mathrm{f}}(\redshift)\dif\redshift$, weighted by the stacked photometric redshift distribution $P_{\mathrm{f}}(\redshift)$ of the source sample in the random field, is used to estimate the critical surface mass density $\Sigma_{\mathrm{c}}$. 
That is, only one parameter---the cluster mass \Mfiveoo---is needed to compute the model ${\Delta_{\mathrm{model}}}_{i}$ for the $i$-th cluster at the redshift ${\redshift}_{i}$.

The final likelihood is the product of all clusters,
\begin{equation}
\label{eq:final_likelihood}
\mathcal{L}(\mathbf{p}) = \prod\limits_{i=1}^{N_{\mathrm{cl}}} \mathcal{L} ( {\Delta_{\mu}}_{i} | {\rich}_{i}, {\redshift}_{i}, \mathbf{p} )  \, ,
\end{equation}
where $N_{\mathrm{cl}}$ is the number of clusters.
We explore the parameter space for our model using Bayesian inference in
which the posterior probability distribution $P(\mathbf{p})$ is
expressed as
\begin{equation}
\label{eq:posteriors}
P(\mathbf{p}) = \mathcal{L(\mathbf{p})}\cdot\mathcal{P(\mathbf{p})} \, ,
\end{equation}
where $\mathcal{P(\mathbf{p})}$ is the prior probability distribution of
the parameters $\mathbf{p}$.
When analyzing the \lowz\ or \hghz\ background sample separately,
we have five parameters in $\mathbf{p}$,
namely the scaling relation parameters $\left(\Arich, \Brich, \Crich, \Drich\right)$ and a
parameter $\delta$ describing the global offset of the estimator for
each background sample (see Section~\ref{sec:magnification_estimator}). 
For a joint analysis of the combined \lowz\ and \hghz\ backgrounds,
we have six parameters, $\mathbf{p} = \left\lbrace\Arich, \Brich,
\Crich, \Drich, \delta_{\mathrm{\lowz}},
\delta_{\mathrm{\hghz}}\right\rbrace$. 
We use \texttt{emcee} \citep{foreman13}, the \texttt{Python} code
employing the Affine Invariant Markov Chain Monte Carlo (MCMC)
algorithm, to obtain the posterior probability distribution for our
model parameters.

We stress that we do not fit the magnification measurements of the clusters at the highest redshift bin ($0.8\leq\redshift<1.1$) derived using the \lowz\ background, because our validation tests suggest that these measurements are highly corrupted (see Figure~\ref{fig:fem}).
The residual bias must be removed (based on the \fem\ quantified in Section~\ref{sec:em}) in order to include this sample in the fitting, otherwise introducing significant bias in the scaling relation parameters.
We carry out the fitting with and without the correction for the residual bias in Section~\ref{sec:sys}, and find that the results are consistent with each other at a level of $1.9\sigma$ (see Table~\ref{tab:params}).

In this work, we cannot constrain the mass, redshift trends and the
log-normal intrinsic scatter using the current data alone. 
We thus focus on constraining the normalization parameter \Arich,
and adopt informative priors on \Brich, \Crich, and \Drich.
Specifically, a Gaussian prior of $\mathcal{N}(0.7, 0.2^2)$ is applied on the mass trend parameter \Brich.
This prior is different from, but statistically consistent with,
the previous results of \cite{oguri14}, \cite{murata18}, and
\cite{murata19}.
We verified that the constraint on \Arich\ is insensitive to the choice of the prior on \Brich.
However, the constraint of \Brich\ will be unbounded toward a non-physically high value with a flat prior, which is the cause of the shift in our final posterior of \Brich\ from the adopted prior.
We assume a conservative Gaussian prior on the redshift trend parameter
\Crich\ of $\mathcal{N}(0, 1^2)$ with a zero mean.
We note that, according to previous studies
\citep{saro15,mcclintock18, capasso19a}, we do not expect a strong
dependence of richness on redshift at fixed mass.
For the intrinsic scatter \Drich, we use the normal distribution
$\mathcal{N}(0.15,0.09^2)$ suggested by \cite{saro15}, with a lower bound of $0.01$.
We apply a uniform prior between 0 and 100 on the normalization $\Arich$.
The priors on  $\delta_{\mathrm{\lowz}}$ and $\delta_{\mathrm{\hghz}}$
are taken to be
$\mathcal{N}(-0.0053,0.0023^2)$ and
$\mathcal{N}(-0.0068,0.0028^2)$, respectively,
which are based on our analysis of random fields (see Section~\ref{sec:magnification_estimator}). 
These priors are summarized in Table~\ref{tab:params}.

\begin{figure*}
\centering
\resizebox{0.33\textwidth}{!}{
\includegraphics[scale=1]{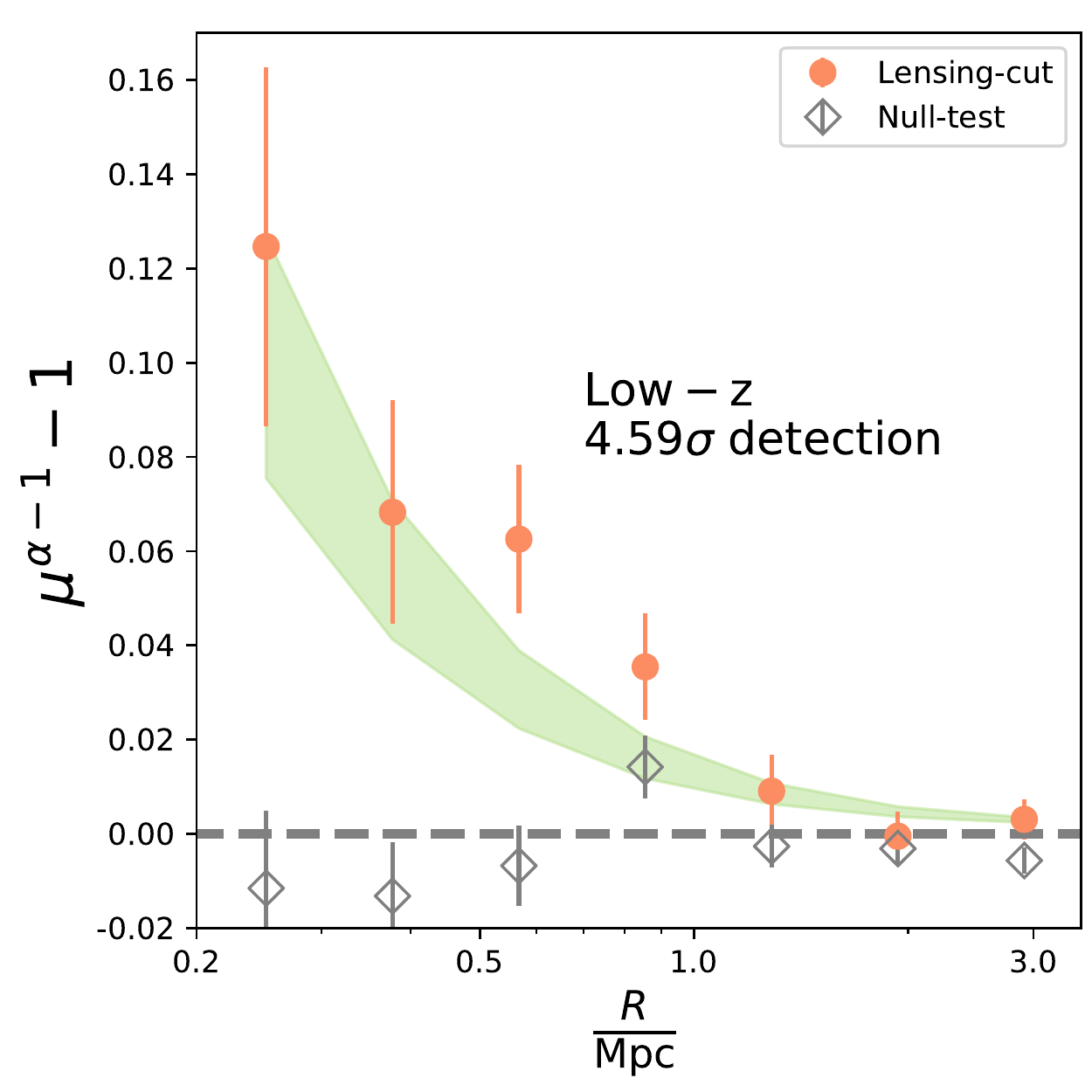}
}
\resizebox{0.33\textwidth}{!}{
\includegraphics[scale=1]{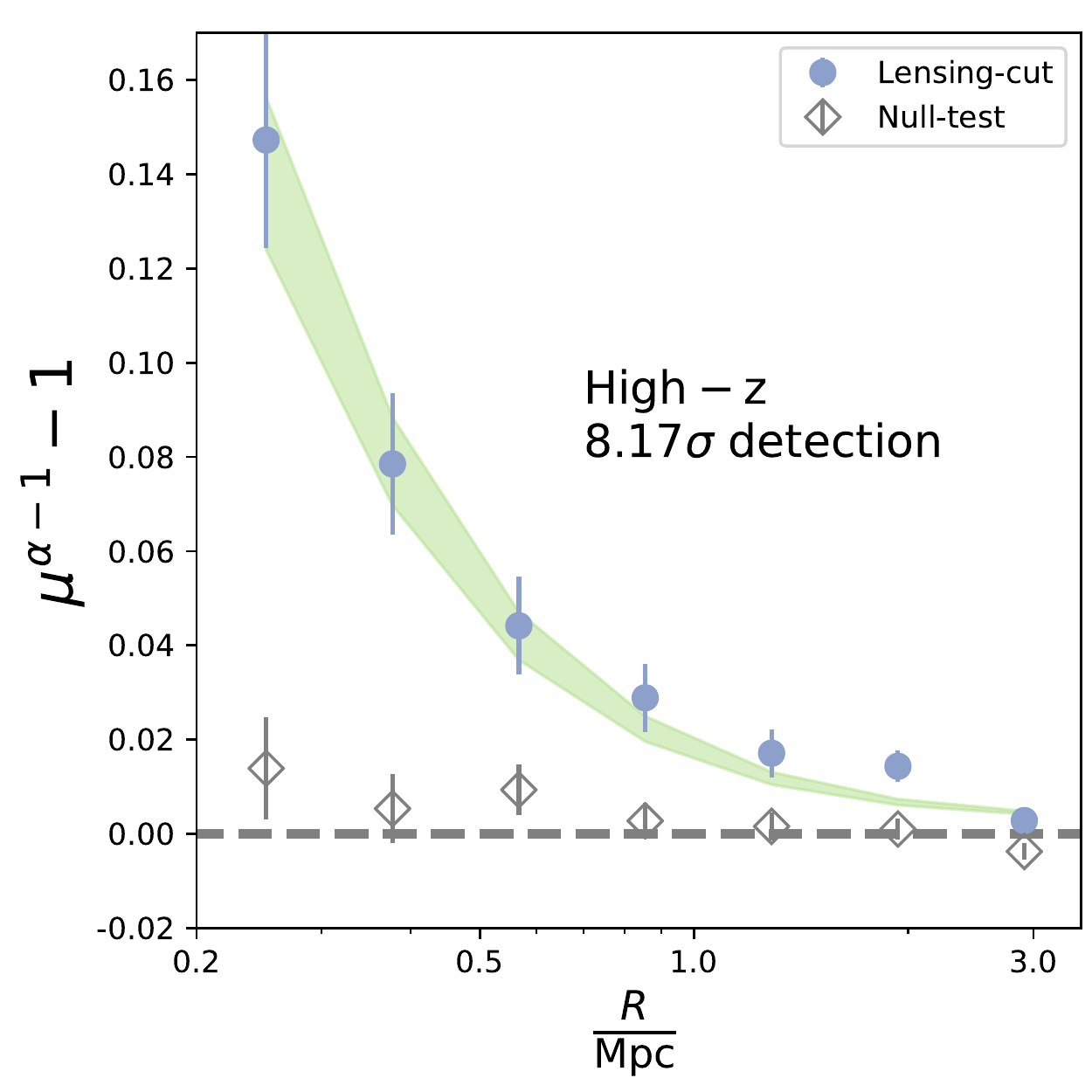}
}
\resizebox{0.33\textwidth}{!}{
\includegraphics[scale=1]{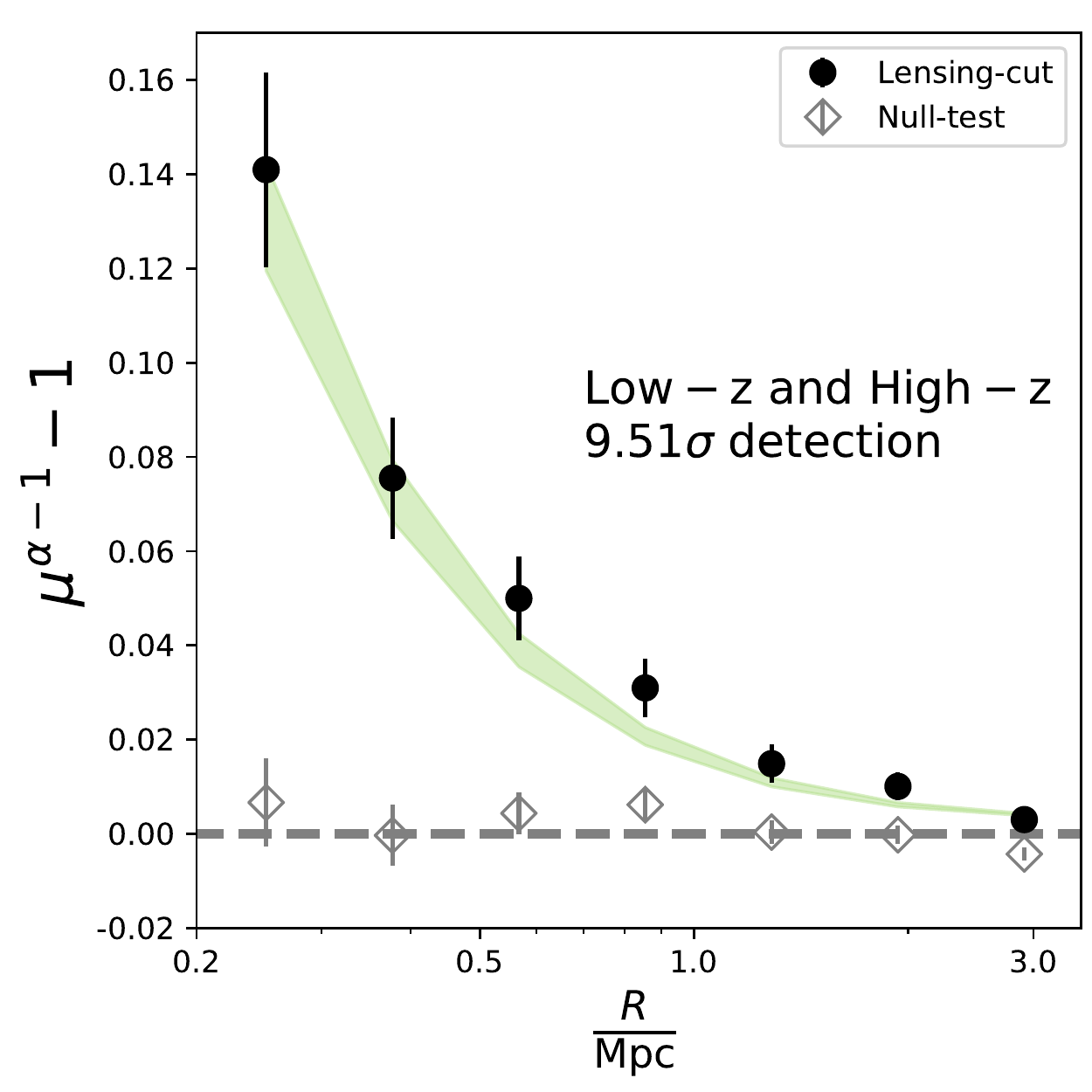}
}
\caption{
The stacked magnification profiles of \lowz\ (left), \hghz\ (middle) and the joint (right) backgrounds.
The circles are the stacked profiles of the ``lensing-cut'' samples, showing the detection significance labelled in each plot.
On the other hand, the open diamonds are the ``null-test'' samples, which are used to validate the magnification measurements and quantify the residual bias \fem; they are all statistically consistent with zero.
The green shaded regions in the left, middle and right panels, are the best-fit with the $1\sigma$ confidence level using the \lowz, \hghz\ and the joint backgrounds, respectively.
}
\label{fig:stackedprofiles}
\end{figure*}
%

%%%%%%%%%%%%%%%%%%%%%%%%%%%%%%%%%%%%%%%%%%
%
% Results
%
%%%%%%%%%%%%%%%%%%%%%%%%%%%%%%%%%%%%%%%%%%

\section{Results and Discussion}
\label{sec:results}

\begin{figure}
\centering
\resizebox{0.48\textwidth}{!}{
\includegraphics[scale=1]{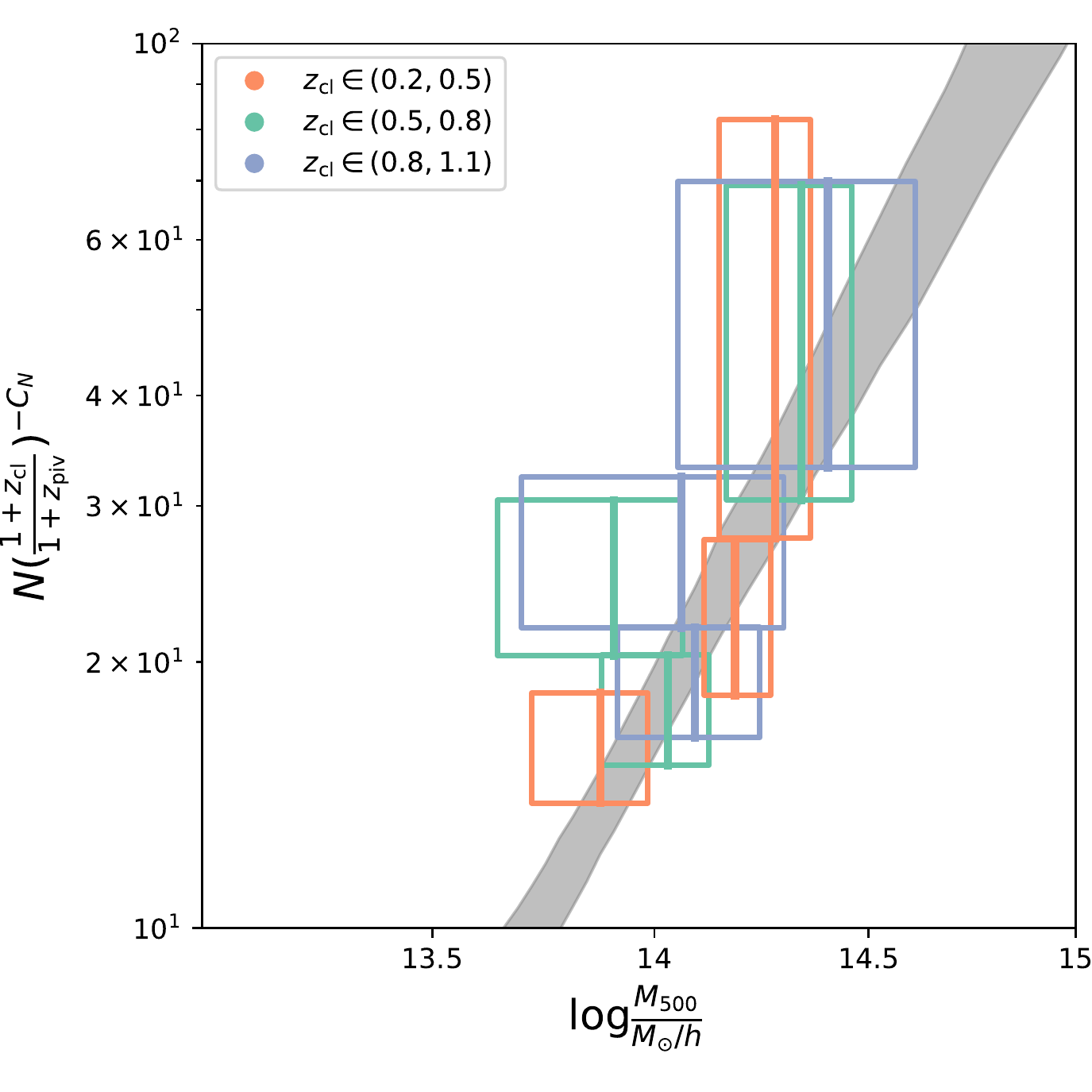}
}
\caption{
The stacked mass estimates of different richness-redshift bins and the resulting \rtm\ scaling relation of the CAMIRA clusters.
The stacked mass in each richness and redshift bin is derived using the constraints of the combined background and is color-coded by the redshift as shown in the figure.
The central and vertical line in each box indicates the peak location of the stacked
 mass probability distribution $P(M_{500})$, while the horizontal width
 of the box indicate the lower and upper
 $68\percent$ confidence limits (see Section~\ref{sec:results}).
The range of the box in the y-axis represents the range of the richness of the clusters in the bin, normalized to the pivot redshift \ZPIV.
The best-fit scaling relation is obtained by marginalizing all parameters constrained by the combined background, as shown in grey.
}
\label{fig:srrelation}
\end{figure}

In this section, we present and discuss the main results of this work, including the parameter constraints, the best-fit \rtm\ scaling relation, the stacked lensing profiles, and cluster mass estimates of individual clusters calibrated by weak lensing magnification.

In Section~\ref{sec:sr_modelling}, we use a forward-modelling technique to simultaneously fit all individual clusters in our CAMIRA sample, effectively stacking all systems in the Bayesian likelihood space, and derive direct constraints on the parameters describing the \rtm\ scaling relation, namely $\left(\Arich,  \Brich, \Crich, \Drich\right)$. 
We first carry out the modelling for the \lowz\ and \hghz\ backgrounds separately.
The results are summarized in Figure~\ref{fig:gtc} and Table~\ref{tab:params}.
We note that we can only constrain the normalization \Arich\ of the scaling relation in this work, and  the posterior distributions of the other parameters are mostly driven by the chosen priors.
As seen in Figure~\ref{fig:gtc}, our constraint on \Arich\ based on the \lowz\ background is in good agreement with that on the \hghz\ background. 
Therefore, we can combine both \lowz\ and \hghz\ backgrounds to obtain joint constraints on the scaling relation, as shown in Figure~\ref{fig:gtc}. 
The resulting normalization $\Arich$ of the joint constraints is $17.72\pm 2.60$, corresponding to an uncertainty at the level of $15\percent$.

Next, we present the stacked profiles of lensing magnification bias. 
We stress again that we have simultaneously modeled all individual clusters in the likelihood space, so that we show the stacked lensing profiles here for visualization purposes only.
Since we predict magnification profiles for all individual clusters in each sampled point of the parameter space, we can stack the best-fit profile of each individual cluster in data space.  
The stacked lensing magnification profiles of the \lowz\ background in different richness and redshift bins are presented in
Figure~\ref{fig:lowz_prof}, where the data points represent the stacked lensing measurements (see Section~\ref{sec:magnification_estimator}).
Similarly, Figure~\ref{fig:hghz_prof} shows the results for the \hghz\ background. 
As seen in these figures, we find broad agreement between the stacked measurements and model predictions.

Furthermore, we stack all clusters together without richness and redshift binning. Since the covariance matrix depends on the cluster properties and the background source populations, we weight the magnification profiles of individual clusters by $\vec{\omega}$.  
The weighting vector $\vec{\omega}$ is defined as the inverse of the diagonal part of the covariance matrix of each cluster. 
That is, $\vec{\omega} \equiv 1/\mathrm{diag}\left(\mathbb{C}_{i, b}\right)$, where $i$ runs over all clusters, and $b$ runs over the background populations, namely the \lowz\ and \hghz\ backgrounds. 
We thus ignore the off-diagonal terms of the covariance matrix here for simplicity.

The resulting magnification profiles of the \lowz\ and \hghz\ backgrounds are presented in the left and middle panels of Figure~\ref{fig:stackedprofiles}, respectively.
In the left and middle panels, we overplot the best-fit model profiles for the \lowz\ and \hghz\ backgrounds with their respective $1\sigma$ confidence range (green shaded regions).
The significance levels of detection after stacking are $4.59\sigma$ and $8.17\sigma$ for the \lowz\ and \hghz\ backgrounds, respectively. 
The significance is calculated as the confidence level to reject the probability (or $p$-value) that the observed profile is caused by random fluctuations.
Specifically, the $p$-value is calculated using the survival function of a $\chi^2$-distribution evaluated at $\chi^2=\Delta_{\mu}(R)^{\mathrm{T}} \cdot \mathfrak{C}^{-1} \cdot \Delta_{\mu}(R)$ with seven degrees of freedom, where $\mathfrak{C}$ is the re-scaled covariance matrix defined in equation~(\ref{eq:magni_prob}).

We also stack the \lowz\ and \hghz\ backgrounds together to derive the combined stacked profile of magnification bias, as shown in the right panel of Figure~\ref{fig:stackedprofiles}. 
In total, stacking the \lowz\ and \hghz\ backgrounds for all 3029 clusters yields a detection significance level of $9.51\sigma$.  
As shown in Figure~\ref{fig:stackedprofiles}, the best-fit models provide reasonably good fits to the observed magnification measurements. 
We also overplot the stacked results of the ``null-test'' samples in Figure~\ref{fig:stackedprofiles}, showing that they are all statistically consistent with zero (i.e., no residual deviations).

In Figure~\ref{fig:srrelation} we summarize our results in the richness--mass space.
Here the grey shaded area represents the joint constraints from the combined \lowz\ and \hghz\ background populations, marginalized over all parameters at the pivot redshift \ZPIV.
The open boxes color-coded by the cluster redshift show the binned stacked constraints in different richness and mass bins.
To be more exact, the stacked mass probability distribution $P(\Mfiveoo)$ in each richness and redshift bin is obtained by computing the joint probability distribution of equation~(\ref{eq:magni_prob}) for all clusters in that bin as a function of \Mfiveoo. 
A clear positive correlation between the richness and cluster mass can be seen in Figure~\ref{fig:srrelation}, albeit with large scatter. 
This \rtm\ relation exhibits little redshift dependence, as we quantified in this work ($\Crich=-0.48 \pm 0.69$).   
As shown in Figures~\ref{fig:lowz_prof}--\ref{fig:srrelation}, our results clearly demonstrate that we obtain significant constraints on the underlying \rtm\ scaling relation by stacking a large sample of clusters together, although each individual cluster measurement is noisy.
In this work, we determine the normalization parameter \Arich\ to a $15\percent$ precision.

Once we determine the underlying \rtm\ scaling relation, we can infer a lensing-calibrated estimate of \Mfiveoo\ for each cluster using its observed richness in an ensemble manner.  
Specifically, we derive the probability distribution $P(\Mfiveoo |\rich,\redshift)$ of $\Mfiveoo$ given the observed richness \rich\ at redshift \redshift\ as  
\[
P(\Mfiveoo | \rich, \redshift, \mathbf{p}) \propto P(\rich|\Mfiveoo, \redshift, \mathbf{p} ) P(\Mfiveoo | \redshift, \mathbf{p}) \, ,
\]
using the Bayes' theorem.
Here we evaluate the second term $P(\Mfiveoo | \redshift, \mathbf{p})$ using the halo mass function.
Next, we randomly draw 500 realizations of the parameters \textbf{p} from the MCMC-sampled posterior distributions and derive the probability distribution of $\Mfiveoo$ for each cluster given the richness and redshift.
We use the posterior distributions from the joint \lowz\ and \hghz\ background constraints.
In this way, we construct $P(\Mfiveoo | \rich, \redshift)$ for each cluster while effectively marginalizing over all parameters. 
Finally, we randomly sample the value of $\Mfiveoo$ for each cluster from their resulting $P(\Mfiveoo | \rich, \redshift)$. 
This approach has been widely used in previous work \citep[e.g.,][]{chiu18a}, which demonstrates a statistically robust method for inferring individual cluster masses from ensemble population modeling \citep{bocquet19}.

\begin{figure}
\centering
\resizebox{0.52\textwidth}{!}{
\includegraphics[scale=1]{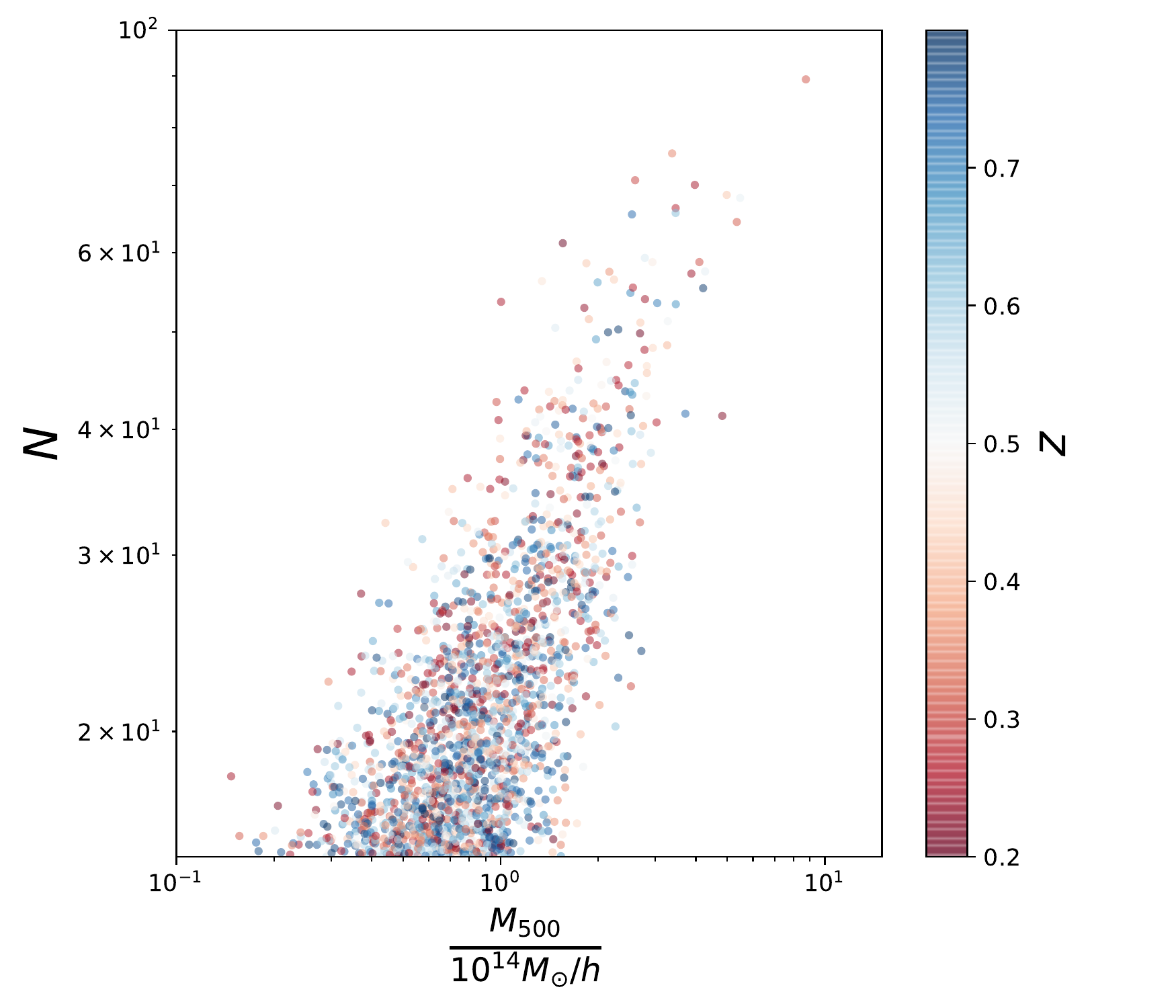}
}
\caption{
The observed richness \rich\ and richness-inferred mass \Mfiveoo\ of individual clusters color-coded by the redshift as shown in the colorbar.
Each cluster mass is sampled from the posterior of the mass distribution marginalizing over the all parameters (see the text in Section~\ref{sec:results}).
}
\label{fig:nmrelation}
\end{figure}

The results are shown in Figure~\ref{fig:nmrelation}, where the
richness-inferred mass of each cluster is plotted with their redshift color-coded. 
For clusters with $N=40$ and $20$, the average fractional uncertainty in \Mfiveoo\ is $\approx32\percent$ and $\approx39\percent$, respectively, after marginalizing over all parameters of the \rtm\ relation.  
If the parameters of the scaling relation are fixed to the best-fit values in Table~\ref{tab:params}---which means that we only consider the measurement uncertainty in richness and the intrinsic scatter---then the average fractional uncertainty of \Mfiveoo\ is reduced to $\approx27\percent$ and $\approx34\percent$ at $N=40$ and $20$, respectively.

\begin{figure}
\centering
\resizebox{0.48\textwidth}{!}{
\includegraphics[scale=1]{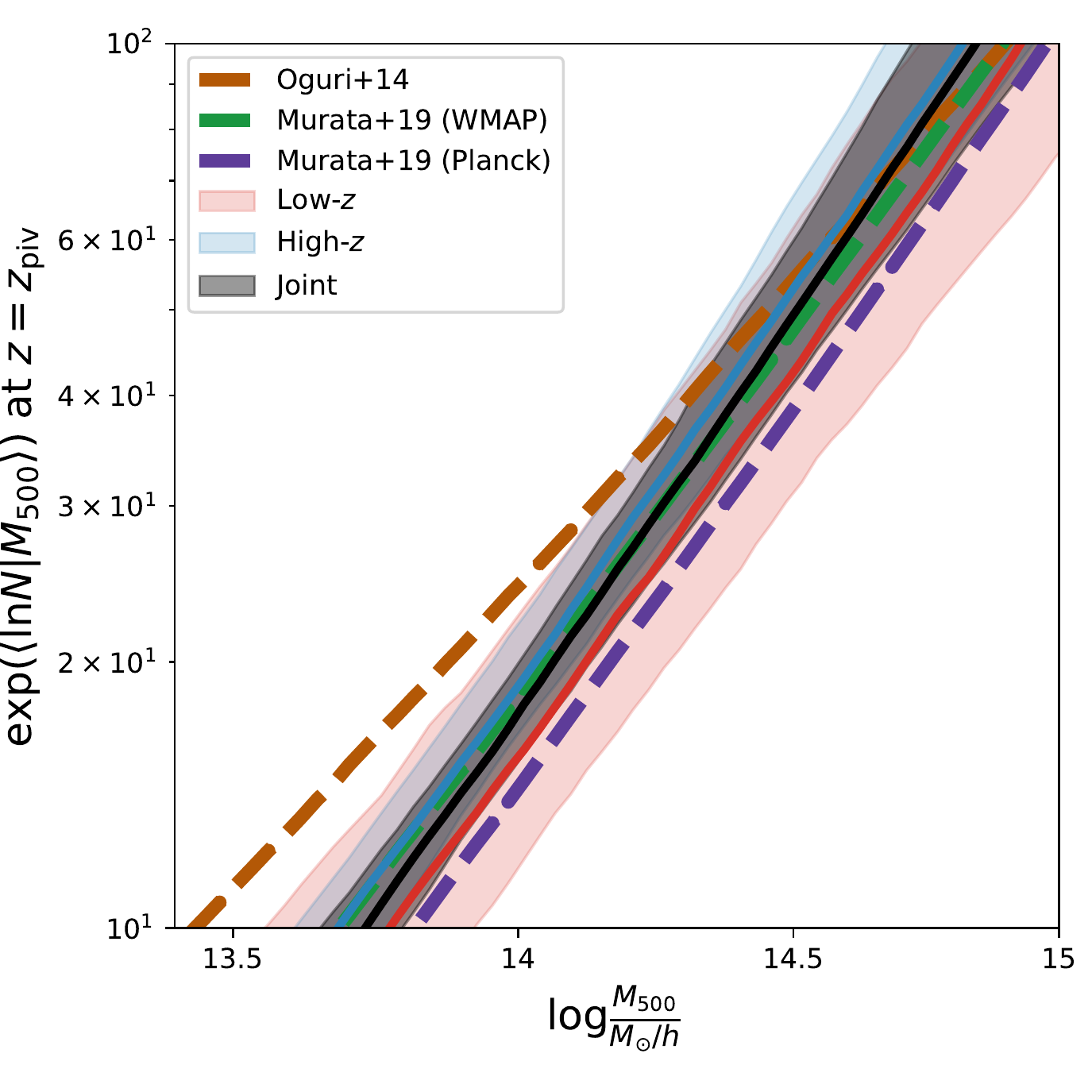}
}
\caption{
The comparison of the \rtm\ relations of the CAMIRA clusters at the pivot redshift \ZPIV.
The results of this work based on the magnification bias using the \lowz, \hghz\ and the combined backgrounds are in red, blue and black, respectively.
The $68\percent$ confidence levels, marginalizing over all other parameters, are all shown by the same color for each result.
The \rtm\ relations of the CAMIRA clusters from \citet{murata19}---using the joint constraints of the weak lensing shearing effect and cluster number counts---are in green and purple under the cosmology with the cosmological parameters fixed to the \textit{WMAP} and \textit{Planck} results, respectively.
Using a sample of CAMIRA clusters constructed by the SDSS data at $0.1<\redshift<0.3$, the resulting \rtm\ relation from \citet{oguri14} is shown in brown.
}
\label{fig:comparison_shear}
\end{figure}

\subsection{Comparison with previous work}
\label{sec:comparison}

Here we compare our derived \rtm\ relation of the CAMIRA sample to that of \cite{murata19} evaluated at the pivot redshift \ZPIV. 
\cite{murata19} independently determined the \rtm\ scaling relation of the CAMIRA sample in a forward-modeling approach by combining the HSC shear measurements and the CAMIRA cluster counts.
Their analysis is based on the HSC first-year data covering a smaller area of 
$\approx140$\,deg$^2$. 
Their cluster sample spans similar ranges of mass and redshift to our study, thus providing an interesting comparison. 

To this end, we first translate the cluster mass definition ($M_\mathrm{200m}$) adopted by \citet{murata19} into \Mfiveoo\ by assuming an NFW density profile with the concentration parameter fixed to $c_{500}\equiv \Rfiveoo/r_\mathrm{s}=2.4$, which is predicted by the \cite{diemer15} $c$--$M$ relation at the pivot mass \MPIV\ and redshift \ZPIV. 
The results are shown in Figure~\ref{fig:comparison_shear}. 
At $\Mfiveoo=10^{14}h^{-1}\Msun$, the corresponding 
normalization of the \rtm\ relation
%richness \rich\ 
of \cite{murata19} is $17.4$ and $13.7$ when using the mass slope of $\Brich=0.83\pm0.03$ and $0.86\pm0.05$ for their reference $WMAP$ and $Planck$ cosmology, respectively. 
Our constraint on \Arich\ yields 
$\Arich=16.93 \pm 7.01$ ($19.35 \pm 3.04$, or $17.72 \pm 2.60$)
according to the \lowz \ (\hghz, or combined) background-based results.  
That is, our magnification-based results are in good agreement with the shear-based ones of \citet{murata19} obtained assuming the $WMAP$ cosmology.   
However, their results assuming the $Planck$ cosmology are discrepant with our results at a level of $1.5\sigma\sim1.8\sigma$.
Much larger samples are thus needed to further investigate the cause of this discrepancy. 
It is worth mentioning that using the weak shear effect alone cannot constrain well the mass and redshift trends either, as shown in \cite{murata19}. 
This suggests that a joint analysis of weak-lensing and additional complementary probes is needed to obtain an adequate constraint on the  \rtm\ scaling relation.

We also compare our results with \cite{oguri14}, who used weak lensing shear to constrain the mass-to-richness relation for the CAMIRA sample at $0.1<\redshift<0.3$ using Sloan Digital Sky Survey (SDSS) data.
A notable difference between our approach and theirs is that \cite{oguri14} constructed the mass-to-richness relation, i.e., $P(M|\rich)$, as opposed to the \rtm\ relation, $P(\rich | M)$, studied in this work. 
Therefore, we need to translate their constraints to $P(\rich | M)$ for a fair comparison. 
Specifically, we use an analytic model described in \cite{evrard14} to translate their $P(M|\rich)$ to $P(\rich|M)$, by accounting for the effect of intrinsic scatter.
In short, this conversion requires the richness distribution and the intrinsic scatter in richness at fixed mass.
Here we rely on the abundance analysis of \cite{murata19} for the former, and we approximate the latter by the best-fit value of $\Drich$ obtained in this work.   
We translate $P(M|\rich)$ of \cite{oguri14} to $P(\rich | M)$ at their pivot richness and redshift, and convert their virial mass definition to
$\Mfiveoo$ by assuming an NFW profile with $c_{500}=2.4$. 
The resulting constraints of \citet{oguri14} are shown by the brown line in Figure~\ref{fig:comparison_shear}. 
As seen in Figure~\ref{fig:comparison_shear}, their results are in good agreement with ours within the errors, suggesting $\Mfiveoo\approx10^{14}h^{-1}\Msun$ at $\rich\approx20$.

%%%%%%%%%%%%%%%%%%%%%%%%%%%%%%%%%%%%%%%%%%
%
% Systematics
%
%%%%%%%%%%%%%%%%%%%%%%%%%%%%%%%%%%%%%%%%%%

\section{Systematics}
\label{sec:sys}

Here we discuss and quantify potential systematics that
could bias our results. 

In this work, the magnification signal has been measured from the projected number density contrast of background source samples with respect to random fields, accounting for the known bias from the cluster contamination and masking effects.  
We have quantified the level of residual bias by repeating the measurements using the ``null-test'' samples. 
It is important to stress that the residual bias correction \fem\ accounts for various systematics that could arise from, for example, the impact of any incorrect assumptions about the $P(\redshift)$-decomposition and the deblending effect in the central cluster regions. 
However, it is likely that the primary source of residual bias is due to systematics in the photometry in the crowded fields, as discussed  in Section~\ref{sec:em}. 
A detailed investigation with extensive simulations and dedicated spectroscopic follow-up observations is required to further clarify the cause of this residual bias.

In this work, we rely on an empirical approach to further quantify the  impact of the residual bias on our final results. 
Specifically, we repeat the whole analysis assuming 
the residual residual bias $\fem(R)$ in our measurements, by replacing equation~(\ref{eq:magni_model}) with the following:
\begin{equation}
\label{eq:correct_em}
\Delta_{\mu}(R) 
= \left(\omega(R) + 1\right)\times \left(\frac{1 - \fcl(R)}{\fem(R)}\right)\fmk(R) - 1 \, .
\end{equation}
That is, we effectively calibrate the magnification signal of the ``lensing-cut'' samples against the ``null-test'' samples, assuming that these two samples share the same systematic errors due to the presence of the clusters.
Using equation~(\ref{eq:correct_em}), we find that the resulting constraints on the parameters  $(\Arich, \Brich, \Crich, \Drich)$ from a joint fit to the combined background population\footnote{We note that we still discard the clusters at  $0.8\leq\redshift<1.1$ for the \lowz\ background.} are  $(21.71 \pm 3.51, 0.86 \pm 0.14, -0.22 \pm 0.72, 0.15 \pm 0.07)$, 
consistent with our fiducial analysis.

Assuming that applying the \fem\ correction can empirically remove the bias present in our magnification measurements of the \lowz\ background for the high-redshift clusters ($0.8\leq\redshift<1.1$), we could further include these measurements into our analysis.
This results in parameter constraints of $(\Arich, \Brich, \Crich, \Drich)=(22.77\pm 3.63, 0.86 \pm 0.14, -0.05 \pm 0.70, 0.14 \pm 0.07)$, 
as also tabulated in Table~\ref{tab:params}. 
That is, the inclusion of the high-redshift clusters at $0.8\leq\redshift<1.1$ in the fitting does not significantly change the resulting scaling relation, if the correction \fem\ is applied. 
Compared to our fiducial analysis, the most significant change in the parameter constraints is in the normalization \Arich, which corresponds to a positive shift in \Arich\ (or a negative offset in the mass \Mfiveoo) of $1.9\sigma$. 
To summarize, the residual bias in the ``null-test'' samples is the most important source of systematics, and the systematic changes in the parameters due to the residual bias correction can be regarded as systematic uncertainties in the present study.

We also estimate the level of potential bias due to systematic errors in the photometric redshift distribution, because a biased estimate of $P(\redshift)$ alters the lensing efficiency $\beta$ and the cluster mass estimate.   
Specifically, we adopt a conservative value for the photo-\redshift\ bias $|(z_{\mathrm{phot}} - z_{\mathrm{spec}})/(1 + z_{\mathrm{phot}})|$ as estimated in \cite{tanaka18}, shift the $P(\redshift)$ distribution by this amount, and estimate the resulting change in the cluster mass given the observed lensing signal.
The photo-\redshift\ bias $|(z_{\mathrm{phot}} - z_{\mathrm{spec}})/(1+z_{\mathrm{phot}})|$ is quantified by \cite{tanaka18} to be $\delta z<0.0005$ and $0.007$ for the \lowz\ and \hghz\ backgrounds, respectively.  
Given a fixed magnification observable $\omega$, the inferred cluster mass is inversely proportional to $\beta$ (see equation~(\ref{eq:kappa_def})). 
As a result, a shift of $\delta z<0.0005$ ($0.007$) in $P(\redshift)$ leads to a change in mass of $\lesssim1\percent$ ($\approx3\percent$) for a cluster at $\redshift=1.1$ when using the \lowz\ (\hghz) background. 
In addition, this change decreases with decreasing cluster redshift.
That is, this effect is negligible in this work.

In this work, we have ignored systematic uncertainties due to  
the projection effect \citep{costanzi19}, 
the triaxiality of clusters \citep{chiu18b}, 
the orientation bias \citep{dietrich14}, 
the presence of ICL \citep{gruen19}, 
the miscentering of clusters \citep{ford13}, 
the halo-modelling systematics \citep{dietrich19}, 
and the intrinsic scatter in the concentration-to-mass and the halo
bias-to-mass relations. 
\cite{costanzi19} suggest that the ignorance of the projection effect, especially for the optically selected clusters, could lead to a systematic bias in the mass calibration at a level of $5\percent$.
With the high-quality lensing data, the difference in terms of cluster mass estimates is $\approx2\percent$ between the modelling with and without triaxiality \citep{chiu18b}.
The stacked weak-lensing mass of optically selected clusters is suggested to be biased at a level of $<6\percent$ if ignoring the orientation bias \citep{dietrich14}.
\cite{ford13} estimated that the miscentering effect leads to a systematic uncertainty at a level of $4.5\percent$ for a magnification-based analysis.
The inaccuracy of the assumed functional form (i.e., the NFW model) for halos is quantified to be $2.8\percent$ using a suite of N-body simulations \citep{dietrich19}.
The impact  from the scatter in the concentration-to-mass relation on cluster mass is estimated to be at a level of $1.5\percent$ \citep{dietrich19}.
Give the current size of the statistical uncertainty, we expect the
systematic errors above to be subdominant in this work.

%%%%%%%%%%%%%%%%%%%%%%%%%%%%%%%%%%%%%%%%%%
%
% Conclusions
%
%%%%%%%%%%%%%%%%%%%%%%%%%%%%%%%%%%%%%%%%%%

\section{Summary and Conclusions}
\label{sec:conclusions}

In this paper, we have detected the density enhancement of background source galaxies due to lensing magnification around a sample of 3029 galaxy clusters with richness $\rich>15$ at $0.2\leq\redshift<1.1$,  which are optically selected in the Subaru HSC survey over the area of $\approx380$\,deg$^2$.  
The lensing magnification effect is measured by using two distinct populations of \lowz\ and \hghz\ background galaxies at mean redshifts of $\left\langle\redshift\right\rangle\approx1.1$ and $\left\langle\redshift\right\rangle\approx1.4$, respectively.   
We carefully correct for contamination by cluster members and the masking effect due to bright objects.
Our magnification measurements are found to be uncontaminated according to validation tests based on the ``null-test'' samples, for which the net magnification effect is expected to vanish. 
The magnification bias effect has been detected at a significance level of $4.59\sigma$, $8.17\sigma$, and $9.51\sigma$ for the \lowz, \hghz\, and combined background sample, respectively. 

With the constraints from lensing magnification alone, we use a forward-modelling approach to constrain the underlying \rtm\ scaling relation, which we characterize by a power-law relation described by four parameters:
the normalization \Arich,
the mass-trend parameter \Brich,
the redshift-trend parameter \Crich,
and the log-normal intrinsic scatter \Drich.
In this work, we can only constrain the normalization \Arich, with an aid of informative priors on \Brich, \Crich\, and \Drich. 
The \Arich\ parameter is constrained by the \lowz\ and \hghz\ background populations separately as 
$16.93 \pm 7.01$ and $19.35\pm3.04$, 
respectively, at the pivot mass $\MPIV=10^{14}h^{-1}\Msun$.
Given this consistency between the measurements with the \lowz\ and \hghz\ backgrounds, we combine them together to derive an improved joint
constraint. 
Using the combined background population, the resulting best-fit parameter \Arich\ is 
$17.72 \pm 2.60$ with a $15\percent$ uncertainty at the pivot mass.

With the derived \rtm\ scaling relation, we infer lensing-calibrated mass estimates for individual clusters based on the measured richness in
an ensemble manner.  
At the characteristic richness of $N\approx40$ ($\approx20$), the fractional uncertainty in mass constrained by the combined background
population  is  $\approx32\percent$ ($\approx39\percent$) when marginalizing over all
parameters.

We find that the most significant source of systematic errors in this
work is the residual bias found with the ``null-test'' samples.
The source of residual bias is likely coming from systematics in the photometry in the crowded fields,  which could result in biased estimates of colors and thus
photo-\redshift,  misidentification between cluster members and background galaxies, or the combination of both.
A further investigation with intensive image simulations and spectroscopic follow-up observations of these clusters is needed to understand the cause of residual systematics.
With the residual correction applied, the resulting normalization \Arich\ from the combined background constraints increases at a significance level of $1.9\sigma$, suggesting a smaller mass scale at fixed richness compared to our fiducial results.

We compared our magnification-based constraints on the \rtm\ relation for the CAMIRA sample with the shear-based results of \cite{murata19} and \cite{oguri14}, finding that our normalization is in good agreement with their results. 
The comparison with \cite{murata19}, who studied a subset of our CAMIRA sample, shows that our magnification-based \rtm\ relation is statistically consistent with their shear-based results obtained assuming the $WMAP$ cosmology.
However, the normalization of our magnification-based \rtm\ relation is higher at 
$\approx 1.5\sigma\sim1.8\sigma$ than their shear-based one obtained assuming the $Planck$ cosmology.
This discrepancy arises because \citet{murata19} included the cluster abundance as a constraint into their shear-based weak-lensing analysis.   
A possible explanation for this discrepancy includes the redshift-dependent intrinsic scatter of richness at fixed mass, as
proposed in \cite{murata19}. 

In this work, we have demonstrated that the cluster mass scale can be robustly  inferred from lensing magnification alone, using well-calibrated deep multi-band photometry across the sky.
Lensing magnification provides an independent way to calibrate an observable-to-mass scaling relation via gravitational lensing, and serves as a unique mass probe complementary to weak lensing shear, CMB lensing, and galaxy kinematics.
One of the most important tasks to carry out such a magnification analysis is to select a clean sample of background galaxies at sufficiently high redshift, with a very low level of contamination.
A possible improvement in future is to include near-Infrared  data to select higher-redshift background populations \citep{schrabback18}.
With the whole HSC-Wide coverage ($\approx 1400$\,deg$^2$), we expect that the normalization of the \rtm\ relation could be determined to better than $8\percent$ precision, which will allow us to place competitive cosmological constraints using galaxy clusters.

%%%%%%%%%%%%%%%%%%%%%%%%%%%%%%%%%%%%%%%%%%
%
% ACKNOWLEDGE
%
%%%%%%%%%%%%%%%%%%%%%%%%%%%%%%%%%%%%%%%%%%

\section*{Acknowledgments}
\label{sec:acknowledgements}

We thank the anonymous referee for the constructive comments that lead to the improvement of this paper.
I-Non Chiu acknowledges fruitful discussions during an international
meeting entitled ``Cosmology with size and flux magnification'' held at
the International Space Science Institute (ISSI), 
which was organized by Alan Heavens and Hendrik Hildebrandt.
This work is supported by the Ministry of Science and Technology of
Taiwan (grant MOST 106-2628-M-001-003-MY3) and by Academia Sinica (grant
AS-IA-107-M01). 
This work is supported in part by World Premier International Research Center Initiative (WPI Initiative), MEXT, Japan, and JSPS KAKENHI Grant Number JP15H05892 and JP18K03693.

The Hyper Suprime-Cam (HSC) collaboration includes the astronomical communities of Japan and Taiwan, and Princeton University.  The HSC instrumentation and software were developed by the National Astronomical Observatory of Japan (NAOJ), the Kavli Institute for the Physics and Mathematics of the Universe (Kavli IPMU), the University of Tokyo, the High Energy Accelerator Research Organization (KEK), the Academia Sinica Institute for Astronomy and Astrophysics in Taiwan (ASIAA), and Princeton University.  Funding was contributed by the FIRST program from Japanese Cabinet Office, the Ministry of Education, Culture, Sports, Science and Technology (MEXT), the Japan Society for the Promotion of Science (JSPS),  Japan Science and Technology Agency  (JST),  the Toray Science  Foundation, NAOJ, Kavli IPMU, KEK, ASIAA,  and Princeton University.

The Pan-STARRS1 Surveys (PS1) have been made possible through contributions of the Institute for Astronomy, the University of Hawaii, the Pan-STARRS Project Office, the Max-Planck Society and its participating institutes, the Max Planck Institute for Astronomy, Heidelberg and the Max Planck Institute for Extraterrestrial Physics, Garching, The Johns Hopkins University, Durham University, the University of Edinburgh, Queen's University Belfast, the Harvard-Smithsonian Center for Astrophysics, the Las Cumbres Observatory Global Telescope Network Incorporated, the National Central University of Taiwan, the Space Telescope Science Institute, the National Aeronautics and Space Administration under Grant No. NNX08AR22G issued through the Planetary Science Division of the NASA Science Mission Directorate, the National Science Foundation under Grant No. AST-1238877, the University of Maryland, and Eotvos Lorand University (ELTE).

This paper makes use of software developed for the Large Synoptic Survey Telescope. We thank the LSST Project for making their code available as free software at \url{http://dm.lsst.org}.
This work is based on data collected at the Subaru Telescope and retrieved from the HSC data archive system, which is operated by Subaru Telescope and Astronomy Data Center at National Astronomical Observatory of Japan.

This work made use of the IPython package \citep{PER-GRA:2007}, SciPy \citep{jones_scipy_2001}, TOPCAT, an interactive graphical viewer and editor for tabular data \citep{2005ASPC..347...29T}, matplotlib, a Python library for publication quality graphics \citep{Hunter:2007}, Astropy, a community-developed core Python package for Astronomy \citep{2013A&A...558A..33A}, NumPy \citep{van2011numpy}. 
This work made use of \citet{bocquet16b} for producing Figure~\ref{fig:gtc}.
This work made use of \texttt{colossus} \citep[][]{diemer17} for calculating the cosmology-dependent quantities.

%%%%%%%%%%%%%%%%%%%%%%%%%%%%%%%%%%%%%%%%%%
%
% Appendix
%
%%%%%%%%%%%%%%%%%%%%%%%%%%%%%%%%%%%%%%%%%%

\appendix

\section{Catalog Query}
\label{sec:catalog_query}

In Table~\ref{tab:sql} we summarize the key \texttt{sql} statements to
query the HSC database.

\begin{table*}
\centering
\begin{tabular}{cccc}
\hline\hline
\multicolumn{3}{c}{Flags} & Meaning \\[3pt]
\hline
       \texttt{isprimary}                                                 &\texttt{is} &\texttt{True}    & Select unique detection only \\
       \texttt{g[rizy]\_pixelflags\_edge}                         &\texttt{is}  &\texttt{False}  & Discard the objects that are outside the usable exposure region \\
       \texttt{g[rizy]\_pixelflags\_interpolatedcenter}     &\texttt{is}  &\texttt{False}  & Discard the objects whose centers are flagged as interpolated \\
       \texttt{g[rizy]\_pixelflags\_crcenter}                    &\texttt{is}  &\texttt{False}  & Discard the objects whose centers are flagged as cosmic ray \\
       \texttt{g[rizy]\_cmodel\_flag}                             &\texttt{is}  &\texttt{False}  & Discard the objects whose final cmodel fits are failed \\
       \texttt{i[z]\_pixelflags\_bright\_object}               &\texttt{is}  &\texttt{False}  & Discard the objects whose footprints contain a pixel flagged as star-masked \\
       \texttt{i\_extendedness\_value}                          &$==$          &1                    & Select galaxies only \\
       \texttt{g[r]countinputs}                                       &$\geq$       &4                    & Select full-depth in $g$ and $r$-band \\
       \texttt{i[zy]countinputs}                                      &$\geq$       &6                    & Select full-depth in $i$, $z$, and $Y$-band \\
       \texttt{z\_cmodel\_mag} $-$ \texttt{a\_z}          &$<$           &26                  & Only use the objects with $z$-band magnitude brighter than $26$~mag \\
\hline\hline
\end{tabular}
\caption{
The \texttt{sql} query to construct the galaxy photometry
 catalog used in this work.
}
\label{tab:sql}
\end{table*}

\section{The effect of the bright-end cut on the slope \slope}
\label{sec:brightendcut_slope}

Applying an additional bright-end cut to a flux-limited sample of
background galaxies modifies
the signal of magnification bias, because magnified source galaxies near the
bright-end cut will be removed from the observed sample.  
As a result, the net change of the source counts due to lensing
magnification includes the contribution from the brighter cut as well as
from the fainter cut. 
Therefore, when applying a brighter magnitude cut, one must account for
this effect in the calculation of the effective count slope.
In what follows, we provide the derivation of the effective count slope
\slope\ to be used for a magnification-bias analysis with a bright-end cut.

We begin by considering the standard analysis of a flux-limited sample
of source galaxies defined with a faint-end magnitude cut $m_\mathrm{Faint}$.
The ratio of the magnified and unlensed source counts,
$N(<m_\mathrm{Faint})$ and
$N_0(<m_\mathrm{Faint})$,
is given by 
\begin{equation}
\label{eq:faint_signal}
\frac{N(<m_{\mathrm{Faint}})}{N_0(<m_{\mathrm{Faint}})} =
\mu^{\alpha(m_{\mathrm{Faint}}) - 1} \approx 1 +
2\left[\alpha(m_{\mathrm{Faint}}) - 1\right]\kappa \, , 
\end{equation}
where we make use of the approximation in the linear regime
($\mu\approx 1+2\kappa$) and define the logarithmic count slope at
$m_\mathrm{Faint}$ as
\begin{equation}
\label{eq:alpha_faint}
\alpha(m_{\mathrm{Faint}}) \equiv 2.5\frac{d\log N(<m_{\mathrm{Faint}})}{dm} \, .
\end{equation}
We further split $N(<m_{\mathrm{Faint}})$ into two
components,
$N(<m_{\mathrm{Bright}})$ and
$N(m_{\mathrm{Bright}}<m<m_{\mathrm{Faint}})$ with
$m_{\mathrm{Bright}}$ the brighter magnitude cut, as
\begin{eqnarray}
\label{eq:split_alpha}
\frac{N(<m_{\mathrm{Faint}})}{N_0(<m_{\mathrm{Faint}})} 
&= 
&\frac{N(<m_{\mathrm{Bright}}) + N(m_{\mathrm{Bright}}<m<m_{\mathrm{Faint}})}
{N_0(<m_{\mathrm{Faint}})}  \nonumber  \\
&= 
&\frac{N(<m_{\mathrm{Bright}})}
{N_0(<m_{\mathrm{Faint}})} + 
\frac{N(m_{\mathrm{Bright}}<m<m_{\mathrm{Faint}})}
{N_0(<m_{\mathrm{Faint}})} \nonumber  \\
&= 
&\frac{N_0(<m_{\mathrm{Bright}})}
{N_0(<m_{\mathrm{Faint}})} 
\frac{N(<m_{\mathrm{Bright}})}
{N_0(<m_{\mathrm{Bright}})}
 + \nonumber  \\
& & \frac{N_0(m_{\mathrm{Bright}}<m<m_{\mathrm{Faint}})}
{N_0(<m_{\mathrm{Faint}})}
\frac{N(m_{\mathrm{Bright}}<m<m_{\mathrm{Faint}})}
{N_0(m_{\mathrm{Bright}}<m<m_{\mathrm{Faint}})} \nonumber  \\
&=
&f_{\mathrm{Bright}} 
\frac{N(<m_{\mathrm{Bright}})}
{N_0(<m_{\mathrm{Bright}})}
+ \nonumber  \\
& &  (1 - f_{\mathrm{Bright}})
\frac{N(m_{\mathrm{Bright}}<m<m_{\mathrm{Faint}})}
{N_0(m_{\mathrm{Bright}}<m<m_{\mathrm{Faint}})}  \, ,
\end{eqnarray}
in which we have defined the quantity $f_{\mathrm{Bright}}$ by
\begin{equation}
\label{eq:fmis}
f_{\mathrm{Bright}} =  \frac{N_0(<m_{\mathrm{Bright}})}
{N_0(<m_{\mathrm{Faint}})}  \, .
\end{equation}

In equation~(\ref{eq:split_alpha}), the term
$N(<m_{\mathrm{Bright}})/N_0(<m_{\mathrm{Bright}})$ corresponds to the
magnification bas signal evaluated for a flux-limited sample at the bright-end cut 
$m_{\mathrm{Bright}}$. 
The second term
$N(m_{\mathrm{Bright}}<m<m_{\mathrm{Faint}})/N_0(m_{\mathrm{Bright}}<m<m_{\mathrm{Faint}})$
is the magnification bias signal expected for the background sample defined in
the magnitude range
$m_{\mathrm{Bright}}<m<m_{\mathrm{Faint}}$.
These two terms are weighted by the respective fractions of the
cumulative source counts ($f_\mathrm{Bright}$ and $1-f_\mathrm{Bright}$)
in each magnitude range. 
We can rewrite equation~(\ref{eq:split_alpha}) in terms of $\kappa$ as
\begin{eqnarray}
\label{eq:split_kappa}
1 + 2\left[\alpha(m_{\mathrm{Faint}}) - 1\right]\kappa &=
&f_{\mathrm{Bright}} 
\left(
1 + 2\left[\alpha(m_{\mathrm{Bright}}) - 1\right]\kappa
\right) \nonumber  \\
&+
&(1 - f_{\mathrm{Bright}})
\left[
1 + 2\left(\alpha_{\mathrm{eff}} - 1\right)\kappa
\right] \, ,
\end{eqnarray}
with $\alpha_{\mathrm{eff}}$ the effective count slope,
\begin{equation}
\begin{aligned}
 \label{eq:split_kappa}
\alpha_{\mathrm{eff}} &= \frac{
\alpha(m_{\mathrm{Faint}}) - f_{\mathrm{Bright}} \alpha(m_{\mathrm{Bright}})
}{
1 - f_{\mathrm{Bright}}
} \, .
 \end{aligned}
\end{equation}
That is, the bright-end cut removes the cumulative contribution of
magnification bias up to $m_{\mathrm{Bright}}$ and results
in a modified slope ${\slope}_{\mathrm{eff}}$ that is required for
interpreting and modelling
the magnification measurements in the magnitude range $m_{\mathrm{Bright}}<m<m_{\mathrm{Faint}}$. 
Moreover, $\alpha_{\mathrm{eff}}$ depends on not only the slopes at
the bright- and faint-end cuts, but also their relative fraction
$f_{\mathrm{Bright}}$.

In typical observations of the magnification bias based on deep
multi-band imaging \citep[e.g.,][]{chiu16b,umetsu14}, we would
require $m_{\mathrm{Bright}}$ to be at least $\approx3$ magnitudes
brighter than $m_{\mathrm{Faint}}$. Then, we have
$f_{\mathrm{Bright}}\rightarrow0$ and
\[
\frac{N(m_{\mathrm{Bright}}<m<m_{\mathrm{Faint}})}
{N_0(m_{\mathrm{Bright}}<m<m_{\mathrm{Faint}})} \rightarrow
\frac{N(<m_{\mathrm{Faint}})}{N_0(<m_{\mathrm{Faint}})} \, ,
\]
for which the correction due to the bright-end cut is negligible.
However, this effect needs to be taken into account when
$m_{\mathrm{Bright}}\approx m_{\mathrm{Faint}}$.
We thus do not impose any bright-end cut in this work when defining 
background samples for magnification-bias measurements.

\bibliographystyle{mn2e}
\bibliography{literature}

\end{document}